\documentclass[12pt]{article}

\usepackage{newtxtext,newtxmath}
\usepackage{graphicx}

\usepackage[letterpaper,margin=1in]{geometry}

\renewenvironment{abstract}
	{\quotation}
	{\endquotation}

\date{}

\makeatletter
\renewcommand{\fnum@figure}{\textbf{Figure \thefigure}}
\renewcommand{\fnum@table}{\textbf{Table \thetable}}
\makeatother

\usepackage{scite}

\usepackage{url}

\def\scititle{
	Atmospheric retrieval evidence for water isotopologue HDO on exoplanet WASP-39~b
}
\title{\bfseries \boldmath \scititle}

\author{
  Fabian Grübel$^{1,2,\ast}$,
  Karan Molaverdikhani$^{2,3}$,
  Barbara Ercolano$^{1,2,3}$,
  Katy L. Chubb$^{4}$,\and
  Paola Caselli$^{3,2}$,
  Tommaso Grassi$^{3,2}$,
  Rosa Arenales-Lope$^{1,2}$,
  and Dwaipayan Dubey$^{1,2}$\and
	\small$^{1}$ Universitäts-Sternwarte, Fakultät für Physik, Ludwig-Maximilians-Universität München, München, Germany.\and\
	\small$^{2}$Exzellenzcluster `Origins’, Garching, Germany.\and
    \small$^{3}$Max-Planck-Institut für Extraterrestrische Physik, Garching, Germany.\and
    \small$^{4}$HH Wills Physics Laboratory, University of Bristol, Bristol, UK.\and
	\small$^\ast$Corresponding author. Email: ercolano@usm.lmu.de
}

\begin{document} 
%TC:ignore

% Insert the title and author list
\maketitle

% Abstract, in bold
% There are strict length limits, and not all formats have abstracts.
% Consult the journal instructions to authors for details.
% Do not cite any references in the abstract.
\begin{abstract} \bfseries \boldmath
% Start with one or two sentences of background
Hydrogen-isotopologues are commonly used to trace the chemical processing and origin of hydrogen-bearing species throughout the Universe, however, their abundance remains unconstrained for extrasolar planets.
% Then summarise the results of your observations, experiments, simulations etc.
Here, we report atmospheric retrieval evidence for the deuterated water molecule HDO in an exoplanet atmosphere, retrieved from James Webb Space Telescope transmission spectra of the hot Jupiter WASP-39~b, resulting in a deuterium-to-hydrogen ratio in water of $(4.0^{+1.3}_{-1.1})\times10^{-3}$. The inferred value is substantially higher than those measured for the Solar System gas giants and overlaps numerically with values reported for some protostellar and inner Solar System environments.
% End with a statement of your main conclusions
This enrichment may reflect either inherited water-rich material accreted beyond the snow line or isotopic processing in the observable atmosphere through transport, photochemistry, and subsequent escape. 
\end{abstract}

% Nor is it indented
\noindent
%TC:endignore
The detection of isotopologues and the subsequent isotopic ratios such as the deuterium-to-hydrogen (D/H) ratio are a powerful tool to identify chemical processes and trace the origin of abundant molecules in the Universe (e.g.~\cite{2009ASPC..417...23L,2011Natur.478..218H,2012AARv..20...56C,2014prpl.conf..859C,2014Sci...345.1590C,2015Sci...347A.387A,2021MNRAS.500.4901D,2021NatAs...5..943A,2023Natur.615..227T}), including those essential for the emergence of life. While the total D/H ratio in the Universe has only moderately changed since the primordial nucleosynthesis (e.g.~\cite{2006ApJ...647.1106L,2025AA...694A.174F}), certain deuterated molecules such as semi-heavy water (HDO) are able to accumulate locally due to their difference in mass and chemical properties (e.g.~\cite{2013Icar..226..256Y,2017RSPTA.37550390H}). In these environments, processes such as isotopic exchange reactions are responsible for distinct D/H ratios in different molecular species. Accordingly, large variations have been observed in and outside of the Solar System. While the gas giants show similarities to the local interstellar medium (ISM), the ratios of the inner planets, comets and asteroids differ by orders of magnitude\cite{2021NatAs...5..943A,2011Natur.478..218H,2017RSPTA.37550390H,2024PNAS..12101638M}. These deviations provide insight into the specific chemistry, formation history, and molecular origin of the observed objects. However, the determination of planetary D/H ratios has remained restricted to the Solar System.

Thousands of extrasolar planets have been confirmed to date, the majority of which deviate significantly from the planets in the Solar System and reside in stellar systems with inherently different architectures (for an updated list, refer to the NASA Exoplanet Archive\footnote{https://exoplanetarchive.ipac.caltech.edu}). Among the most commonly observed are hot Jupiters, giant, high-temperature exoplanets that orbit their stars on the order of a few days. Although the detection of deuterated molecules on exoplanets has been considered in previous studies (e.g.~\cite{2019AJ....158...26L, 2019ApJ...882L..29M,2019AA...622A.139M,2023ASPC..534.1075N}), it has only become observationally feasible with the launch of the James Webb Space Telescope (JWST). 

The hot Jupiter WASP-39~b is widely regarded as a benchmark exoplanet owing to its role in a JWST early release science\cite{2016PASP..128i4401S,2018PASP..130k4402B}  programme\footnote{ERS 1366, PIs: N. M. Batalha, J. L. Bean and K. B. Stevenson}. JWST has observed WASP-39~b across the near and mid-infrared using four different instruments: the Near-InfraRed Imager and Slitless Spectrograph (NIRISS)\cite{2023Natur.614..670F}, the Near-InfraRed Camera (NIRCam)\cite{2023Natur.614..653A}, the Near-InfraRed Spectrograph (NIRSpec) with the G395H grating\cite{2023Natur.614..664A} and the PRISM mode\cite{2023Natur.614..659R}, as well as the Mid-Infrared Instrument (MIRI)\cite{2024Natur.626..979P}. This leaves us with the most detailed transmission spectrum of any exoplanet to date. Nevertheless, our understanding of its atmospheric chemistry remains incomplete (e.g.~\cite{2024AA...687A.110L,2025arXiv250407823M}). Measuring the water D/H ratio of this planet would offer invaluable constraints on its chemical environment and formation pathway\cite{2018ARAA..56..175D}.

\subsection*{Atmospheric Retrieval Framework}
In this work, we constrain the HDO abundance on WASP-39 b by applying the well-established atmospheric retrieval framework (e.g.~\cite{2018haex.bookE.104M}) based on the radiative transfer code petitRADTRANS\cite{2019AA...627A..67M, Nasedkin2024,2024JOSS....9.7028B}. The retrieval  generates synthetic spectra from a predefined model atmosphere and iteratively compares them to the data and optimises them using Bayesian inference to estimate the  statistical preference for and abundances of molecular species. The majority of the transmission spectra used in this work originate from the most recent release of reduced spectra\cite{2024NatAs...8.1008C,carter_2024_10161743}, where four of the five datasets were jointly fit. The authors find that after the consideration of additional offsets, this analysis yields the highest consistency between the observations yet. The MIRI data were obtained from a separate study\footnote{ERS 2783, PI: D. Powell} where we adopted the Eureka! reduction\cite{powell_2023_10055845,2024Natur.626..979P} and removed any wavelengths above 10~$\mu m$, a region subjected to large scatter/uncertainties and possible circumstellar debris disk contamination (e.g.~\cite{2024ApJ...969L..19F,2025arXiv250407823M}).

We conduct retrievals using freely varying, vertically constant molecular abundances,  independent of chemical equilibrium processes, also known as "free chemistry retrievals". We chose the temperature profile described in~\cite{2010AA...520A..27G} (hereafter also referred to as a Guillot profile), including a combination of different cloud parameters (see Materials and Methods). In our extended analysis, we additionally tested an isothermal temperature profile to further explore the effect of model dependency on our detection, the results of which are shown in the Supplementary Materials. Our model atmosphere included both H$_2$O and HDO as tracers of the D/H ratio, alongside several major infrared absorbers and continuum opacities (refer to Materials and Methods). The forward modelling of the transmission spectra was performed at medium spectral resolution (R = 1,000) using the correlated-k treatment.

In the main analysis, the fully joint JWST dataset was considered. According to the data source, we included fixed vertical offsets for NIRISS, NIRCam, NIRSpec PRISM, and NIRSpec G395H and a freely varying offset for MIRI. The datasets were each considered at a resolution of R~$\approx$~100, and the retrievals were conducted with and without HDO present in the molecular inventory of the model atmosphere (model 1 and 0, respectively). To further ensure that the adopted data properties do not strongly bias our results, we carried out additional tests on several joint and individual datasets at R $\approx$ 100 and higher resolution, the results of which are given in the Supplementary Materials. As described in ref.~\cite{2024NatAs...8.1008C}, the NIRSpec PRISM spectrum is partially affected by saturation, and thus we exclude all data points flagged with saturation. Note that our science case is at first order unaffected by the saturation since the main absorption feature of HDO is located outside of this regime (refer to Supplementary Fig.~\ref{fig:HDO_opacities}) where other datasets provide more information. Finally, the Bayesian evidence ($\ln(Z)$) was estimated using the nested sampling algorithm PyMultiNest\cite{2014AA...564A.125B}, allowing us to compute the Bayes factor ($B_{10}$) and the detection significance $n_\sigma$.

\subsection*{Results}
Fig.~\ref{fig:Spectrum} shows the observed data including offsets alongside the highest likelihood spectrum from the HDO-inclusive retrieval, with residuals given in units of parts-per-million (ppm). Additionally, the contributions of each molecule, except for the CIA and Rayleigh scattering species, are presented. The influence of HDO is most significant from 3 to 4~$\mu m$, a region covered by 3 out of the 5 instrument modes. Table~\ref{tab:Main_Results} provides details on the retrieved results, including the Bayesian evidence $\ln(Z)$ and reduced $\chi^2$, medians and uncertainties for all free parameters, the derived HDO/H$_2$O and D/H ratios, the resulting Bayes factor $B_{10}$ in support of HDO, as well as the corresponding frequentist detection significance ($n_\sigma$-value). Hence, the joint dataset retrieval shows strong evidence for the presence of HDO in our model comparison, with the Bayes Factor falling within the "decisive evidence" regime of Jeffreys' scale\cite{Jeffreys1939}. In the frequentist interpretation, the detection significance corresponds to a value of 4.81$\sigma$. The calculation and interpretation of the $n_\sigma$-values are further discussed in the Materials and Methods. To investigate the impact of model dependency on this detection, as mentioned, we additionally considered an isothermal temperature profile in our extended analysis. The results yield a value within the decisive evidence regime i.e. around 5.37 for $\ln(B_{10})$ (3.54$\sigma$) and a similar D/H ratio, despite providing an overall much worse fit to the data ($\Delta\ln{Z}\approx30$, $\Delta\chi^2/\mathrm{dof}\approx0.2$, refer to Supplementary  Tables~\ref{tab:Results1} and~\ref{tab:Results2}). 

The retrieved posterior (see Fig.~\ref{fig:Post_Temp}) shows a Gaussian-like distribution for the HDO log volume-mixing-ratio (VMR) with a median value of around -4.1. The cut-off in the H$_2$O posterior is marking the upper limit of the prior at 10$^{-1}$ (internal mass-mixing-ratio), which is enforced for the atmosphere to remain in the hydrogen-helium-dominated regime. Due to the ratio being governed by the relative spectral feature strengths, this physics-informed limit does not have a major impact on the retrieved D/H ratio, which is calculated a posteriori from the molecule pairs of each posterior sample. Tests including wider priors produced D/H ratios in agreement with the presented results. Fig.~\ref{fig:Post_Temp} additionally shows the retrieved Guillot\cite{2010AA...520A..27G} pressure-temperature profile as well as the spectrally weighted contribution of each pressure layer. Accordingly, we see a strong gradient at high pressures (P $>10^{-2}$ bar) that results in extreme temperatures at the lower limit of the modelling regime. However, these altitudes do not contribute significantly to the transmission spectrum and temperatures are thus able to reach these values. In the pressure range that has the strongest impact on the spectrum (around $10^{-7}<$ P $<10^{-3}$ bar) we observe a temperature inversion, a phenomenon that has been reported in recent analyses of WASP-39~b (e.g.~\cite{2025arXiv250407823M}). In the highest layers of the atmosphere, the temperature settles at values around 1,000 - 1,300 Kelvin, which is approximately equal to the equilibrium temperature of the exoplanet\cite{2011AA...531A..40F}. Finally, we show the isothermal solution from the extended retrieval analysis, which roughly agrees with the average Guillot profile temperature across the spectrally contributing layers.

As presented in the Supplementary Materials, the support for HDO is not uniform across the individual datasets. NIRISS and MIRI do not show a preference for HDO, as expected from their wavelength coverage (see Supplementary Tables~\ref{tab:Results1} and~\ref{tab:Results2}). The strongest single-instrument support comes from NIRSpec PRISM. NIRSpec G395H shows a posterior peak at similar HDO abundances, but does not yield a formal Bayes factor preference on its own, and NIRCam provides only an upper limit.

We attribute this behaviour to the limited wavelength ranges of both instrument modes, which may not allow them to constrain molecular abundances of other line species and therefore HDO adequately. Furthermore, their datasets are inherently different to the one from NIRSpec PRISM, showing increased scatter and larger uncertainties at the benefit of a higher spectral resolution. This elevated resolution appears insufficient to break the degeneracies caused by the narrow wavelength coverage, as an additional test using a finer binning scale demonstrates (see Materials and Methods). Follow-up observations with NIRSpec G395H could thus improve the detection significance. To test whether NIRSpec PRISM alone is driving this detection, we additionally conducted retrievals on a joint dataset excluding NIRSpec PRISM. These results are weakly supportive, with only a minor preference for HDO, which is encouraging but not independently decisive. Note that these complementary retrievals were carried out using an isothermal temperature profile.

The tests on the individual instruments indicate that a strong bias introduced by the fixed offsets is not expected. Nevertheless, to further support this claim, we conducted an additional set of retrievals. After determining the relative data offsets between NIRSpec PRISM, NIRISS, NIRCam, and NIRSpec G395H using two complementary methods, we carried out eight retrievals, including both fixed offsets at the newly determined medians and free offsets with Gaussian priors, with and without HDO, over the wavelength ranges 3--5 $\mu$m and 2--5 $\mu$m. The results (see Supplementary Material) show a statistical preference for the HDO model in the 2--5 $\mu$m retrievals, while no HDO signal is recovered in the 3--5 $\mu$m range. These findings support the conclusion that the main results are not primarily driven by the adopted fixed offsets, while also illustrating the importance of a broader spectral coverage.

We want to further highlight that in multiple retrievals, signals of other trace species are weakened or removed when including HDO (see Table~\ref{tab:Main_Results} and Supplementary Fig.~\ref{fig:Full_Post2}). This is especially prominent in the case of H$_2$S, which is constrained in the HDO-exclusive retrieval, but not found when HDO is added.  Photochemical modelling\cite{2023Natur.617..483T} has shown that, due to the detection of SO$_2$, H$_2$S is expected to be present in the atmosphere. However, it populates higher pressure regimes and its abundance is expected to decrease at pressures lower than approximately $10^{-3}$ bar. Since this limit marks the beginning of the main opacity contribution in our retrievals, the results seem consistent with this trend.

\subsection*{Discussion}
\subsubsection*{Comparison to Known Astrophysical Objects/Environments}
Using the retrieved HDO and H$_2$O abundances as tracers, we derive a HDO/H$_2$O ratio of $(8.0^{+2.7}_{-2.2})\times10^{-3}$, corresponding to a D/H value in water of $(4.0^{+1.3}_{-1.1})\times10^{-3}$ (factor 1/2). In Fig.~\ref{fig:Ratio} we compare these values to measurements from the Solar System\cite{2011Natur.478..218H,2021NatAs...5..943A,2024PNAS..12101638M,2025ApJ...986L..19S}, protostellar envelopes and protoplanetary disks\cite{2025ApJ...986L..19S}, the ISM\cite{2025ApJ...986L..19S}, and a brown dwarf\cite{2024ApJ...977L..49R}. The D/H ratios for the gas giants, as well as the ISM and protosolar values, were determined from hydrogen gas measurements, while the brown dwarf study used methane as a tracer. The remaining values were inferred from the HDO abundances of the specific object (refer to Materials and Methods).

Fig.~\ref{fig:Ratio} indicates that the D/H ratio we find is significantly elevated compared to the Solar System gas giants, the objects which might be expected to be the closest local analogues to WASP 39 b. It also exceeds the values measured in comets and chondrites. Instead, the inferred atmospheric water D/H on WASP-39~b overlaps  numerically with values reported for some inner Solar System and protostellar environments. Note that, these comparisons are illustrative in magnitude only, and the physical environments and enrichment pathways need not be the same. In the following sections, we give a brief overview on how the retrieved value could be explained.

\subsubsection*{Atmospheric Escape}
We regard atmospheric processing in the observable atmosphere as a plausible pathway. The high D/H ratios  in the inner Solar System planets are generated by a combination of photolysis and atmospheric escape, where low gravity and/or high temperatures allow the heavier HDO molecules to accumulate relative to H$_2$O (e.g.~\cite{2008JGRE..113.0B22F,2021NatAs...5..943A}). Specifically on Venus, sulfur compounds have been proposed to enhance atmospheric escape via vertical transport mechanisms\cite{2024PNAS..12101638M}. 

The atmospheric properties of rocky planets are in numerous ways different from gas giants. However, WASP-39~b itself is a highly inflated, low-gravity, and high-temperature exoplanet. Recent studies have detected photochemically produced sulfur-bearing molecules such as SO$_2$ in its photosphere\cite{2023Natur.614..659R,2024Natur.626..979P}, indicating that photochemical processes are likely significant. Therefore, this raises the question of whether atmospheric escape mechanisms may enhance the D/H ratio to observed values. 

Numerical modelling of low-temperature sub-Neptunes has shown that a D/H enrichment is expected for atmospheres with significant atmospheric escape\cite{2024ApJ...967..139C}. Accordingly, deuterium-enhanced planets tend to be methane-depleted with more prominent contributions from CO and CO$_2$ which is consistent with previous analyses of WASP-39 b and the results presented in this work. However, at high temperatures, small differences in the D/H ratio may equilibrate rapidly due to strong atmospheric mixing. Furthermore, the planet's bulk H$_2$ gas envelope could balance out the relative loss rates and therefore only evolved exoplanets with reduced atmospheric masses are expected to exhibit higher values.

Nevertheless, a recent study reports a tentative signal of HDO in a high-resolution observation of the hot Jupiter HIP 67522 b\cite{2026arXiv260203498L} with a statistical significance of approximately 2$\sigma$. Their analysis is probing higher altitudes of a hot Jupiter atmosphere and predicts a D/H ratio that is significantly elevated even compared to WASP-39 b. This suggests the presence of a possible vertical gradient in the D/H ratio, a state that could hint at strong, ongoing atmospheric escape. Detailed photochemical modelling is necessary in the future to test this possible D/H enrichment pathway for the considered objects.

\subsubsection*{Giant Planet Formation}
In protostellar environments, various scenarios have been proposed to explain the observed D/H ratios. At low temperatures, isotopic exchange rates are able to significantly increase the deuteriation of molecular species (e.g.~\cite{2017RSPTA.37550390H}). While this process could be substantial within protostellar envelopes themselves, the elevated D/H ratios are presumably inherited from the conditions in the prestellar cloud phase (e.g.~\cite{2003AA...403L..37C,2004AA...418.1035W,2012AARv..20...56C,2014prpl.conf..859C,2014Sci...345.1590C,2022ApJ...929...13C,2023Natur.615..227T}) and represent a chemical link to these environments. Ultimately, this enhancement is thought to be responsible for the high HDO content in protoplanetary disks, asteroids, and comets (e.g.~\cite{2023ASPC..534.1075N}). 

Gas giants primarily accrete the bulk H$_2$ gas from the disks, which is not significantly affected by the processes described above due to the large hydrogen abundance, resulting in expected bulk D/H values similar to those of the ISM and protosolar nebula. At high temperatures, isotopic exchange reaches an equilibrium where molecules retain similar D/H ratios (e.g.~\cite{2013Icar..226..256Y}). This is consistent with the brown dwarf value\cite{2024ApJ...977L..49R}, which shows a D/H ratio in methane that is comparable to the ISM value. Due to its high temperature, the same may apply to WASP-39~b's atmosphere, however, we observed much greater D/H ratios in water. 

A scenario in which a substantial planetary D/H ratio observed in water is inherited from disks would require the accretion of a significant fraction of the planet's mass from water-rich protostellar material, which sublimates in the high-temperature environment and drives the atmospheric deuterium enrichment. If the measured atmospheric water D/H reflects inherited material, the planet must have formed beyond the "snow line" of its protoplanetary disk\cite{2018ARAA..56..175D} and later migrated inwards towards its current position, which has been proposed in previous analyses of WASP-39 b\cite{2023Natur.614..670F,2024A&A...685A..64K}. Fig.~\ref{fig:Ratio} includes observations of protostellar ices from this region that show a degree of deuterium-enrichment consistent with the D/H ratio inferred from WASP-39 b (e.g.~\cite{2023Natur.615..227T}). Nevertheless, the key difficulty is that such a signal must survive dilution by the H$_2$/He envelope and isotopic re-equilibration at high temperature. Determining whether this scenario could produce the observed values requires detailed chemical modelling, which is beyond the scope of this analysis.

\subsubsection*{Analysis Limitations}
In any case, the inferred abundances are affected by limitations inherent to the atmospheric retrieval approach and transit spectroscopy. In particular, the present analysis adopts a 1D free-chemistry framework with vertically constant abundances and fixed inter-instrument offsets for the non-MIRI datasets. Specific model choices can lead to notable variations in absolute abundances, and detection significances (refer to Supplementary Tables~\ref{tab:Results1} and~\ref{tab:Results2}). The choice of the analysis design is an essential feature of Bayesian inference methods, and our results are set in the framework we created. However, regardless of this context, the values of molecular/atomic ratios and therefore the reliability of our findings benefit from the necessity of constant relative spectral feature strengths in the data fitting process of the atmospheric retrievals. Furthermore, low-resolution transit spectra probe only a narrow region of the exoplanet's atmosphere and therefore the measured D/H ratios may not represent the object's bulk chemistry. Independent constraints with complementary methods, such as high-resolution spectroscopy, are essential to further test the accuracy of our results. Nevertheless, even if the measured D/H ratio is only characteristic of the observable part of the atmosphere rather than the bulk planet, it would allow for a unique access to atmospheric processes.

\subsection*{Conclusion}
This work presents retrieval evidence for a deuterium-bearing molecule in an exoplanetary atmosphere. Future studies should aim to detect multiple D/H ratio tracers in an exoplanet population study to constrain their atmospheric chemistry and formation history more robustly. HDO, however, currently remains the most accessible tracer due to its distinct spectral features, and its detection may even be possible on temperate rocky exoplanets in the near future, where it could serve as a potential habitability marker (e.g.~\cite{2019AJ....158...26L,2025ApJ...979..137C}). 

% If your text is very short you might need to uncomment the following line to avoid
% layout problems with the figures and tables.
%\newpage

%%%%%%%%%%%%%%%% MAIN TEXT FIGURES %%%%%%%%%%%%%%%

%TC:ignore
\begin{figure}
    \centering
    \includegraphics[width=\textwidth]{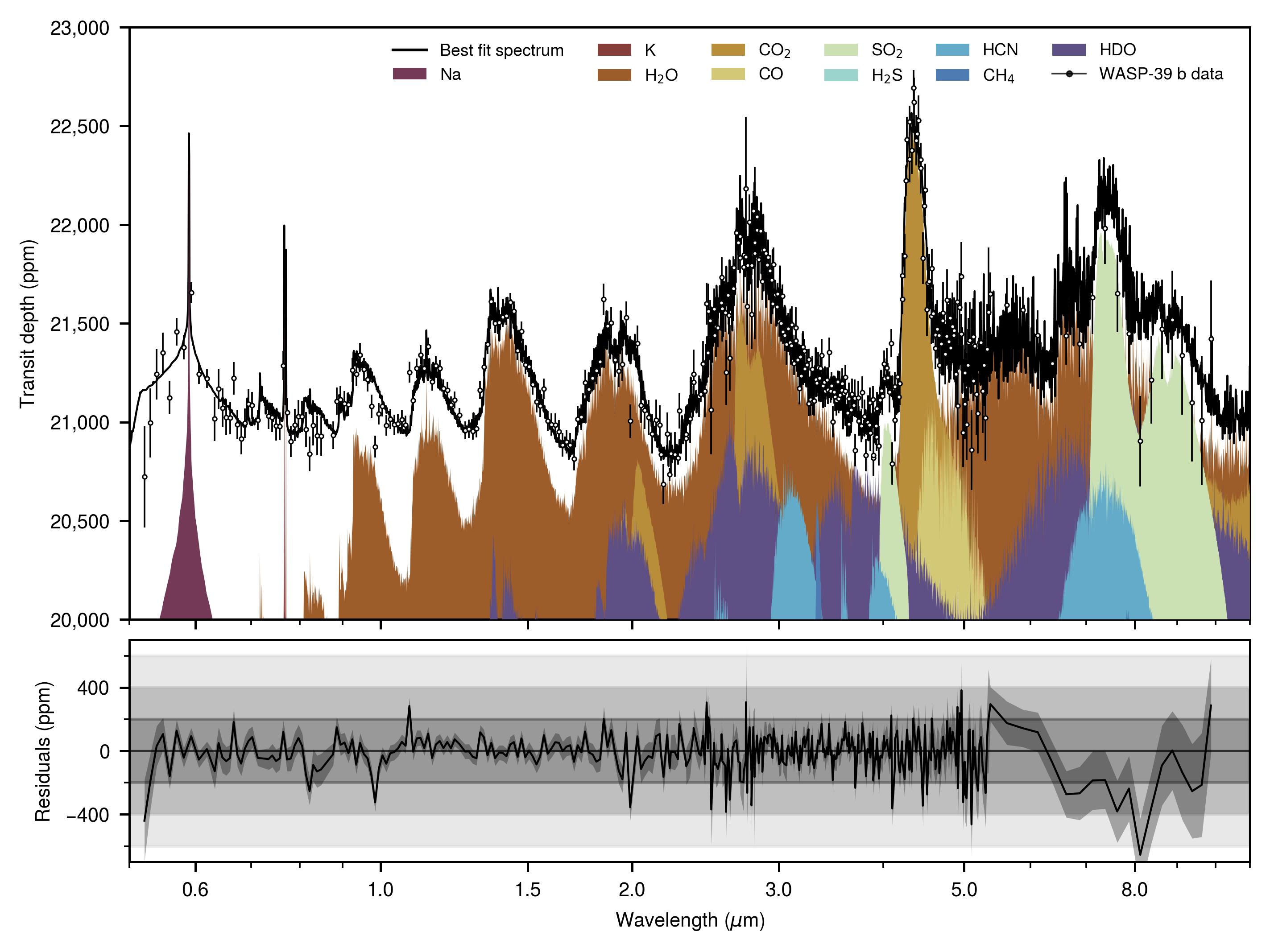}
    \setlength{\unitlength}{1pt}
    \begin{picture}(0,0)
        \put(-180,355){\fontsize{8}{9.6}\selectfont\textbf{A}}
        \put(-180,123){\fontsize{8}{9.6}\selectfont\textbf{B}}
    \end{picture}
    \vspace{-2em}
    \caption{\textbf{Observational data and retrieved transmission spectrum.} \textbf{A}, Retrieved highest likelihood transmission spectrum including semi-heavy water (HDO) alongside the full dataset (R~$\approx$~100) with uncertainties in units of transit depth. The respective contributions from the line species at their specific abundance level are shown. \textbf{B}, Residuals between the data and the binned model spectrum in parts-per-million (ppm) with shaded regions indicating $\pm$200, $\pm$400, and $\pm$600 ppm. Data uncertainties are shown as transparent bands.}
    \label{fig:Spectrum}
\end{figure}
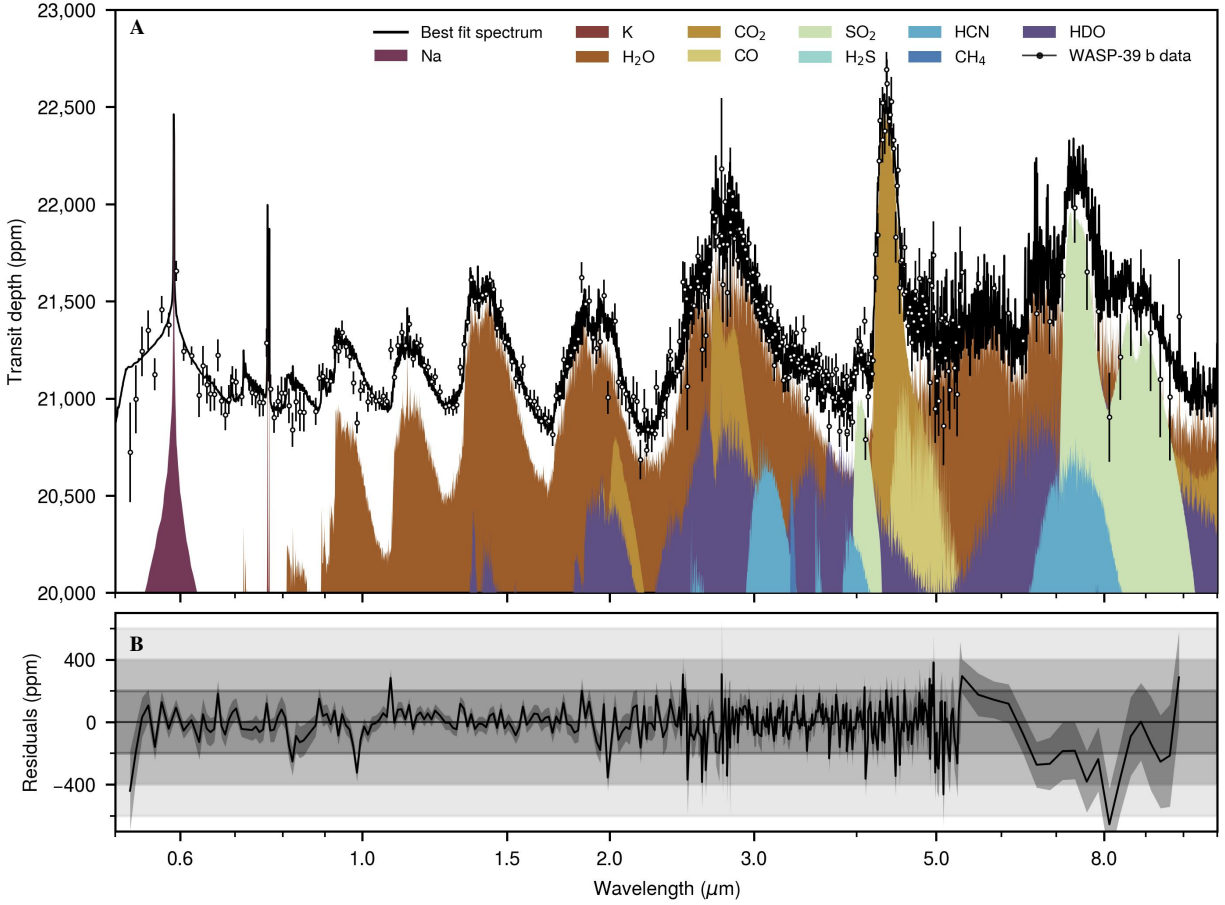

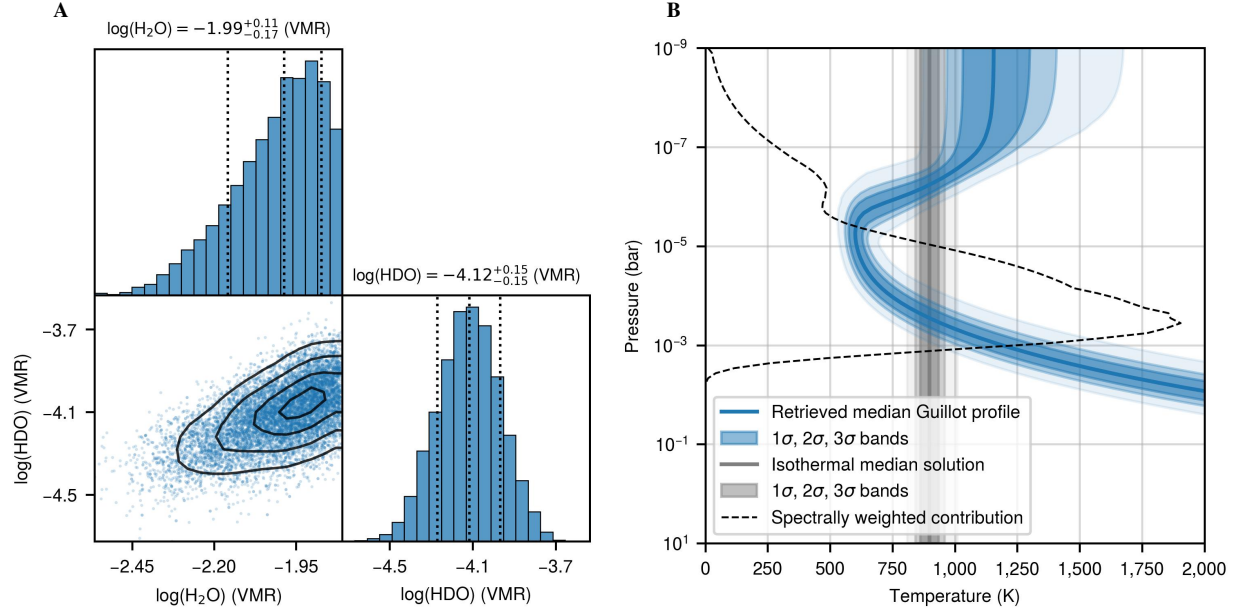
\begin{figure}
  \centering
  \makebox[\textwidth][c]{%
    \begin{minipage}[t]{0.49\textwidth}
  \centering
      \includegraphics[width=\linewidth]{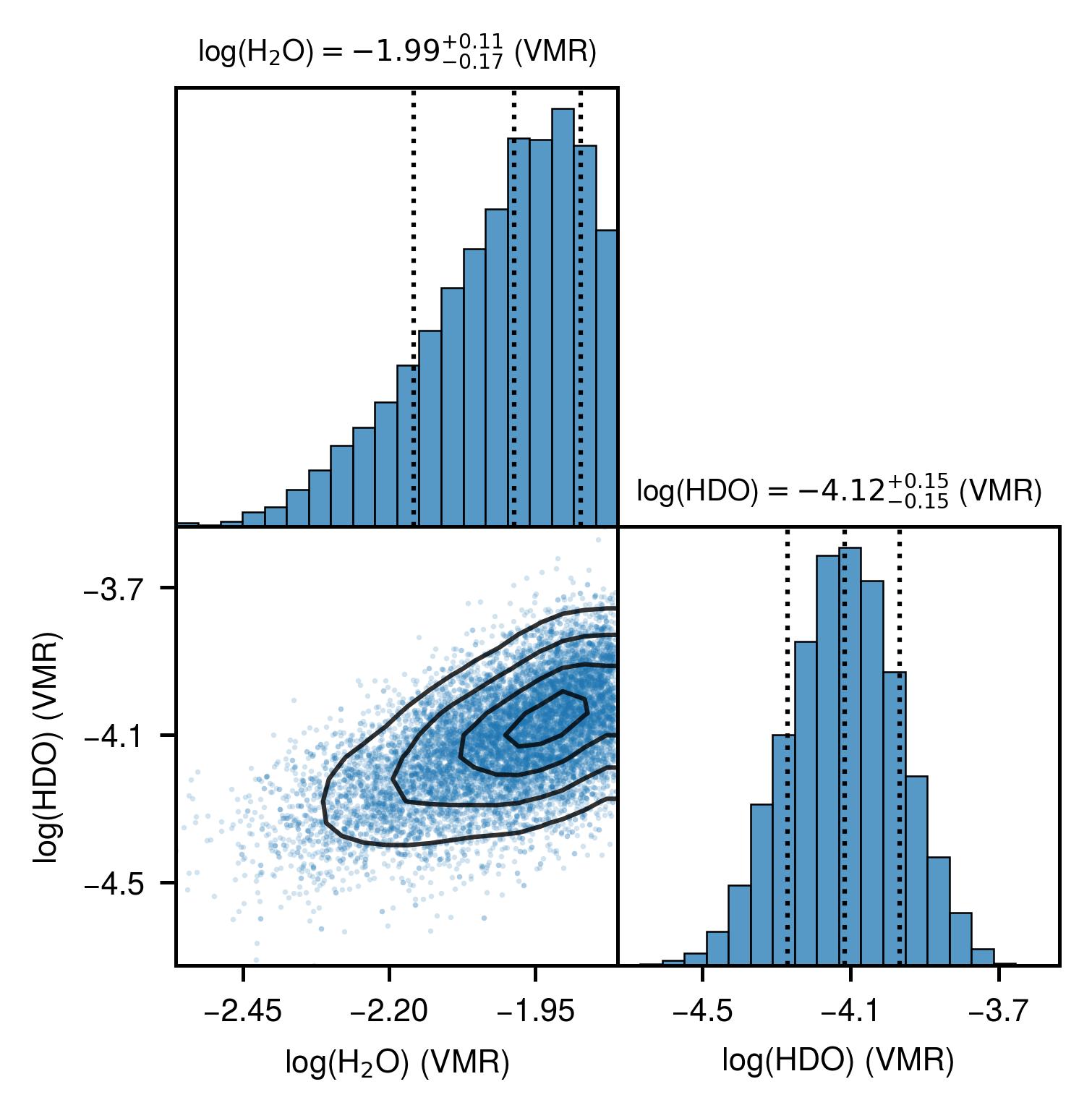}
      \setlength{\unitlength}{1pt}
      \begin{picture}(0,0)
        \put(-92,250){\fontsize{8}{9.6}\selectfont\textbf{A}}
      \end{picture}
      \label{fig:Post_Temp1}
    \end{minipage}\hfill
    \begin{minipage}[t]{0.51\textwidth}
      \centering
      \includegraphics[width=\linewidth]{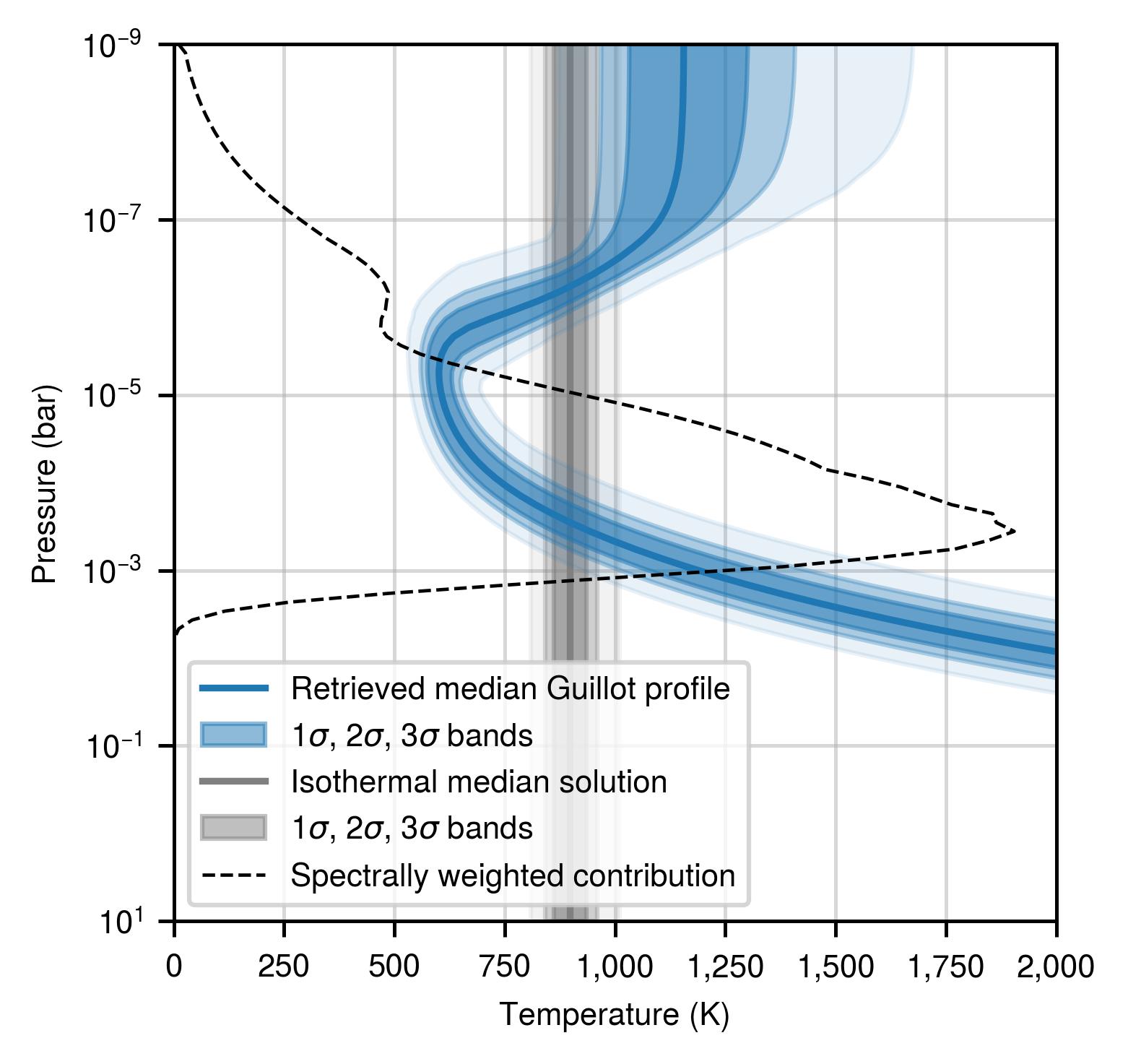}
      \setlength{\unitlength}{1pt}
      \begin{picture}(0,0)
        \put(-96,250){\fontsize{8}{9.6}\selectfont\textbf{B}}
      \end{picture}
      \label{fig:Post_Temp2}
    \end{minipage}%
  }
  \vspace{-2em}
  \caption{\textbf{Retrieved molecular posterior distribution and pressure-temperature profile.} \textbf{A},  Posterior distribution of the HDO and H$_2$O abundances on WASP-39~b from the joint instrument retrieval. The distributions are given as marginalised 1D histograms. Dotted lines indicate the medians and 16th and 84th percentiles. Additionally, the 2D samples are displayed, including reference contours. \textbf{B}, Retrieved Guillot\cite{2010AA...520A..27G} temperature profile. Shown are the median, as well as the 1$\sigma$, 2$\sigma$, and 3$\sigma$ ranges. For reference, the retrieved isothermal profile from the joint complementary retrieval is given. Additionally, the dashed line represents the spectrally weighted contribution of each pressure layer. Note: log indicates the base 10 logarithms.}
  \label{fig:Post_Temp}
\end{figure}

\begin{figure}
    \centering
    \includegraphics[width=1\textwidth]{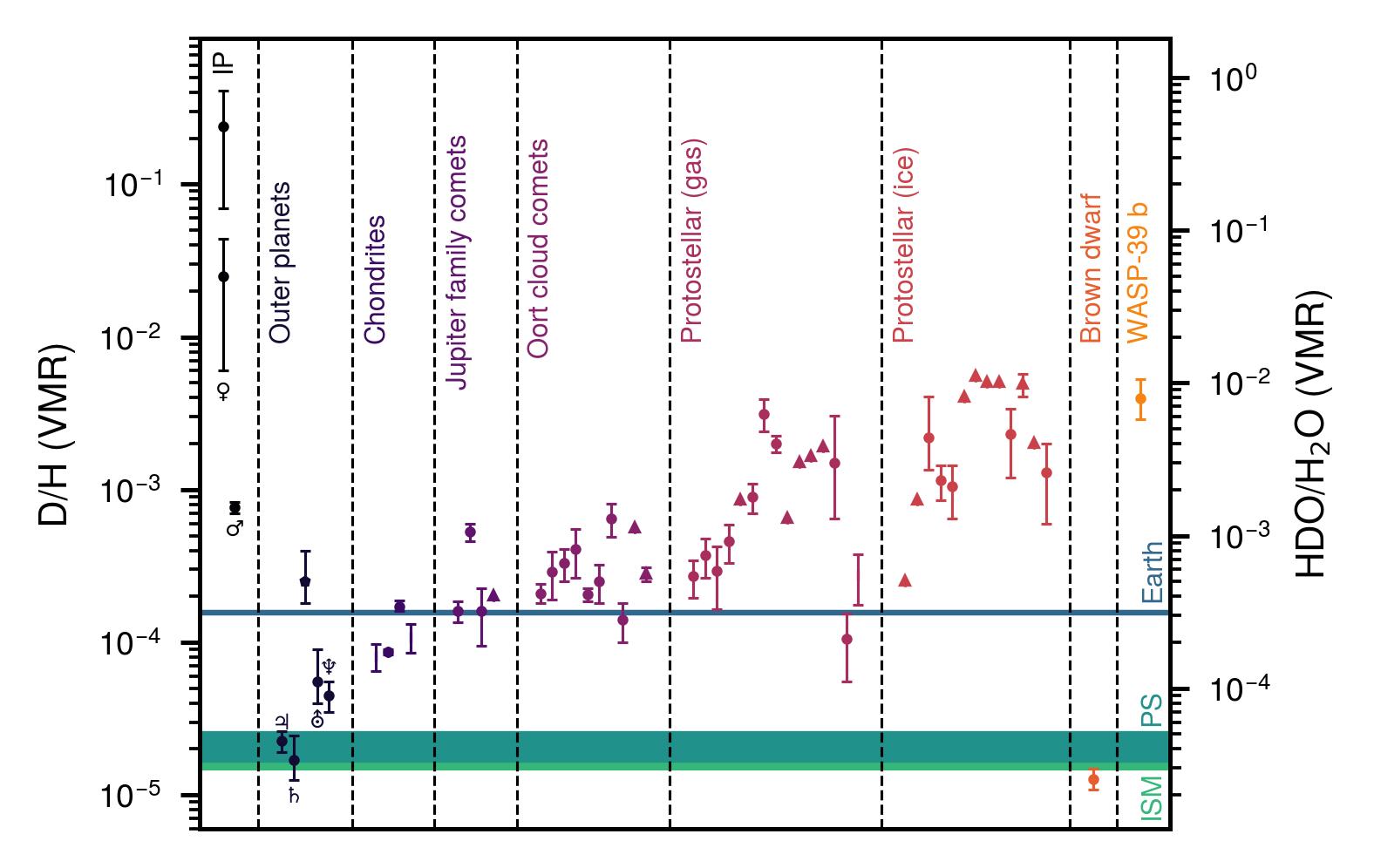}
    \caption{\textbf{D/H and HDO/H$_2$O ratio comparison between this work and known solar and extrasolar values.} Reference D/H ratios including uncertainties are taken from (summaries of) five previous works\cite{2011Natur.478..218H,2021NatAs...5..943A,2024PNAS..12101638M,2024ApJ...977L..49R,2025ApJ...986L..19S}. The individual sources for each object are listed in Supplementary Tables~\ref{tab:Source1} and~\ref{tab:Source2}. The HDO/H$_2$O ratio axis assumes a conversion factor of 1/2. The interstellar medium (ISM), protosolar (PS), and Earth ocean D/H ratios are shown as horizontal bars. The uncertainties on the WASP-39~b data point follow from the derived 16th and 84th percentiles of the paired H$_2$O and HDO posterior distributions. Note: IP group (inner planets) includes Venus (at 70km and 108 km altitudes), and Mars; outer planets group includes Jupiter, Saturn, Enceladus (pentagon), Uranus, and Neptune. Planets are additionally represented by their respective astronomical symbols, and triangles indicate upper limits.}
    \label{fig:Ratio}
\end{figure}

%%%%%%%%%%%%%%%% MAIN TEXT TABLES %%%%%%%%%%%%%%%
\begin{table}
\centering
\small
  \caption{\textbf{WASP-39~b atmospheric retrieval results.} Bayesian evidence $\ln(Z)$, $\chi^2$/dof (degrees of freedom), median and 16th/84th percentiles for all retrieved parameters (for details, refer to main text), $\mathrm{HDO/H_2O}$ and $\mathrm{D/H}$ ratios, Bayes factor $\ln(B_{10})$ and detection significance $n_\sigma$ (one- and two-sided, refer to Materials and Methods). Model 0 and Model 1 represent the HDO-exclusive and HDO-inclusive retrievals. Note: log and ln indicate base 10 and the natural logarithms, respectively. Units: ppm: parts-per-million, VMR: volume-mixing-ratio. 
  }
   \vspace{1em}
\begin{minipage}{0.55\textwidth}  % adjust width as you like (e.g. 0.7, 0.8)
  \label{tab:Main_Results}
  \setlength{\tabcolsep}{4.5pt}   % default is 6pt
  \renewcommand{\arraystretch}{0.9}
  \begin{tabular}{lcr}
  \hline
  \textbf{Results} & \textbf{Model 0} & \textbf{Model 1} \\
  \hline
  $\ln(Z)$&$2780.69\pm0.23$&$2791.14\pm0.23$\\
  $\chi^2/\mathrm{dof}$&1.78&1.74\\
  \hline
  $\log P_\mathrm{ref}$ (bar) & $-2.54^{+0.14}_{-0.14}$ & $-2.58^{+0.13}_{-0.15}$ \\
  $\mathrm{\Delta MIRI}\,(\mathrm{ppm})$ & $-297^{+38}_{-38}$ & $-324^{+38}_{-36}$ \\
  \hline
  $T_\mathrm{equ}\,(\mathrm{K})$ & $562^{+92}_{-113}$ & $554^{+79}_{-93}$ \\
  $T_\mathrm{int}\,(\mathrm{K})$ & $595^{+97}_{-108}$ & $584^{+81}_{-86}$ \\
  $\log \kappa_\mathrm{IR}$ (cm$^2$/g) & $0.93^{+0.37}_{-0.31}$ & $1.00^{+0.31}_{-0.28}$ \\
  $\log \gamma$ & $1.41^{+0.45}_{-0.35}$ & $1.62^{+0.40}_{-0.32}$ \\
  \hline
  $\log P_\mathrm{cloud}$ (bar) & $-0.8^{+1.2}_{-1.2}$ & $-0.8^{+1.2}_{-1.2}$ \\
  $\log \kappa_{350\,\mathrm{nm}}$ (cm$^2$/g) & $-0.11^{+0.22}_{-0.23}$ & $-0.11^{+0.21}_{-0.22}$ \\
  $\alpha_\mathrm{PL}$ & $-2.60^{+0.30}_{-0.32}$ & $-2.50^{+0.25}_{-0.28}$ \\
  $f_\mathrm{cloud}$ & $0.682^{+0.037}_{-0.040}$ & $0.695^{+0.033}_{-0.036}$ \\
  \hline
  $\log(\mathrm{Na})$ (VMR) & $-4.50^{+0.23}_{-0.26}$ & $-4.56^{+0.27}_{-0.31}$ \\
  $\log(\mathrm{K})$ (VMR) & $-7.04^{+0.46}_{-0.66}$ & $-7.14^{+0.50}_{-0.64}$ \\
  $\log(\mathrm{H_2O})$ (VMR) & $-1.90^{+0.067}_{-0.127}$ & $-1.99^{+0.11}_{-0.17}$ \\
  $\log(\mathrm{CO_2})$ (VMR) & $-3.35^{+0.11}_{-0.13}$ & $-3.37^{+0.13}_{-0.17}$ \\
  $\log(\mathrm{CO})$ (VMR) & $-2.048^{+0.031}_{-0.063}$ & $-2.077^{+0.044}_{-0.088}$ \\
  $\log(\mathrm{SO_2})$ (VMR) & $-4.95^{+0.13}_{-0.14}$ & $-4.95^{+0.14}_{-0.14}$ \\
  $\log(\mathrm{H_2S})$ (VMR) & $-3.22^{+0.13}_{-0.14}$ & $-8.7^{+2.9}_{-2.8}$ \\
  $\log(\mathrm{HCN})$ (VMR) & $-4.84^{+0.20}_{-0.24}$ & $-4.76^{+0.19}_{-0.22}$ \\
  $\log(\mathrm{CH_4})$ (VMR) & $-5.73^{+0.14}_{-0.15}$ & $-6.3^{+0.29}_{-1.22}$ \\
  $\log(\mathrm{HDO})$ (VMR) & -- & $-4.12^{+0.15}_{-0.15}$ \\
  \hline
  $\mathrm{HDO/H_2O}$ (VMR) & -- &$(8.0^{+2.7}_{-2.2})\times10^{-3}$ \\
  $\mathrm{D/H}$ (VMR) & -- &$(4.0^{+1.3}_{-1.1})\times10^{-3}$ \\
  \hline
  $\ln(B_{10})$                 & \multicolumn{2}{c}{$10.45\pm0.33$} \\
  $n_\sigma$                    & \multicolumn{2}{c}{4.81 (4.95)} \\
  \hline
  \end{tabular}

\end{minipage}
\end{table}

%%%%%%%%%%%%%%%% REFERENCES %%%%%%%%%%%%%%%

\clearpage % Clear all remaining figures and tables then start a new page

% The list of references goes after the main text and before the acknowledgements
% When preparing an initial submission, we recommend you use BibTeX, like this:
%
\bibliography{Bib} % for a file named s_template.bib

@STRING{jan = "January"}

@STRING{feb = "February"}

@STRING{mar = "March"}

@STRING{apr = "April"}

@STRING{may = "May"}

@STRING{jun = "June"}

@STRING{jul = "July"}

@STRING{aug = "August"}

@STRING{sep = "September"}

@STRING{oct = "October"}

@STRING{nov = "November"}

@STRING{dec = "December"}

@STRING{jgr        = "J. Geophys. Res."}

@STRING{ptrsla     = "Philos. Trans. R. Soc. Lond. A"}

@STRING{pnas       = "Proc. Natl. Acad. Sci. USA"}

@STRING{mnras      = "Mon. Not. R. Astron. Soc."}

@STRING{apj        = "Astrophys. J."}

@STRING{apjl       = "Astrophys. J. Lett."}

@STRING{aap        = "Astron. Astrophys."}

@STRING{nature     = "Nature"}

@STRING{nat        = "Nature"}

@STRING{science    = "Science"}

@STRING{icarus     = "Icarus"}

@STRING{aj         = "Astron. J."}

@STRING{planss     = "Planet. Space Sci."}

@STRING{apjs       = "Astrophys. J. Suppl. Ser."}

@STRING{jqsrt      = "J. Quant. Spectrosc. Radiat. Transfer"}

@STRING{pasp       = "Publ. Astron. Soc. Pac."}

@STRING{araa       = "Annu. Rev. Astron. Astrophys."}

@STRING{aapr       = "Astron. Astrophys. Rev."}

@STRING{jms        = "J. Mol. Spectrosc."}

@STRING{pps        = "Proc. Phys. Soc."}

@STRING{ojap       = "Open J. Astrophys."}

@STRING{cse        = "Comput. Sci. Eng."}

@STRING{joss       = "J. Open Source Softw."}

@STRING{jpdc       = "J. Parallel Distrib. Comput."}

@STRING{awr        = "Adv. Water Resour."}

@STRING{amstat     = "Am. Stat."}

@STRING{pac        = "Pure Appl. Chem."}

@STRING{ApJS       = "Astrophys. J. Suppl. Ser."}

@STRING{APJS       = "Astrophys. J. Suppl. Ser."}

@STRING{APJ        = "Astrophys. J."}

@STRING{APJL       = "Astrophys. J. Lett."}

@STRING{ApJ        = "Astrophys. J."}

@STRING{ApJL       = "Astrophys. J. Lett."}

@INPROCEEDINGS{2023ASPC..534.1075N,
       author = {{Nomura}, H. and {Furuya}, K. and {Cordiner}, M.~A. and {Charnley}, S.~B. and {Alexander}, C.~M. O'D. and {Nixon}, C.~A. and {Guzman}, V.~V. and {Yurimoto}, H. and {Tsukagoshi}, T. and {Iino}, T.},
        title = "{The Isotopic Links from Planet Forming Regions to the Solar System}",
    booktitle = {Protostars and Planets VII},
         year = 2023,
       editor = {{Inutsuka}, S. and {Aikawa}, Y. and {Muto}, T. and {Tomida}, K. and {Tamura}, M.},
       series = {Astronomical Society of the Pacific Conference Series},
       volume = {534},
        month = jul,
        pages = {1075},
       adsurl = {https://ui.adsabs.harvard.edu/abs/2023ASPC..534.1075N}
}

@ARTICLE{2023Natur.615..227T,
       author = {{Tobin}, John J. and {van't Hoff}, Merel L.~R. and {Leemker}, Margot and {van Dishoeck}, Ewine F. and {Paneque-Carre{\~n}o}, Teresa and {Furuya}, Kenji and {Harsono}, Daniel and {Persson}, Magnus V. and {Cleeves}, L. Ilsedore and {Sheehan}, Patrick D. and {Cieza}, Lucas},
        title = "{Deuterium-enriched water ties planet-forming disks to comets and protostars}",
      journal = nat,
         year = 2023,
        month = mar,
       volume = {615},
       number = {7951},
        pages = {227-230},
          doi = {10.1038/s41586-022-05676-z},
       adsurl = {https://ui.adsabs.harvard.edu/abs/2023Natur.615..227T}
}

@ARTICLE{2014Sci...345.1590C,
       author = {{Cleeves}, L. Ilsedore and {Bergin}, Edwin A. and {Alexander}, Conel M.~O. 'D. and {Du}, Fujun and {Graninger}, Dawn and {{\"O}berg}, Karin I. and {Harries}, Tim J.},
        title = "{The ancient heritage of water ice in the solar system}",
      journal = {Science},
         year = 2014,
        month = sep,
       volume = {345},
       number = {6204},
        pages = {1590-1593},
          doi = {10.1126/science.1258055},
archivePrefix = arXiv,
       eprint = {1409.7398},
 primaryClass = {astro-ph.SR},
       adsurl = {https://ui.adsabs.harvard.edu/abs/2014Sci...345.1590C}
}

@ARTICLE{2025ApJ...986L..19S,
       author = {{Slavicinska}, Katerina and {Tychoniec}, {\L}ukasz and {Navarro}, Mar{\'\i}a Gabriela and {van Dishoeck}, Ewine F. and {Tobin}, John J. and {van Gelder}, Martijn L. and {Chen}, Yuan and {Boogert}, A.~C. Adwin and {Drechsler}, W. Blake and {Beuther}, Henrik and {Caratti o Garatti}, Alessio and {Megeath}, S. Thomas and {Klaassen}, Pamela and {Looney}, Leslie W. and {Kavanagh}, Patrick J. and {Brunken}, Nashanty G.~C. and {Sheehan}, Patrick and {Fischer}, William J.},
        title = "{HDO Ice Detected toward an Isolated Low-mass Protostar with JWST}",
      journal = apjl,
         year = 2025,
        month = jun,
       volume = {986},
       number = {2},
          eid = {L19},
        pages = {L19},
          doi = {10.3847/2041-8213/addb45},
archivePrefix = {arXiv},
       eprint = {2505.14686},
 primaryClass = {astro-ph.SR},
       adsurl = {https://ui.adsabs.harvard.edu/abs/2025ApJ...986L..19S}
}

@ARTICLE{2013Icar..226..256Y,
       author = {{Yang}, Le and {Ciesla}, Fred J. and {Alexander}, Conel M.~O. 'D.},
        title = "{The D/H ratio of water in the solar nebula during its formation and evolution}",
      journal = icarus,
         year = 2013,
        month = sep,
       volume = {226},
       number = {1},
        pages = {256-267},
          doi = {10.1016/j.icarus.2013.05.027},
       adsurl = {https://ui.adsabs.harvard.edu/abs/2013Icar..226..256Y}
}

@ARTICLE{2019AA...622A.139M,
       author = {{Molli{\`e}re}, P. and {Snellen}, I.~A.~G.},
        title = "{Detecting isotopologues in exoplanet atmospheres using ground-based high-dispersion spectroscopy}",
      journal = aap,
         year = 2019,
        month = feb,
       volume = {622},
          eid = {A139},
        pages = {A139},
          doi = {10.1051/0004-6361/201834169},
archivePrefix = {arXiv},
       eprint = {1809.01156},
 primaryClass = {astro-ph.EP},
       adsurl = {https://ui.adsabs.harvard.edu/abs/2019AA...622A.139M}
}

@ARTICLE{2025AA...694A.174F,
       author = {{Francis}, L. and {van Dishoeck}, E.~F. and {Caratti o Garatti}, A. and {van Gelder}, M.~L. and {Gieser}, C. and {Beuther}, H. and {Ray}, T.~P. and {Tychoniec}, L. and {Nazari}, P. and {Reyes}, S. and {Kavanagh}, P.~J. and {Klaassen}, P. and {G{\"u}del}, M. and {Henning}, T.},
        title = "{JOYS: The [D/H] abundance derived from protostellar outflows across the Galactic disk measured with JWST}",
      journal = aap,
         year = 2025,
        month = feb,
       volume = {694},
          eid = {A174},
        pages = {A174},
          doi = {10.1051/0004-6361/202451629},
archivePrefix = {arXiv},
       eprint = {2501.02085},
 primaryClass = {astro-ph.GA},
       adsurl = {https://ui.adsabs.harvard.edu/abs/2025AA...694A.174F}
}

@ARTICLE{2025ApJ...979..137C,
       author = {{Catling}, David C. and {Krissansen-Totton}, Joshua and {Robinson}, Tyler D.},
        title = "{Potential Technosignature from Anomalously Low Deuterium/Hydrogen in Planetary Water Depleted by Nuclear Fusion Technology}",
      journal = apj,
         year = 2025,
        month = feb,
       volume = {979},
       number = {2},
          eid = {137},
        pages = {137},
          doi = {10.3847/1538-4357/ad99a9},
archivePrefix = {arXiv},
       eprint = {2411.18595},
 primaryClass = {astro-ph.EP},
       adsurl = {https://ui.adsabs.harvard.edu/abs/2025ApJ...979..137C}
}

@ARTICLE{2008JGRE..113.0B22F,
       author = {{Fedorova}, A. and {Korablev}, O. and {Vandaele}, A. -C. and {Bertaux}, J. -L. and {Belyaev}, D. and {Mahieux}, A. and {Neefs}, E. and {Wilquet}, W.~V. and {Drummond}, R. and {Montmessin}, F. and {Villard}, E.},
        title = "{HDO and H$_{2}$O vertical distributions and isotopic ratio in the Venus mesosphere by Solar Occultation at Infrared spectrometer on board Venus Express}",
      journal = jgr,
         year = 2008,
        month = dec,
       volume = {113},
       number = {E12},
          eid = {E00B22},
        pages = {E00B22},
          doi = {10.1029/2008JE003146},
       adsurl = {https://ui.adsabs.harvard.edu/abs/2008JGRE..113.0B22F}
}

@ARTICLE{2017RSPTA.37550390H,
       author = {{Hallis}, L.~J.},
        title = "{D/H ratios of the inner Solar System}",
      journal = ptrsla,
         year = 2017,
        month = apr,
       volume = {375},
       number = {2094},
          eid = {20150390},
        pages = {20150390},
          doi = {10.1098/rsta.2015.0390},
       adsurl = {https://ui.adsabs.harvard.edu/abs/2017RSPTA.37550390H}
}

@ARTICLE{2019AJ....158...26L,
       author = {{Lincowski}, Andrew P. and {Lustig-Yaeger}, Jacob and {Meadows}, Victoria S.},
        title = "{Observing Isotopologue Bands in Terrestrial Exoplanet Atmospheres with the James Webb Space Telescope: Implications for Identifying Past Atmospheric and Ocean Loss}",
      journal = aj,
         year = 2019,
        month = jul,
       volume = {158},
       number = {1},
          eid = {26},
        pages = {26},
          doi = {10.3847/1538-3881/ab2385},
archivePrefix = {arXiv},
       eprint = {1905.12821},
 primaryClass = {astro-ph.EP},
       adsurl = {https://ui.adsabs.harvard.edu/abs/2019AJ....158...26L}
}

@ARTICLE{2006ApJ...647.1106L,
       author = {{Linsky}, Jeffrey L. and {Draine}, Bruce T. and {Moos}, H.~W. and {Jenkins}, Edward B. and {Wood}, Brian E. and {Oliveira}, Cristina and {Blair}, William P. and {Friedman}, Scott D. and {Gry}, Cecile and {Knauth}, David and {Kruk}, Jeffrey W. and {Lacour}, Sylvestre and {Lehner}, Nicolas and {Redfield}, Seth and {Shull}, J. Michael and {Sonneborn}, George and {Williger}, Gerard M.},
        title = "{What Is the Total Deuterium Abundance in the Local Galactic Disk?}",
      journal = apj,
         year = 2006,
        month = aug,
       volume = {647},
       number = {2},
        pages = {1106-1124},
          doi = {10.1086/505556},
archivePrefix = {arXiv},
       eprint = {astro-ph/0608308},
 primaryClass = {astro-ph},
       adsurl = {https://ui.adsabs.harvard.edu/abs/2006ApJ...647.1106L}
}

@INPROCEEDINGS{2009ASPC..417...23L,
       author = {{Lis}, D.~C. and {Goldsmith}, P.~F. and {Bergin}, E.~A. and {Falgarone}, E. and {Gerin}, M. and {Roueff}, E.},
        title = "{Hydrides in Space: Past, Present, and Future}",
    booktitle = {Submillimeter Astrophysics and Technology: a Symposium Honoring Thomas G. Phillips},
         year = 2009,
       editor = {{Lis}, D.~C. and {Vaillancourt}, J.~E. and {Goldsmith}, P.~F. and {Bell}, T.~A. and {Scoville}, N.~Z. and {Zmuidzinas}, J.},
       series = {Astronomical Society of the Pacific Conference Series},
       volume = {417},
        month = dec,
        pages = {23},
       adsurl = {https://ui.adsabs.harvard.edu/abs/2009ASPC..417...23L}
}

@ARTICLE{2024PNAS..12101638M,
       author = {{Mahieux}, Arnaud and {Viscardy}, S{\'e}bastien and {Yelle}, Roger Vincent and {Karyu}, Hiroki and {Chamberlain}, Sarah and {Robert}, S{\'e}verine and {Piccialli}, Arianna and {Trompet}, Lo{\"\i}c and {Erwin}, Justin Tyler and {Ubukata}, Soma and {Nakagawa}, Hiromu and {Koyama}, Shungo and {Maggiolo}, Romain and {Pereira}, Nuno and {Cessateur}, Ga{\"e}l and {Willame}, Yannick and {Vandaele}, Ann Carine},
        title = "{Unexpected increase of the deuterium to hydrogen ratio in the Venus mesosphere}",
      journal = pnas,
         year = 2024,
        month = aug,
       volume = {121},
       number = {34},
          eid = {e2401638121},
        pages = {e2401638121},
          doi = {10.1073/pnas.2401638121},
       adsurl = {https://ui.adsabs.harvard.edu/abs/2024PNAS..12101638M}
}

@ARTICLE{2019ApJ...882L..29M,
       author = {{Morley}, Caroline V. and {Skemer}, Andrew J. and {Miles}, Brittany E. and {Line}, Michael R. and {Lopez}, Eric D. and {Brogi}, Matteo and {Freedman}, Richard S. and {Marley}, Mark S.},
        title = "{Measuring the D/H Ratios of Exoplanets and Brown Dwarfs}",
      journal = apjl,
         year = 2019,
        month = sep,
       volume = {882},
       number = {2},
          eid = {L29},
        pages = {L29},
          doi = {10.3847/2041-8213/ab3c65},
archivePrefix = {arXiv},
       eprint = {1810.04241},
 primaryClass = {astro-ph.EP},
       adsurl = {https://ui.adsabs.harvard.edu/abs/2019ApJ...882L..29M}
}

@ARTICLE{2011Natur.478..218H,
       author = {{Hartogh}, Paul and {Lis}, Dariusz C. and {Bockel{\'e}e-Morvan}, Dominique and {de Val-Borro}, Miguel and {Biver}, Nicolas and {K{\"u}ppers}, Michael and {Emprechtinger}, Martin and {Bergin}, Edwin A. and {Crovisier}, Jacques and {Rengel}, Miriam and {Moreno}, Raphael and {Szutowicz}, Slawomira and {Blake}, Geoffrey A.},
        title = "{Ocean-like water in the Jupiter-family comet 103P/Hartley 2}",
      journal = nat,
         year = 2011,
        month = oct,
       volume = {478},
       number = {7368},
        pages = {218-220},
          doi = {10.1038/nature10519},
       adsurl = {https://ui.adsabs.harvard.edu/abs/2011Natur.478..218H}
}

@ARTICLE{2024ApJ...977L..49R,
       author = {{Rowland}, Melanie J. and {Morley}, Caroline V. and {Miles}, Brittany E. and {Suarez}, Genaro and {Faherty}, Jacqueline K. and {Skemer}, Andrew J. and {Beiler}, Samuel A. and {Line}, Michael R. and {Bjoraker}, Gordon L. and {Fortney}, Jonathan J. and {Vos}, Johanna M. and {Alejandro Merchan}, Sherelyn and {Marley}, Mark and {Burningham}, Ben and {Freedman}, Richard and {Gharib-Nezhad}, Ehsan and {Batalha}, Natasha and {Lupu}, Roxana and {Visscher}, Channon and {Schneider}, Adam C. and {Geballe}, T.~R. and {Carter}, Aarynn and {Allers}, Katelyn and {Mang}, James and {Apai}, D{\'a}niel and {Limbach}, Mary Anne and {Wilson}, Mikayla J.},
        title = "{Protosolar D-to-H Abundance and One Part per Billion PH$_{3}$ in the Coldest Brown Dwarf}",
      journal = apjl,
         year = 2024,
        month = dec,
       volume = {977},
       number = {2},
          eid = {L49},
        pages = {L49},
          doi = {10.3847/2041-8213/ad9744},
archivePrefix = {arXiv},
       eprint = {2411.14541},
 primaryClass = {astro-ph.SR},
       adsurl = {https://ui.adsabs.harvard.edu/abs/2024ApJ...977L..49R}
}

@ARTICLE{2021NatAs...5..943A,
       author = {{Alday}, Juan and {Trokhimovskiy}, Alexander and {Irwin}, Patrick G.~J. and {Wilson}, Colin F. and {Montmessin}, Franck and {Lef{\'e}vre}, Franck and {Fedorova}, Anna A. and {Belyaev}, Denis A. and {Olsen}, Kevin S. and {Korablev}, Oleg and {Vals}, Margaux and {Rossi}, Lo{\"\i}c and {Baggio}, Lucio and {Bertaux}, Jean-Loup and {Patrakeev}, Andrey and {Shakun}, Alexey},
        title = "{Isotopic fractionation of water and its photolytic products in the atmosphere of Mars}",
      journal = {Nature Astronomy},
         year = 2021,
        month = jun,
       volume = {5},
        pages = {943-950},
          doi = {10.1038/s41550-021-01389-x},
       adsurl = {https://ui.adsabs.harvard.edu/abs/2021NatAs...5..943A}
}

@ARTICLE{2010AA...520A..27G,
       author = {{Guillot}, T.},
        title = "{On the radiative equilibrium of irradiated planetary atmospheres}",
      journal = aap,
         year = 2010,
        month = sep,
       volume = {520},
          eid = {A27},
        pages = {A27},
          doi = {10.1051/0004-6361/200913396},
archivePrefix = {arXiv},
       eprint = {1006.4702},
 primaryClass = {astro-ph.EP},
       adsurl = {https://ui.adsabs.harvard.edu/abs/2010AA...520A..27G}
}

@ARTICLE{2024JOSS....9.7028B,
       author = {{Blain}, Doriann and {Molli{\`e}re}, Paul and {Nasedkin}, Evert},
        title = "{SpectralModel: a high-resolution framework for petitRADTRANS 3}",
      journal = joss,
         year = 2024,
        month = oct,
       volume = {9},
       number = {102},
          eid = {7028},
        pages = {7028},
          doi = {10.21105/joss.07028},
       adsurl = {https://ui.adsabs.harvard.edu/abs/2024JOSS....9.7028B}
}

@ARTICLE{1998Sci...279..842M,
       author = {{Meier}, Roland and {Owen}, Tobias C. and {Matthews}, Henry E. and {Jewitt}, David C. and {Bockelee-Morvan}, Dominique and {Biver}, Nicolas and {Crovisier}, Jacques and {Gautier}, Daniel},
        title = "{A Determination of the HDO/H2O Ratio in Comet C/1995 O1 (Hale-Bopp)}",
      journal = {Science},
         year = 1998,
        month = feb,
       volume = {279},
        pages = {842},
          doi = {10.1126/science.279.5352.842},
       adsurl = {https://ui.adsabs.harvard.edu/abs/1998Sci...279..842M}
}

@ARTICLE{2015Sci...347A.387A,
       author = {{Altwegg}, K. and {Balsiger}, H. and {Bar-Nun}, A. and {Berthelier}, J.~J. and {Bieler}, A. and {Bochsler}, P. and {Briois}, C. and {Calmonte}, U. and {Combi}, M. and {De Keyser}, J. and {Eberhardt}, P. and {Fiethe}, B. and {Fuselier}, S. and {Gasc}, S. and {Gombosi}, T.~I. and {Hansen}, K.~C. and {H{\"a}ssig}, M. and {J{\"a}ckel}, A. and {Kopp}, E. and {Korth}, A. and {LeRoy}, L. and {Mall}, U. and {Marty}, B. and {Mousis}, O. and {Neefs}, E. and {Owen}, T. and {R{\`e}me}, H. and {Rubin}, M. and {S{\'e}mon}, T. and {Tzou}, C. -Y. and {Waite}, H. and {Wurz}, P.},
        title = "{67P/Churyumov-Gerasimenko, a Jupiter family comet with a high D/H ratio}",
      journal = {Science},
         year = 2015,
        month = jan,
       volume = {347},
       number = {6220},
          eid = {1261952},
        pages = {1261952},
          doi = {10.1126/science.1261952},
       adsurl = {https://ui.adsabs.harvard.edu/abs/2015Sci...347A.387A}
}

@ARTICLE{2018ARAA..56..175D,
       author = {{Dawson}, Rebekah I. and {Johnson}, John Asher},
        title = "{Origins of Hot Jupiters}",
      journal = araa,
         year = 2018,
        month = sep,
       volume = {56},
        pages = {175-221},
          doi = {10.1146/annurev-astro-081817-051853},
archivePrefix = {arXiv},
       eprint = {1801.06117},
 primaryClass = {astro-ph.EP},
       adsurl = {https://ui.adsabs.harvard.edu/abs/2018ARAA..56..175D}
}

@ARTICLE{2016PASP..128i4401S,
       author = {{Stevenson}, Kevin B. and {Lewis}, Nikole K. and {Bean}, Jacob L. and {Beichman}, Charles and {Fraine}, Jonathan and {Kilpatrick}, Brian M. and {Krick}, J.~E. and {Lothringer}, Joshua D. and {Mandell}, Avi M. and {Valenti}, Jeff A. and {Agol}, Eric and {Angerhausen}, Daniel and {Barstow}, Joanna K. and {Birkmann}, Stephan M. and {Burrows}, Adam and {Charbonneau}, David and {Cowan}, Nicolas B. and {Crouzet}, Nicolas and {Cubillos}, Patricio E. and {Curry}, S.~M. and {Dalba}, Paul A. and {de Wit}, Julien and {Deming}, Drake and {D{\'e}sert}, Jean-Michel and {Doyon}, Ren{\'e} and {Dragomir}, Diana and {Ehrenreich}, David and {Fortney}, Jonathan J. and {Garc{\'\i}a Mu{\~n}oz}, Antonio and {Gibson}, Neale P. and {Gizis}, John E. and {Greene}, Thomas P. and {Harrington}, Joseph and {Heng}, Kevin and {Kataria}, Tiffany and {Kempton}, Eliza M. -R. and {Knutson}, Heather and {Kreidberg}, Laura and {Lafreni{\`e}re}, David and {Lagage}, Pierre-Olivier and {Line}, Michael R. and {Lopez-Morales}, Mercedes and {Madhusudhan}, Nikku and {Morley}, Caroline V. and {Rocchetto}, Marco and {Schlawin}, Everett and {Shkolnik}, Evgenya L. and {Shporer}, Avi and {Sing}, David K. and {Todorov}, Kamen O. and {Tucker}, Gregory S. and {Wakeford}, Hannah R.},
        title = "{Transiting Exoplanet Studies and Community Targets for JWST's Early Release Science Program}",
      journal = pasp,
         year = 2016,
        month = sep,
       volume = {128},
       number = {967},
        pages = {094401},
          doi = {10.1088/1538-3873/128/967/094401},
archivePrefix = {arXiv},
       eprint = {1602.08389},
 primaryClass = {astro-ph.EP},
       adsurl = {https://ui.adsabs.harvard.edu/abs/2016PASP..128i4401S}
}

@ARTICLE{2018PASP..130k4402B,
       author = {{Bean}, Jacob L. and {Stevenson}, Kevin B. and {Batalha}, Natalie M. and {Berta-Thompson}, Zachory and {Kreidberg}, Laura and {Crouzet}, Nicolas and {Benneke}, Bj{\"o}rn and {Line}, Michael R. and {Sing}, David K. and {Wakeford}, Hannah R. and {Knutson}, Heather A. and {Kempton}, Eliza M. -R. and {D{\'e}sert}, Jean-Michel and {Crossfield}, Ian and {Batalha}, Natasha E. and {de Wit}, Julien and {Parmentier}, Vivien and {Harrington}, Joseph and {Moses}, Julianne I. and {Lopez-Morales}, Mercedes and {Alam}, Munazza K. and {Blecic}, Jasmina and {Bruno}, Giovanni and {Carter}, Aarynn L. and {Chapman}, John W. and {Decin}, Leen and {Dragomir}, Diana and {Evans}, Thomas M. and {Fortney}, Jonathan J. and {Fraine}, Jonathan D. and {Gao}, Peter and {Garc{\'\i}a Mu{\~n}oz}, Antonio and {Gibson}, Neale P. and {Goyal}, Jayesh M. and {Heng}, Kevin and {Hu}, Renyu and {Kendrew}, Sarah and {Kilpatrick}, Brian M. and {Krick}, Jessica and {Lagage}, Pierre-Olivier and {Lendl}, Monika and {Louden}, Tom and {Madhusudhan}, Nikku and {Mandell}, Avi M. and {Mansfield}, Megan and {May}, Erin M. and {Morello}, Giuseppe and {Morley}, Caroline V. and {Nikolov}, Nikolay and {Redfield}, Seth and {Roberts}, Jessica E. and {Schlawin}, Everett and {Spake}, Jessica J. and {Todorov}, Kamen O. and {Tsiaras}, Angelos and {Venot}, Olivia and {Waalkes}, William C. and {Wheatley}, Peter J. and {Zellem}, Robert T. and {Angerhausen}, Daniel and {Barrado}, David and {Carone}, Ludmila and {Casewell}, Sarah L. and {Cubillos}, Patricio E. and {Damiano}, Mario and {de Val-Borro}, Miguel and {Drummond}, Benjamin and {Edwards}, Billy and {Endl}, Michael and {Espinoza}, Nestor and {France}, Kevin and {Gizis}, John E. and {Greene}, Thomas P. and {Henning}, Thomas K. and {Hong}, Yucian and {Ingalls}, James G. and {Iro}, Nicolas and {Irwin}, Patrick G.~J. and {Kataria}, Tiffany and {Lahuis}, Fred and {Leconte}, J{\'e}r{\'e}my and {Lillo-Box}, Jorge and {Lines}, Stefan and {Lothringer}, Joshua D. and {Mancini}, Luigi and {Marchis}, Franck and {Mayne}, Nathan and {Palle}, Enric and {Rauscher}, Emily and {Roudier}, Ga{\"e}l and {Shkolnik}, Evgenya L. and {Southworth}, John and {Swain}, Mark R. and {Taylor}, Jake and {Teske}, Johanna and {Tinetti}, Giovanna and {Tremblin}, Pascal and {Tucker}, Gregory S. and {van Boekel}, Roy and {Waldmann}, Ingo P. and {Weaver}, Ian C. and {Zingales}, Tiziano},
        title = "{The Transiting Exoplanet Community Early Release Science Program for JWST}",
      journal = pasp,
         year = 2018,
        month = nov,
       volume = {130},
       number = {993},
        pages = {114402},
          doi = {10.1088/1538-3873/aadbf3},
archivePrefix = {arXiv},
       eprint = {1803.04985},
 primaryClass = {astro-ph.EP},
       adsurl = {https://ui.adsabs.harvard.edu/abs/2018PASP..130k4402B}
}

@ARTICLE{2025arXiv250407823M,
       author = {{Ma}, Sushuang and {Saba}, Arianna and {Faris Al-Refaie}, Ahmed and {Tinetti}, Giovanna and {Yurchenko}, Sergei N. and {Tennyson}, Jonathan and {Cecchi Pestellini}, Cesare},
        title = "{A new look into the atmospheric composition of WASP-39 b}",
         year = 2025,
        month = apr,
          eid = {arXiv:2504.07823},
        pages = {arXiv:2504.07823},
          doi = {10.48550/arXiv.2504.07823},
archivePrefix = {arXiv},
       eprint = {2504.07823},
 primaryClass = {astro-ph.EP},
       adsurl = {https://ui.adsabs.harvard.edu/abs/2025arXiv250407823M}
}

@ARTICLE{2024AA...687A.110L,
       author = {{Lueber}, Anna and {Novais}, Aline and {Fisher}, Chloe and {Heng}, Kevin},
        title = "{Information content of JWST spectra of WASP-39b}",
      journal = aap,
         year = 2024,
        month = jul,
       volume = {687},
          eid = {A110},
        pages = {A110},
          doi = {10.1051/0004-6361/202348802},
archivePrefix = {arXiv},
       eprint = {2405.02656},
 primaryClass = {astro-ph.EP},
       adsurl = {https://ui.adsabs.harvard.edu/abs/2024AA...687A.110L}
}

@ARTICLE{2023Natur.614..670F,
       author = {{Feinstein}, Adina D. and {Radica}, Michael and {Welbanks}, Luis and {Murray}, Catriona Anne and {Ohno}, Kazumasa and {Coulombe}, Louis-Philippe and {Espinoza}, N{\'e}stor and {Bean}, Jacob L. and {Teske}, Johanna K. and {Benneke}, Bj{\"o}rn and {Line}, Michael R. and {Rustamkulov}, Zafar and {Saba}, Arianna and {Tsiaras}, Angelos and {Barstow}, Joanna K. and {Fortney}, Jonathan J. and {Gao}, Peter and {Knutson}, Heather A. and {MacDonald}, Ryan J. and {Mikal-Evans}, Thomas and {Rackham}, Benjamin V. and {Taylor}, Jake and {Parmentier}, Vivien and {Batalha}, Natalie M. and {Berta-Thompson}, Zachory K. and {Carter}, Aarynn L. and {Changeat}, Quentin and {dos Santos}, Leonardo A. and {Gibson}, Neale P. and {Goyal}, Jayesh M. and {Kreidberg}, Laura and {L{\'o}pez-Morales}, Mercedes and {Lothringer}, Joshua D. and {Miguel}, Yamila and {Molaverdikhani}, Karan and {Moran}, Sarah E. and {Morello}, Giuseppe and {Mukherjee}, Sagnick and {Sing}, David K. and {Stevenson}, Kevin B. and {Wakeford}, Hannah R. and {Ahrer}, Eva-Maria and {Alam}, Munazza K. and {Alderson}, Lili and {Allen}, Natalie H. and {Batalha}, Natasha E. and {Bell}, Taylor J. and {Blecic}, Jasmina and {Brande}, Jonathan and {Caceres}, Claudio and {Casewell}, S.~L. and {Chubb}, Katy L. and {Crossfield}, Ian J.~M. and {Crouzet}, Nicolas and {Cubillos}, Patricio E. and {Decin}, Leen and {D{\'e}sert}, Jean-Michel and {Harrington}, Joseph and {Heng}, Kevin and {Henning}, Thomas and {Iro}, Nicolas and {Kempton}, Eliza M. -R. and {Kendrew}, Sarah and {Kirk}, James and {Krick}, Jessica and {Lagage}, Pierre-Olivier and {Lendl}, Monika and {Mancini}, Luigi and {Mansfield}, Megan and {May}, E.~M. and {Mayne}, N.~J. and {Nikolov}, Nikolay K. and {Palle}, Enric and {Petit dit de la Roche}, Dominique J.~M. and {Piaulet}, Caroline and {Powell}, Diana and {Redfield}, Seth and {Rogers}, Laura K. and {Roman}, Michael T. and {Roy}, Pierre-Alexis and {Nixon}, Matthew C. and {Schlawin}, Everett and {Tan}, Xianyu and {Tremblin}, P. and {Turner}, Jake D. and {Venot}, Olivia and {Waalkes}, William C. and {Wheatley}, Peter J. and {Zhang}, Xi},
        title = "{Early Release Science of the exoplanet WASP-39b with JWST NIRISS}",
      journal = nat,
         year = 2023,
        month = feb,
       volume = {614},
       number = {7949},
        pages = {670-675},
          doi = {10.1038/s41586-022-05674-1},
archivePrefix = {arXiv},
       eprint = {2211.10493},
 primaryClass = {astro-ph.EP},
       adsurl = {https://ui.adsabs.harvard.edu/abs/2023Natur.614..670F}
}

@ARTICLE{2023Natur.614..659R,
       author = {{Rustamkulov}, Z. and {Sing}, D.~K. and {Mukherjee}, S. and {May}, E.~M. and {Kirk}, J. and {Schlawin}, E. and {Line}, M.~R. and {Piaulet}, C. and {Carter}, A.~L. and {Batalha}, N.~E. and {Goyal}, J.~M. and {L{\'o}pez-Morales}, M. and {Lothringer}, J.~D. and {MacDonald}, R.~J. and {Moran}, S.~E. and {Stevenson}, K.~B. and {Wakeford}, H.~R. and {Espinoza}, N. and {Bean}, J.~L. and {Batalha}, N.~M. and {Benneke}, B. and {Berta-Thompson}, Z.~K. and {Crossfield}, I.~J.~M. and {Gao}, P. and {Kreidberg}, L. and {Powell}, D.~K. and {Cubillos}, P.~E. and {Gibson}, N.~P. and {Leconte}, J. and {Molaverdikhani}, K. and {Nikolov}, N.~K. and {Parmentier}, V. and {Roy}, P. and {Taylor}, J. and {Turner}, J.~D. and {Wheatley}, P.~J. and {Aggarwal}, K. and {Ahrer}, E. and {Alam}, M.~K. and {Alderson}, L. and {Allen}, N.~H. and {Banerjee}, A. and {Barat}, S. and {Barrado}, D. and {Barstow}, J.~K. and {Bell}, T.~J. and {Blecic}, J. and {Brande}, J. and {Casewell}, S. and {Changeat}, Q. and {Chubb}, K.~L. and {Crouzet}, N. and {Daylan}, T. and {Decin}, L. and {D{\'e}sert}, J. and {Mikal-Evans}, T. and {Feinstein}, A.~D. and {Flagg}, L. and {Fortney}, J.~J. and {Harrington}, J. and {Heng}, K. and {Hong}, Y. and {Hu}, R. and {Iro}, N. and {Kataria}, T. and {Kempton}, E.~M. -R. and {Krick}, J. and {Lendl}, M. and {Lillo-Box}, J. and {Louca}, A. and {Lustig-Yaeger}, J. and {Mancini}, L. and {Mansfield}, M. and {Mayne}, N.~J. and {Miguel}, Y. and {Morello}, G. and {Ohno}, K. and {Palle}, E. and {Petit dit de la Roche}, D.~J.~M. and {Rackham}, B.~V. and {Radica}, M. and {Ramos-Rosado}, L. and {Redfield}, S. and {Rogers}, L.~K. and {Shkolnik}, E.~L. and {Southworth}, J. and {Teske}, J. and {Tremblin}, P. and {Tucker}, G.~S. and {Venot}, O. and {Waalkes}, W.~C. and {Welbanks}, L. and {Zhang}, X. and {Zieba}, S.},
        title = "{Early Release Science of the exoplanet WASP-39b with JWST NIRSpec PRISM}",
      journal = nat,
         year = 2023,
        month = feb,
       volume = {614},
       number = {7949},
        pages = {659-663},
          doi = {10.1038/s41586-022-05677-y},
archivePrefix = {arXiv},
       eprint = {2211.10487},
 primaryClass = {astro-ph.EP},
       adsurl = {https://ui.adsabs.harvard.edu/abs/2023Natur.614..659R}
}

@ARTICLE{2023Natur.614..664A,
       author = {{Alderson}, Lili and {Wakeford}, Hannah R. and {Alam}, Munazza K. and {Batalha}, Natasha E. and {Lothringer}, Joshua D. and {Adams Redai}, Jea and {Barat}, Saugata and {Brande}, Jonathan and {Damiano}, Mario and {Daylan}, Tansu and {Espinoza}, N{\'e}stor and {Flagg}, Laura and {Goyal}, Jayesh M. and {Grant}, David and {Hu}, Renyu and {Inglis}, Julie and {Lee}, Elspeth K.~H. and {Mikal-Evans}, Thomas and {Ramos-Rosado}, Lakeisha and {Roy}, Pierre-Alexis and {Wallack}, Nicole L. and {Batalha}, Natalie M. and {Bean}, Jacob L. and {Benneke}, Bj{\"o}rn and {Berta-Thompson}, Zachory K. and {Carter}, Aarynn L. and {Changeat}, Quentin and {Col{\'o}n}, Knicole D. and {Crossfield}, Ian J.~M. and {D{\'e}sert}, Jean-Michel and {Foreman-Mackey}, Daniel and {Gibson}, Neale P. and {Kreidberg}, Laura and {Line}, Michael R. and {L{\'o}pez-Morales}, Mercedes and {Molaverdikhani}, Karan and {Moran}, Sarah E. and {Morello}, Giuseppe and {Moses}, Julianne I. and {Mukherjee}, Sagnick and {Schlawin}, Everett and {Sing}, David K. and {Stevenson}, Kevin B. and {Taylor}, Jake and {Aggarwal}, Keshav and {Ahrer}, Eva-Maria and {Allen}, Natalie H. and {Barstow}, Joanna K. and {Bell}, Taylor J. and {Blecic}, Jasmina and {Casewell}, Sarah L. and {Chubb}, Katy L. and {Crouzet}, Nicolas and {Cubillos}, Patricio E. and {Decin}, Leen and {Feinstein}, Adina D. and {Fortney}, Joanthan J. and {Harrington}, Joseph and {Heng}, Kevin and {Iro}, Nicolas and {Kempton}, Eliza M. -R. and {Kirk}, James and {Knutson}, Heather A. and {Krick}, Jessica and {Leconte}, J{\'e}r{\'e}my and {Lendl}, Monika and {MacDonald}, Ryan J. and {Mancini}, Luigi and {Mansfield}, Megan and {May}, Erin M. and {Mayne}, Nathan J. and {Miguel}, Yamila and {Nikolov}, Nikolay K. and {Ohno}, Kazumasa and {Palle}, Enric and {Parmentier}, Vivien and {Petit dit de la Roche}, Dominique J.~M. and {Piaulet}, Caroline and {Powell}, Diana and {Rackham}, Benjamin V. and {Redfield}, Seth and {Rogers}, Laura K. and {Rustamkulov}, Zafar and {Tan}, Xianyu and {Tremblin}, P. and {Tsai}, Shang-Min and {Turner}, Jake D. and {de Val-Borro}, Miguel and {Venot}, Olivia and {Welbanks}, Luis and {Wheatley}, Peter J. and {Zhang}, Xi},
        title = "{Early Release Science of the exoplanet WASP-39b with JWST NIRSpec G395H}",
      journal = nat,
         year = 2023,
        month = feb,
       volume = {614},
       number = {7949},
        pages = {664-669},
          doi = {10.1038/s41586-022-05591-3},
archivePrefix = {arXiv},
       eprint = {2211.10488},
 primaryClass = {astro-ph.EP},
       adsurl = {https://ui.adsabs.harvard.edu/abs/2023Natur.614..664A}
}

@ARTICLE{2023Natur.614..653A,
       author = {{Ahrer}, Eva-Maria and {Stevenson}, Kevin B. and {Mansfield}, Megan and {Moran}, Sarah E. and {Brande}, Jonathan and {Morello}, Giuseppe and {Murray}, Catriona A. and {Nikolov}, Nikolay K. and {Petit dit de la Roche}, Dominique J.~M. and {Schlawin}, Everett and {Wheatley}, Peter J. and {Zieba}, Sebastian and {Batalha}, Natasha E. and {Damiano}, Mario and {Goyal}, Jayesh M. and {Lendl}, Monika and {Lothringer}, Joshua D. and {Mukherjee}, Sagnick and {Ohno}, Kazumasa and {Batalha}, Natalie M. and {Battley}, Matthew P. and {Bean}, Jacob L. and {Beatty}, Thomas G. and {Benneke}, Bj{\"o}rn and {Berta-Thompson}, Zachory K. and {Carter}, Aarynn L. and {Cubillos}, Patricio E. and {Daylan}, Tansu and {Espinoza}, N{\'e}stor and {Gao}, Peter and {Gibson}, Neale P. and {Gill}, Samuel and {Harrington}, Joseph and {Hu}, Renyu and {Kreidberg}, Laura and {Lewis}, Nikole K. and {Line}, Michael R. and {L{\'o}pez-Morales}, Mercedes and {Parmentier}, Vivien and {Powell}, Diana K. and {Sing}, David K. and {Tsai}, Shang-Min and {Wakeford}, Hannah R. and {Welbanks}, Luis and {Alam}, Munazza K. and {Alderson}, Lili and {Allen}, Natalie H. and {Anderson}, David R. and {Barstow}, Joanna K. and {Bayliss}, Daniel and {Bell}, Taylor J. and {Blecic}, Jasmina and {Bryant}, Edward M. and {Burleigh}, Matthew R. and {Carone}, Ludmila and {Casewell}, S.~L. and {Changeat}, Quentin and {Chubb}, Katy L. and {Crossfield}, Ian J.~M. and {Crouzet}, Nicolas and {Decin}, Leen and {D{\'e}sert}, Jean-Michel and {Feinstein}, Adina D. and {Flagg}, Laura and {Fortney}, Jonathan J. and {Gizis}, John E. and {Heng}, Kevin and {Iro}, Nicolas and {Kempton}, Eliza M. -R. and {Kendrew}, Sarah and {Kirk}, James and {Knutson}, Heather A. and {Komacek}, Thaddeus D. and {Lagage}, Pierre-Olivier and {Leconte}, J{\'e}r{\'e}my and {Lustig-Yaeger}, Jacob and {MacDonald}, Ryan J. and {Mancini}, Luigi and {May}, E.~M. and {Mayne}, N.~J. and {Miguel}, Yamila and {Mikal-Evans}, Thomas and {Molaverdikhani}, Karan and {Palle}, Enric and {Piaulet}, Caroline and {Rackham}, Benjamin V. and {Redfield}, Seth and {Rogers}, Laura K. and {Roy}, Pierre-Alexis and {Rustamkulov}, Zafar and {Shkolnik}, Evgenya L. and {Sotzen}, Kristin S. and {Taylor}, Jake and {Tremblin}, P. and {Tucker}, Gregory S. and {Turner}, Jake D. and {de Val-Borro}, Miguel and {Venot}, Olivia and {Zhang}, Xi},
        title = "{Early Release Science of the exoplanet WASP-39b with JWST NIRCam}",
      journal = nat,
         year = 2023,
        month = feb,
       volume = {614},
       number = {7949},
        pages = {653-658},
          doi = {10.1038/s41586-022-05590-4},
archivePrefix = {arXiv},
       eprint = {2211.10489},
 primaryClass = {astro-ph.EP},
       adsurl = {https://ui.adsabs.harvard.edu/abs/2023Natur.614..653A}
}

@ARTICLE{2024Natur.626..979P,
       author = {{Powell}, Diana and {Feinstein}, Adina D. and {Lee}, Elspeth K.~H. and {Zhang}, Michael and {Tsai}, Shang-Min and {Taylor}, Jake and {Kirk}, James and {Bell}, Taylor and {Barstow}, Joanna K. and {Gao}, Peter and {Bean}, Jacob L. and {Blecic}, Jasmina and {Chubb}, Katy L. and {Crossfield}, Ian J.~M. and {Jordan}, Sean and {Kitzmann}, Daniel and {Moran}, Sarah E. and {Morello}, Giuseppe and {Moses}, Julianne I. and {Welbanks}, Luis and {Yang}, Jeehyun and {Zhang}, Xi and {Ahrer}, Eva-Maria and {Bello-Arufe}, Aaron and {Brande}, Jonathan and {Casewell}, S.~L. and {Crouzet}, Nicolas and {Cubillos}, Patricio E. and {Demory}, Brice-Olivier and {Dyrek}, Achr{\`e}ne and {Flagg}, Laura and {Hu}, Renyu and {Inglis}, Julie and {Jones}, Kathryn D. and {Kreidberg}, Laura and {L{\'o}pez-Morales}, Mercedes and {Lagage}, Pierre-Olivier and {Meier Vald{\'e}s}, Erik A. and {Miguel}, Yamila and {Parmentier}, Vivien and {Piette}, Anjali A.~A. and {Rackham}, Benjamin V. and {Radica}, Michael and {Redfield}, Seth and {Stevenson}, Kevin B. and {Wakeford}, Hannah R. and {Aggarwal}, Keshav and {Alam}, Munazza K. and {Batalha}, Natalie M. and {Batalha}, Natasha E. and {Benneke}, Bj{\"o}rn and {Berta-Thompson}, Zach K. and {Brady}, Ryan P. and {Caceres}, Claudio and {Carter}, Aarynn L. and {D{\'e}sert}, Jean-Michel and {Harrington}, Joseph and {Iro}, Nicolas and {Line}, Michael R. and {Lothringer}, Joshua D. and {MacDonald}, Ryan J. and {Mancini}, Luigi and {Molaverdikhani}, Karan and {Mukherjee}, Sagnick and {Nixon}, Matthew C. and {Oza}, Apurva V. and {Palle}, Enric and {Rustamkulov}, Zafar and {Sing}, David K. and {Steinrueck}, Maria E. and {Venot}, Olivia and {Wheatley}, Peter J. and {Yurchenko}, Sergei N.},
        title = "{Sulfur dioxide in the mid-infrared transmission spectrum of WASP-39b}",
      journal = nat,
         year = 2024,
        month = feb,
       volume = {626},
       number = {8001},
        pages = {979-983},
          doi = {10.1038/s41586-024-07040-9},
archivePrefix = {arXiv},
       eprint = {2407.07965},
 primaryClass = {astro-ph.EP},
       adsurl = {https://ui.adsabs.harvard.edu/abs/2024Natur.626..979P}
}

@misc{carter_2024_10161743,
  author       = {Carter, Aarynn L. and
                  May, Erin M.},
  title        = {Products and Models for "A benchmark JWST near-
                   infrared spectrum for the exoplanet WASP-39 b"
                  },
  month        = jul,
  year         = 2024,
  publisher    = {Zenodo},
  version      = {1.0},
  doi          = {10.5281/zenodo.10161743},
  url          = {https://doi.org/10.5281/zenodo.10161743},
}

@ARTICLE{2024NatAs...8.1008C,
       author = {{Carter}, A.~L. and {May}, E.~M. and {Espinoza}, N. and {Welbanks}, L. and {Ahrer}, E. and {Alderson}, L. and {Brahm}, R. and {Feinstein}, A.~D. and {Grant}, D. and {Line}, M. and {Morello}, G. and {O'Steen}, R. and {Radica}, M. and {Rustamkulov}, Z. and {Stevenson}, K.~B. and {Turner}, J.~D. and {Alam}, M.~K. and {Anderson}, D.~R. and {Batalha}, N.~M. and {Battley}, M.~P. and {Bayliss}, D. and {Bean}, J.~L. and {Benneke}, B. and {Berta-Thompson}, Z.~K. and {Brande}, J. and {Bryant}, E.~M. and {Burleigh}, M.~R. and {Coulombe}, L. and {Crossfield}, I.~J.~M. and {Damiano}, M. and {D{\'e}sert}, J. -M. and {Flagg}, L. and {Gill}, S. and {Inglis}, J. and {Kirk}, J. and {Knutson}, H. and {Kreidberg}, L. and {L{\'o}pez Morales}, M. and {Mansfield}, M. and {Moran}, S.~E. and {Murray}, C.~A. and {Nixon}, M.~C. and {Petit dit de la Roche}, D.~J.~M. and {Rackham}, B.~V. and {Schlawin}, E. and {Sing}, D.~K. and {Wakeford}, H.~R. and {Wallack}, N.~L. and {Wheatley}, P.~J. and {Zieba}, S. and {Aggarwal}, K. and {Barstow}, J.~K. and {Bell}, T.~J. and {Blecic}, J. and {Caceres}, C. and {Crouzet}, N. and {Cubillos}, P.~E. and {Daylan}, T. and {de Val-Borro}, M. and {Decin}, L. and {Fortney}, J.~J. and {Gibson}, N.~P. and {Heng}, K. and {Hu}, R. and {Kempton}, E.~M. -R. and {Lagage}, P. and {Lothringer}, J.~D. and {Lustig-Yaeger}, J. and {Mancini}, L. and {Mayne}, N.~J. and {Mayorga}, L.~C. and {Molaverdikhani}, K. and {Nasedkin}, E. and {Ohno}, K. and {Parmentier}, V. and {Powell}, D. and {Redfield}, S. and {Roy}, P. and {Taylor}, J. and {Zhang}, X.},
        title = "{A benchmark JWST near-infrared spectrum for the exoplanet WASP-39 b}",
      journal = {Nature Astronomy},
         year = 2024,
        month = aug,
       volume = {8},
        pages = {1008-1019},
          doi = {10.1038/s41550-024-02292-x},
archivePrefix = {arXiv},
       eprint = {2407.13893},
 primaryClass = {astro-ph.EP},
       adsurl = {https://ui.adsabs.harvard.edu/abs/2024NatAs...8.1008C}
}

@misc{powell_2023_10055845,
  author       = {Powell, Diana and
                  Feinstein, Adina and
                  Lee, Elspeth and
                  Zhang, Michael and
                  Tsai, Shang-Min and
                  Taylor, Jake and
                  Kirk, James and
                  Bell, Taylor and
                  Barstow, Joanna and
                  Gao, Peter},
  title        = {Products and Models for "Sulphur Dioxide in the
                   Mid-Infrared Transmission Spectrum of WASP-39b"
                  },
  month        = oct,
  year         = 2023,
  publisher    = {Zenodo},
  doi          = {10.5281/zenodo.10055845},
  url          = {https://doi.org/10.5281/zenodo.10055845},
}

@INCOLLECTION{2018haex.bookE.104M,
       author = {{Madhusudhan}, Nikku},
        title = "{Atmospheric Retrieval of Exoplanets}",
    booktitle = {Handbook of Exoplanets},
         year = 2018,
       editor = {{Deeg}, Hans J. and {Belmonte}, Juan Antonio},
          eid = {104},
        pages = {104},
          doi = {10.1007/978-3-319-55333-7_104},
       adsurl = {https://ui.adsabs.harvard.edu/abs/2018haex.bookE.104M},
      publisher = {Springer},
}

@article{Nasedkin2024, doi = {10.21105/joss.05875}, url = {https://doi.org/10.21105/joss.05875}, year = {2024}, publisher = {The Open Journal}, volume = {9}, number = {96}, pages = {5875}, author = {Evert Nasedkin and Paul Mollière and Doriann Blain}, title = {Atmospheric Retrievals with petitRADTRANS}, journal = joss }

@ARTICLE{2019AA...627A..67M,
       author = {{Molli{\`e}re}, P. and {Wardenier}, J.~P. and {van Boekel}, R. and {Henning}, Th. and {Molaverdikhani}, K. and {Snellen}, I.~A.~G.},
        title = "{petitRADTRANS. A Python radiative transfer package for exoplanet characterization and retrieval}",
      journal = aap,
         year = 2019,
        month = jul,
       volume = {627},
          eid = {A67},
        pages = {A67},
          doi = {10.1051/0004-6361/201935470},
archivePrefix = {arXiv},
       eprint = {1904.11504},
 primaryClass = {astro-ph.EP},
       adsurl = {https://ui.adsabs.harvard.edu/abs/2019AA...627A..67M}
}

@book{Jeffreys1939,
  author    = {Jeffreys, Harold},
  title     = {Theory of Probability},
  year      = {1939},
  publisher = {Clarendon Press},
  edition   = {1},
  address   = {Oxford, England}
}

@ARTICLE{2024ApJ...969L..19F,
       author = {{Flagg}, Laura and {Weinberger}, Alycia J. and {Bell}, Taylor J. and {Welbanks}, Luis and {Morello}, Giuseppe and {Powell}, Diana and {Bean}, Jacob L. and {Blecic}, Jasmina and {Crouzet}, Nicolas and {Gao}, Peter and {Inglis}, Julie and {Kirk}, James and {L{\'o}pez-Morales}, Mercedes and {Molaverdikhani}, Karan and {Nikolov}, Nikolay and {Oza}, Apurva V. and {Rackham}, Benjamin V. and {Redfield}, Seth and {Tsai}, Shang-Min and {Jayawardhana}, Ray and {Kreidberg}, Laura and {Nixon}, Matthew C. and {Stevenson}, Kevin B. and {Turner}, Jake D.},
        title = "{Debris Disks Can Contaminate Mid-infrared Exoplanet Spectra: Evidence for a Circumstellar Debris Disk around Exoplanet Host WASP-39}",
      journal = apjl,
         year = 2024,
        month = jul,
       volume = {969},
       number = {1},
          eid = {L19},
        pages = {L19},
          doi = {10.3847/2041-8213/ad4649},
archivePrefix = {arXiv},
       eprint = {2406.02305},
 primaryClass = {astro-ph.EP},
       adsurl = {https://ui.adsabs.harvard.edu/abs/2024ApJ...969L..19F}
}

@ARTICLE{2023Natur.617..483T,
       author = {{Tsai}, Shang-Min and {Lee}, Elspeth K.~H. and {Powell}, Diana and {Gao}, Peter and {Zhang}, Xi and {Moses}, Julianne and {H{\'e}brard}, Eric and {Venot}, Olivia and {Parmentier}, Vivien and {Jordan}, Sean and {Hu}, Renyu and {Alam}, Munazza K. and {Alderson}, Lili and {Batalha}, Natalie M. and {Bean}, Jacob L. and {Benneke}, Bj{\"o}rn and {Bierson}, Carver J. and {Brady}, Ryan P. and {Carone}, Ludmila and {Carter}, Aarynn L. and {Chubb}, Katy L. and {Inglis}, Julie and {Leconte}, J{\'e}r{\'e}my and {Line}, Michael and {L{\'o}pez-Morales}, Mercedes and {Miguel}, Yamila and {Molaverdikhani}, Karan and {Rustamkulov}, Zafar and {Sing}, David K. and {Stevenson}, Kevin B. and {Wakeford}, Hannah R. and {Yang}, Jeehyun and {Aggarwal}, Keshav and {Baeyens}, Robin and {Barat}, Saugata and {de Val-Borro}, Miguel and {Daylan}, Tansu and {Fortney}, Jonathan J. and {France}, Kevin and {Goyal}, Jayesh M. and {Grant}, David and {Kirk}, James and {Kreidberg}, Laura and {Louca}, Amy and {Moran}, Sarah E. and {Mukherjee}, Sagnick and {Nasedkin}, Evert and {Ohno}, Kazumasa and {Rackham}, Benjamin V. and {Redfield}, Seth and {Taylor}, Jake and {Tremblin}, Pascal and {Visscher}, Channon and {Wallack}, Nicole L. and {Welbanks}, Luis and {Youngblood}, Allison and {Ahrer}, Eva-Maria and {Batalha}, Natasha E. and {Behr}, Patrick and {Berta-Thompson}, Zachory K. and {Blecic}, Jasmina and {Casewell}, S.~L. and {Crossfield}, Ian J.~M. and {Crouzet}, Nicolas and {Cubillos}, Patricio E. and {Decin}, Leen and {D{\'e}sert}, Jean-Michel and {Feinstein}, Adina D. and {Gibson}, Neale P. and {Harrington}, Joseph and {Heng}, Kevin and {Henning}, Thomas and {Kempton}, Eliza M. -R. and {Krick}, Jessica and {Lagage}, Pierre-Olivier and {Lendl}, Monika and {Lothringer}, Joshua D. and {Mansfield}, Megan and {Mayne}, N.~J. and {Mikal-Evans}, Thomas and {Palle}, Enric and {Schlawin}, Everett and {Shorttle}, Oliver and {Wheatley}, Peter J. and {Yurchenko}, Sergei N.},
        title = "{Photochemically produced SO$_{2}$ in the atmosphere of WASP-39b}",
      journal = nat,
         year = 2023,
        month = may,
       volume = {617},
       number = {7961},
        pages = {483-487},
          doi = {10.1038/s41586-023-05902-2},
archivePrefix = {arXiv},
       eprint = {2211.10490},
 primaryClass = {astro-ph.EP},
       adsurl = {https://ui.adsabs.harvard.edu/abs/2023Natur.617..483T}
}

@ARTICLE{2011AA...531A..40F,
       author = {{Faedi}, F. and {Barros}, S.~C.~C. and {Anderson}, D.~R. and {Brown}, D.~J.~A. and {Collier Cameron}, A. and {Pollacco}, D. and {Boisse}, I. and {H{\'e}brard}, G. and {Lendl}, M. and {Lister}, T.~A. and {Smalley}, B. and {Street}, R.~A. and {Triaud}, A.~H.~M.~J. and {Bento}, J. and {Bouchy}, F. and {Butters}, O.~W. and {Enoch}, B. and {Haswell}, C.~A. and {Hellier}, C. and {Keenan}, F.~P. and {Miller}, G.~R.~M. and {Moulds}, V. and {Moutou}, C. and {Norton}, A.~J. and {Queloz}, D. and {Santerne}, A. and {Simpson}, E.~K. and {Skillen}, I. and {Smith}, A.~M.~S. and {Udry}, S. and {Watson}, C.~A. and {West}, R.~G. and {Wheatley}, P.~J.},
        title = "{WASP-39b: a highly inflated Saturn-mass planet orbiting a late G-type star}",
      journal = aap,
         year = 2011,
        month = jul,
       volume = {531},
          eid = {A40},
        pages = {A40},
          doi = {10.1051/0004-6361/201116671},
archivePrefix = {arXiv},
       eprint = {1102.1375},
 primaryClass = {astro-ph.EP},
       adsurl = {https://ui.adsabs.harvard.edu/abs/2011AA...531A..40F}
}

@ARTICLE{2014AA...564A.125B,
       author = {{Buchner}, J. and {Georgakakis}, A. and {Nandra}, K. and {Hsu}, L. and {Rangel}, C. and {Brightman}, M. and {Merloni}, A. and {Salvato}, M. and {Donley}, J. and {Kocevski}, D.},
        title = "{X-ray spectral modelling of the AGN obscuring region in the CDFS: Bayesian model selection and catalogue}",
      journal = aap,
         year = 2014,
        month = apr,
       volume = {564},
          eid = {A125},
        pages = {A125},
          doi = {10.1051/0004-6361/201322971},
archivePrefix = {arXiv},
       eprint = {1402.0004},
 primaryClass = {astro-ph.HE},
       adsurl = {https://ui.adsabs.harvard.edu/abs/2014AA...564A.125B}
}

@ARTICLE{2012AARv..20...56C,
       author = {{Caselli}, Paola and {Ceccarelli}, Cecilia},
        title = "{Our astrochemical heritage}",
      journal = aapr,
         year = 2012,
        month = oct,
       volume = {20},
          eid = {56},
        pages = {56},
          doi = {10.1007/s00159-012-0056-x},
archivePrefix = {arXiv},
       eprint = {1210.6368},
 primaryClass = {astro-ph.GA},
       adsurl = {https://ui.adsabs.harvard.edu/abs/2012A&ARv..20...56C}
}

@INPROCEEDINGS{2014prpl.conf..859C,
       author = {{Ceccarelli}, C. and {Caselli}, P. and {Bockel{\'e}e-Morvan}, D. and {Mousis}, O. and {Pizzarello}, S. and {Robert}, F. and {Semenov}, D.},
        title = "{Deuterium Fractionation: The Ariadne's Thread from the Precollapse Phase to Meteorites and Comets Today}",
    booktitle = {Protostars and Planets VI},
         year = 2014,
       editor = {{Beuther}, Henrik and {Klessen}, Ralf S. and {Dullemond}, Cornelis P. and {Henning}, Thomas},
        month = jan,
        pages = {859-882},
          doi = {10.2458/azu_uapress_9780816531240-ch037},
archivePrefix = {arXiv},
       eprint = {1403.7143},
 primaryClass = {astro-ph.EP},
       adsurl = {https://ui.adsabs.harvard.edu/abs/2014prpl.conf..859C}
}

@ARTICLE{2021MNRAS.500.4901D,
       author = {{Drozdovskaya}, Maria N. and {Schroeder I}, Isaac R.~H.~G. and {Rubin}, Martin and {Altwegg}, Kathrin and {van Dishoeck}, Ewine F. and {Kulterer}, Beatrice M. and {De Keyser}, Johan and {Fuselier}, Stephen A. and {Combi}, Michael},
        title = "{Prestellar grain-surface origins of deuterated methanol in comet 67P/Churyumov-Gerasimenko}",
      journal = mnras,
         year = 2021,
        month = feb,
       volume = {500},
       number = {4},
        pages = {4901-4920},
          doi = {10.1093/mnras/staa3387},
archivePrefix = {arXiv},
       eprint = {2010.12489},
 primaryClass = {astro-ph.EP},
       adsurl = {https://ui.adsabs.harvard.edu/abs/2021MNRAS.500.4901D}
}

@ARTICLE{2022ApJ...929...13C,
       author = {{Caselli}, Paola and {Pineda}, Jaime E. and {Sipil{\"a}}, Olli and {Zhao}, Bo and {Redaelli}, Elena and {Spezzano}, Silvia and {Maureira}, Maria Jos{\'e} and {Alves}, Felipe and {Bizzocchi}, Luca and {Bourke}, Tyler L. and {Chac{\'o}n-Tanarro}, Ana and {Friesen}, Rachel and {Galli}, Daniele and {Harju}, Jorma and {Jim{\'e}nez-Serra}, Izaskun and {Keto}, Eric and {Li}, Zhi-Yun and {Padovani}, Marco and {Schmiedeke}, Anika and {Tafalla}, Mario and {Vastel}, Charlotte},
        title = "{The Central 1000 au of a Prestellar Core Revealed with ALMA. II. Almost Complete Freeze-out}",
      journal = apj,
         year = 2022,
        month = apr,
       volume = {929},
       number = {1},
          eid = {13},
        pages = {13},
          doi = {10.3847/1538-4357/ac5913},
archivePrefix = {arXiv},
       eprint = {2202.13374},
 primaryClass = {astro-ph.SR},
       adsurl = {https://ui.adsabs.harvard.edu/abs/2022ApJ...929...13C}
}

@ARTICLE{2003AA...403L..37C,
       author = {{Caselli}, P. and {van der Tak}, F.~F.~S. and {Ceccarelli}, C. and {Bacmann}, A.},
        title = "{Abundant H$_{2}$D$^{+}$ in the pre-stellar core L1544}",
      journal = aap,
         year = 2003,
        month = may,
       volume = {403},
        pages = {L37-L41},
          doi = {10.1051/0004-6361:20030526},
archivePrefix = {arXiv},
       eprint = {astro-ph/0304103},
 primaryClass = {astro-ph},
       adsurl = {https://ui.adsabs.harvard.edu/abs/2003A&A...403L..37C}
}

@ARTICLE{2004AA...418.1035W,
       author = {{Walmsley}, C.~M. and {Flower}, D.~R. and {Pineau des For{\^e}ts}, G.},
        title = "{Complete depletion in prestellar cores}",
      journal = aap,
         year = 2004,
        month = may,
       volume = {418},
        pages = {1035-1043},
          doi = {10.1051/0004-6361:20035718},
archivePrefix = {arXiv},
       eprint = {astro-ph/0402493},
 primaryClass = {astro-ph},
       adsurl = {https://ui.adsabs.harvard.edu/abs/2004A&A...418.1035W}
}

@ARTICLE{2024A&A...685A..64K,
       author = {{Khorshid}, N. and {Min}, M. and {Polman}, J. and {Waters}, L.~B.~F.~M.},
        title = "{Constraining the formation of WASP-39b using JWST transit spectroscopy}",
      journal = aap,
         year = 2024,
        month = may,
       volume = {685},
          eid = {A64},
        pages = {A64},
          doi = {10.1051/0004-6361/202347124},
archivePrefix = {arXiv},
       eprint = {2405.12061},
 primaryClass = {astro-ph.EP},
       adsurl = {https://ui.adsabs.harvard.edu/abs/2024A&A...685A..64K}
}

@ARTICLE{2026arXiv260203498L,
       author = {{Lavail}, A. and {Debras}, F. and {Klein}, B. and {Chabrol}, E. and {Vinatier}, S. and {Hood}, T. and {Masson}, A. and {Seidel}, J.~V. and {Moutou}, C. and {Aigrain}, S. and {Meech}, A. and {Barrag{\'a}n}, O.},
        title = "{Atmospheric characterization of HIP 67522 b with VLT/CRIRES+. VLT/CRIRES+ suggests a heavier planet and hints at deuterium fractionation.}",
         year = 2026,
        month = feb,
          eid = {arXiv:2602.03498},
        pages = {arXiv:2602.03498},
          doi = {10.48550/arXiv.2602.03498},
archivePrefix = {arXiv},
       eprint = {2602.03498},
 primaryClass = {astro-ph.EP},
       adsurl = {https://ui.adsabs.harvard.edu/abs/2026arXiv260203498L}
}

@ARTICLE{2024ApJ...967..139C,
       author = {{Cherubim}, Collin and {Wordsworth}, Robin and {Hu}, Renyu and {Shkolnik}, Evgenya},
        title = "{Strong Fractionation of Deuterium and Helium in Sub-Neptune Atmospheres along the Radius Valley}",
      journal = apj,
         year = 2024,
        month = jun,
       volume = {967},
       number = {2},
          eid = {139},
        pages = {139},
          doi = {10.3847/1538-4357/ad3e77},
archivePrefix = {arXiv},
       eprint = {2402.10690},
 primaryClass = {astro-ph.EP},
       adsurl = {https://ui.adsabs.harvard.edu/abs/2024ApJ...967..139C}
}

@ARTICLE{2010MNRAS.402..492V,
       author = {{Voronin}, B.~A. and {Tennyson}, J. and {Tolchenov}, R.~N. and {Lugovskoy}, A.~A. and {Yurchenko}, S.~N.},
        title = "{A high accuracy computed line list for the HDO molecule}",
      journal = mnras,
         year = 2010,
        month = feb,
       volume = {402},
       number = {1},
        pages = {492-496},
          doi = {10.1111/j.1365-2966.2009.15904.x},
       adsurl = {https://ui.adsabs.harvard.edu/abs/2010MNRAS.402..492V}
}

@ARTICLE{2016MNRAS.459.3890U,
       author = {{Underwood}, Daniel S. and {Tennyson}, Jonathan and {Yurchenko}, Sergei N. and {Huang}, Xinchuan and {Schwenke}, David W. and {Lee}, Timothy J. and {Clausen}, S{\o}nnik and {Fateev}, Alexander},
        title = "{ExoMol molecular line lists - XIV. The rotation-vibration spectrum of hot SO$_{2}$}",
      journal = mnras,
         year = 2016,
        month = jul,
       volume = {459},
       number = {4},
        pages = {3890-3899},
          doi = {10.1093/mnras/stw849},
archivePrefix = {arXiv},
       eprint = {1603.04065},
 primaryClass = {astro-ph.EP},
       adsurl = {https://ui.adsabs.harvard.edu/abs/2016MNRAS.459.3890U}
}

@ARTICLE{2024MNRAS.528.3719Y,
       author = {{Yurchenko}, Sergei N. and {Owens}, Alec and {Kefala}, Kyriaki and {Tennyson}, Jonathan},
        title = "{ExoMol line lists - LVII. High accuracy ro-vibrational line list for methane (CH$_{4}$)}",
      journal = mnras,
         year = 2024,
        month = feb,
       volume = {528},
       number = {2},
        pages = {3719-3729},
          doi = {10.1093/mnras/stae148},
       adsurl = {https://ui.adsabs.harvard.edu/abs/2024MNRAS.528.3719Y}
}

@ARTICLE{2016MNRAS.460.4063A,
       author = {{Azzam}, Ala'a. A.~A. and {Tennyson}, Jonathan and {Yurchenko}, Sergei N. and {Naumenko}, Olga V.},
        title = "{ExoMol molecular line lists - XVI. The rotation-vibration spectrum of hot H$_{2}$S}",
      journal = mnras,
         year = 2016,
        month = aug,
       volume = {460},
       number = {4},
        pages = {4063-4074},
          doi = {10.1093/mnras/stw1133},
archivePrefix = {arXiv},
       eprint = {1607.00499},
 primaryClass = {astro-ph.EP},
       adsurl = {https://ui.adsabs.harvard.edu/abs/2016MNRAS.460.4063A}
}

@ARTICLE{2014MNRAS.437.1828B,
       author = {{Barber}, R.~J. and {Strange}, J.~K. and {Hill}, C. and {Polyansky}, O.~L. and {Mellau}, G. Ch. and {Yurchenko}, S.~N. and {Tennyson}, Jonathan},
        title = "{ExoMol line lists - III. An improved hot rotation-vibration line list for HCN and HNC}",
      journal = mnras,
         year = 2014,
        month = jan,
       volume = {437},
       number = {2},
        pages = {1828-1835},
          doi = {10.1093/mnras/stt2011},
archivePrefix = {arXiv},
       eprint = {1311.1328},
 primaryClass = {astro-ph.SR},
       adsurl = {https://ui.adsabs.harvard.edu/abs/2014MNRAS.437.1828B}
}

@ARTICLE{2021AA...646A..21C,
       author = {{Chubb}, Katy L. and {Rocchetto}, Marco and {Yurchenko}, Sergei N. and {Min}, Michiel and {Waldmann}, Ingo and {Barstow}, Joanna K. and {Molli{\`e}re}, Paul and {Al-Refaie}, Ahmed F. and {Phillips}, Mark W. and {Tennyson}, Jonathan},
        title = "{The ExoMolOP database: Cross sections and k-tables for molecules of interest in high-temperature exoplanet atmospheres}",
      journal = aap,
         year = 2021,
        month = feb,
       volume = {646},
          eid = {A21},
        pages = {A21},
          doi = {10.1051/0004-6361/202038350},
archivePrefix = {arXiv},
       eprint = {2009.00687},
 primaryClass = {astro-ph.EP},
       adsurl = {https://ui.adsabs.harvard.edu/abs/2021AA...646A..21C}
}

@ARTICLE{2016AA...589A..21A,
       author = {{Allard}, N.~F. and {Spiegelman}, F. and {Kielkopf}, J.~F.},
        title = "{K-H$_{2}$ line shapes for the spectra of cool brown dwarfs}",
      journal = aap,
         year = 2016,
        month = may,
       volume = {589},
          eid = {A21},
        pages = {A21},
          doi = {10.1051/0004-6361/201628270},
       adsurl = {https://ui.adsabs.harvard.edu/abs/2016AA...589A..21A}
}

@ARTICLE{2019AA...628A.120A,
       author = {{Allard}, N.~F. and {Spiegelman}, F. and {Leininger}, T. and {Molliere}, P.},
        title = "{New study of the line profiles of sodium perturbed by H$_{2}$}",
      journal = aap,
         year = 2019,
        month = aug,
       volume = {628},
          eid = {A120},
        pages = {A120},
          doi = {10.1051/0004-6361/201935593},
archivePrefix = {arXiv},
       eprint = {1908.01989},
 primaryClass = {astro-ph.SR},
       adsurl = {https://ui.adsabs.harvard.edu/abs/2019AA...628A.120A}
}

@ARTICLE{2016JMoSp.327...73T,
       author = {{Tennyson}, Jonathan and {Yurchenko}, Sergei N. and {Al-Refaie}, Ahmed F. and {Barton}, Emma J. and {Chubb}, Katy L. and {Coles}, Phillip A. and {Diamantopoulou}, S. and {Gorman}, Maire N. and {Hill}, Christian and {Lam}, Aden Z. and {Lodi}, Lorenzo and {McKemmish}, Laura K. and {Na}, Yueqi and {Owens}, Alec and {Polyansky}, Oleg L. and {Rivlin}, Tom and {Sousa-Silva}, Clara and {Underwood}, Daniel S. and {Yachmenev}, Andrey and {Zak}, Emil},
        title = "{The ExoMol database: Molecular line lists for exoplanet and other hot atmospheres}",
      journal = jms,
         year = 2016,
        month = sep,
       volume = {327},
        pages = {73-94},
          doi = {10.1016/j.jms.2016.05.002},
archivePrefix = {arXiv},
       eprint = {1603.05890},
 primaryClass = {astro-ph.GA},
       adsurl = {https://ui.adsabs.harvard.edu/abs/2016JMoSp.327...73T}
}

@ARTICLE{2018MNRAS.480.2597P,
       author = {{Polyansky}, Oleg L. and {Kyuberis}, Aleksandra A. and {Zobov}, Nikolai F. and {Tennyson}, Jonathan and {Yurchenko}, Sergei N. and {Lodi}, Lorenzo},
        title = "{ExoMol molecular line lists XXX: a complete high-accuracy line list for water}",
      journal = mnras,
         year = 2018,
        month = oct,
       volume = {480},
       number = {2},
        pages = {2597-2608},
          doi = {10.1093/mnras/sty1877},
archivePrefix = {arXiv},
       eprint = {1807.04529},
 primaryClass = {astro-ph.EP},
       adsurl = {https://ui.adsabs.harvard.edu/abs/2018MNRAS.480.2597P}
}

@ARTICLE{2020MNRAS.496.5282Y,
       author = {{Yurchenko}, S.~N. and {Mellor}, Thomas M. and {Freedman}, Richard S. and {Tennyson}, J.},
        title = "{ExoMol line lists - XXXIX. Ro-vibrational molecular line list for CO$_{2}$}",
      journal = mnras,
         year = 2020,
        month = aug,
       volume = {496},
       number = {4},
        pages = {5282-5291},
          doi = {10.1093/mnras/staa1874},
archivePrefix = {arXiv},
       eprint = {2007.02122},
 primaryClass = {astro-ph.EP},
       adsurl = {https://ui.adsabs.harvard.edu/abs/2020MNRAS.496.5282Y}
}

@ARTICLE{1965PPS....85..227C,
       author = {{Chan}, Y.~M. and {Dalgarno}, A.},
        title = "{The refractive index of helium}",
      journal = pps,
         year = 1965,
        month = feb,
       volume = {85},
       number = {2},
        pages = {227-230},
          doi = {10.1088/0370-1328/85/2/304},
       adsurl = {https://ui.adsabs.harvard.edu/abs/1965PPS....85..227C}
}

@ARTICLE{1962ApJ...136..690D,
       author = {{Dalgarno}, A. and {Williams}, D.~A.},
        title = "{Rayleigh Scattering by Molecular Hydrogen.}",
      journal = apj,
         year = 1962,
        month = sep,
       volume = {136},
        pages = {690-692},
          doi = {10.1086/147428},
       adsurl = {https://ui.adsabs.harvard.edu/abs/1962ApJ...136..690D}
}

@ARTICLE{2001JQSRT..68..235B,
       author = {{Borysow}, Aleksandra and {Jorgensen}, Uffe G. and {Fu}, Yi},
        title = "{High-temperature (1000-7000 K) collision-induced absorption of H''2 pairs computed from the first principles, with application to cool and dense stellar atmospheres}",
      journal = jqsrt,
         year = 2001,
        month = feb,
       volume = {68},
        pages = {235-255},
          doi = {10.1016/S0022-4073(00)00023-6},
       adsurl = {https://ui.adsabs.harvard.edu/abs/2001JQSRT..68..235B}
}

@ARTICLE{2002AA...390..779B,
       author = {{Borysow}, A.},
        title = "{Collision-induced absorption coefficients of H$_{2}$ pairs at temperatures from 60 K to 1000 K}",
      journal = aap,
         year = 2002,
        month = aug,
       volume = {390},
        pages = {779-782},
          doi = {10.1051/0004-6361:20020555},
       adsurl = {https://ui.adsabs.harvard.edu/abs/2002AA...390..779B}
}

@ARTICLE{1988ApJ...326..509B,
       author = {{Borysow}, Jacek and {Frommhold}, Lothar and {Birnbaum}, George},
        title = "{Collision-induced Rototranslational Absorption Spectra of H 2-He Pairs at Temperatures from 40 to 3000 K}",
      journal = apj,
         year = 1988,
        month = mar,
       volume = {326},
        pages = {509},
          doi = {10.1086/166112},
       adsurl = {https://ui.adsabs.harvard.edu/abs/1988ApJ...326..509B}
}

@ARTICLE{1989ApJ...336..495B,
       author = {{Borysow}, Aleksandra and {Frommhold}, Lothar and {Moraldi}, Massimo},
        title = "{Collision-induced Infrared Spectra of H 2-He Pairs Involving 0 1 Vibrational Transitions and Temperatures from 18 to 7000 K}",
      journal = apj,
         year = 1989,
        month = jan,
       volume = {336},
        pages = {495},
          doi = {10.1086/167027},
       adsurl = {https://ui.adsabs.harvard.edu/abs/1989ApJ...336..495B}
}

@ARTICLE{1989ApJ...341..549B,
       author = {{Borysow}, Aleksandra and {Frommhold}, Lothar},
        title = "{Collision-induced Infrared Spectra of H 2-He Pairs at Temperatures from 18 to 7000 K. II. Overtone and Hot Bands}",
      journal = apj,
         year = 1989,
        month = jun,
       volume = {341},
        pages = {549},
          doi = {10.1086/167515},
       adsurl = {https://ui.adsabs.harvard.edu/abs/1989ApJ...341..549B}
}

@ARTICLE{2019OJAp....2E..10F,
       author = {{Feroz}, Farhan and {Hobson}, Michael P. and {Cameron}, Ewan and {Pettitt}, Anthony N.},
        title = "{Importance Nested Sampling and the MultiNest Algorithm}",
      journal = ojap,
         year = 2019,
        month = nov,
       volume = {2},
       number = {1},
          eid = {10},
        pages = {10},
          doi = {10.21105/astro.1306.2144},
archivePrefix = {arXiv},
       eprint = {1306.2144},
 primaryClass = {astro-ph.IM},
       adsurl = {https://ui.adsabs.harvard.edu/abs/2019OJAp....2E..10F}
}

@ARTICLE{2009MNRAS.398.1601F,
       author = {{Feroz}, F. and {Hobson}, M.~P. and {Bridges}, M.},
        title = "{MULTINEST: an efficient and robust Bayesian inference tool for cosmology and particle physics}",
      journal = mnras,
         year = 2009,
        month = oct,
       volume = {398},
       number = {4},
        pages = {1601-1614},
          doi = {10.1111/j.1365-2966.2009.14548.x},
archivePrefix = {arXiv},
       eprint = {0809.3437},
 primaryClass = {astro-ph},
       adsurl = {https://ui.adsabs.harvard.edu/abs/2009MNRAS.398.1601F}
}

@ARTICLE{2008MNRAS.384..449F,
       author = {{Feroz}, F. and {Hobson}, M.~P.},
        title = "{Multimodal nested sampling: an efficient and robust alternative to Markov Chain Monte Carlo methods for astronomical data analyses}",
      journal = mnras,
         year = 2008,
        month = feb,
       volume = {384},
       number = {2},
        pages = {449-463},
          doi = {10.1111/j.1365-2966.2007.12353.x},
archivePrefix = {arXiv},
       eprint = {0704.3704},
 primaryClass = {astro-ph},
       adsurl = {https://ui.adsabs.harvard.edu/abs/2008MNRAS.384..449F}
}

@ARTICLE{2015ApJS..216...15L,
       author = {{Li}, Gang and {Gordon}, Iouli E. and {Rothman}, Laurence S. and {Tan}, Yan and {Hu}, Shui-Ming and {Kassi}, Samir and {Campargue}, Alain and {Medvedev}, Emile S.},
        title = "{Rovibrational Line Lists for Nine Isotopologues of the CO Molecule in the X $^{1}${\ensuremath{\Sigma}}$^{+}$ Ground Electronic State}",
      journal = apjs,
         year = 2015,
        month = jan,
       volume = {216},
       number = {1},
          eid = {15},
        pages = {15},
          doi = {10.1088/0067-0049/216/
1/15},
       adsurl = {https://ui.adsabs.harvard.edu/abs/2015ApJS..216...15L}
}

@ARTICLE{2020Natur.585..357H,
       author = {{Harris}, Charles R. and {Millman}, K. Jarrod and {van der Walt}, St{\'e}fan J. and {Gommers}, Ralf and {Virtanen}, Pauli and {Cournapeau}, David and {Wieser}, Eric and {Taylor}, Julian and {Berg}, Sebastian and {Smith}, Nathaniel J. and {Kern}, Robert and {Picus}, Matti and {Hoyer}, Stephan and {van Kerkwijk}, Marten H. and {Brett}, Matthew and {Haldane}, Allan and {del R{\'\i}o}, Jaime Fern{\'a}ndez and {Wiebe}, Mark and {Peterson}, Pearu and {G{\'e}rard-Marchant}, Pierre and {Sheppard}, Kevin and {Reddy}, Tyler and {Weckesser}, Warren and {Abbasi}, Hameer and {Gohlke}, Christoph and {Oliphant}, Travis E.},
        title = "{Array programming with NumPy}",
      journal = nat,
         year = 2020,
        month = sep,
       volume = {585},
       number = {7825},
        pages = {357-362},
          doi = {10.1038/s41586-020-2649-2},
archivePrefix = {arXiv},
       eprint = {2006.10256},
 primaryClass = {cs.MS},
       adsurl = {https://ui.adsabs.harvard.edu/abs/2020Natur.585..357H}
}

@ARTICLE{2007CSE.....9...90H,
       author = {{Hunter}, John D.},
        title = "{Matplotlib: A 2D Graphics Environment}",
      journal = cse,
         year = 2007,
        month = may,
       volume = {9},
       number = {3},
        pages = {90-95},
          doi = {10.1109/MCSE.2007.55},
       adsurl = {https://ui.adsabs.harvard.edu/abs/2007CSE.....9...90H}
}

@ARTICLE{2021JOSS....6.3021W,
       author = {{Waskom}, Michael},
        title = "{seaborn: statistical data visualization}",
      journal = joss,
         year = 2021,
        month = apr,
       volume = {6},
       number = {60},
          eid = {3021},
        pages = {3021},
          doi = {10.21105/joss.03021},
       adsurl = {https://ui.adsabs.harvard.edu/abs/2021JOSS....6.3021W}
}

@ARTICLE{9965751,
  author={Rogowski, Marcin and Aseeri, Samar and Keyes, David and Dalcin, Lisandro},
  journal={IEEE Transactions on Parallel and Distributed Systems}, 
  title={mpi4py.futures: MPI-Based Asynchronous Task Execution for Python}, 
  year={2023},
  volume={34},
  number={2},
  pages={611-622},
  doi={10.1109/TPDS.2022.3225481}}

@ARTICLE{9439927,
  author={Dalcin, Lisandro and Fang, Yao-Lung L.},
  journal={Computing in Science \& Engineering}, 
  title={mpi4py: Status Update After 12 Years of Development}, 
  year={2021},
  volume={23},
  number={4},
  pages={47-54},
  doi={10.1109/MCSE.2021.3083216}}

@article{DALCIN20111124,
title = {Parallel distributed computing using Python},
journal = awr,
volume = {34},
number = {9},
pages = {1124-1139},
year = {2011},
note = {New Computational Methods and Software Tools},
issn = {0309-1708},
doi = {https://doi.org/10.1016/j.advwatres.2011.04.013},
url = {https://www.sciencedirect.com/science/article/pii/S0309170811000777},
author = {Lisandro D. Dalcin and Rodrigo R. Paz and Pablo A. Kler and Alejandro Cosimo}
}

@article{DALCIN20051108,
title = {MPI for Python},
journal = jpdc,
volume = {65},
number = {9},
pages = {1108-1115},
year = {2005},
issn = {0743-7315},
doi = {https://doi.org/10.1016/j.jpdc.2005.03.010},
url = {https://www.sciencedirect.com/science/article/pii/S0743731505000560},
author = {Lisandro Dalcín and Rodrigo Paz and Mario Storti}
}

@article{DALCIN2008655,
title = {MPI for Python: Performance improvements and MPI-2 extensions},
journal = jpdc,
volume = {68},
number = {5},
pages = {655-662},
year = {2008},
issn = {0743-7315},
doi = {https://doi.org/10.1016/j.jpdc.2007.09.005},
url = {https://www.sciencedirect.com/science/article/pii/S0743731507001712},
author = {Lisandro Dalcín and Rodrigo Paz and Mario Storti and Jorge D’Elía}
}

@ARTICLE{2020SciPy-NMeth,
  author  = {Virtanen, Pauli and Gommers, Ralf and Oliphant, Travis E. and
            Haberland, Matt and Reddy, Tyler and Cournapeau, David and
            Burovski, Evgeni and Peterson, Pearu and Weckesser, Warren and
            Bright, Jonathan and {van der Walt}, St{\'e}fan J. and
            Brett, Matthew and Wilson, Joshua and Millman, K. Jarrod and
            Mayorov, Nikolay and Nelson, Andrew R. J. and Jones, Eric and
            Kern, Robert and Larson, Eric and Carey, C J and
            Polat, {\.I}lhan and Feng, Yu and Moore, Eric W. and
            {VanderPlas}, Jake and Laxalde, Denis and Perktold, Josef and
            Cimrman, Robert and Henriksen, Ian and Quintero, E. A. and
            Harris, Charles R. and Archibald, Anne M. and
            Ribeiro, Ant{\^o}nio H. and Pedregosa, Fabian and
            {van Mulbregt}, Paul and {SciPy 1.0 Contributors}},
  title   = {{{SciPy} 1.0: Fundamental Algorithms for Scientific
            Computing in Python}},
  journal = {Nature Methods},
  year    = {2020},
  volume  = {17},
  pages   = {261--272},
  adsurl  = {https://rdcu.be/b08Wh},
  doi     = {10.1038/s41592-019-0686-2},
}

@misc{reback2020pandas,
    author       = {{The pandas development team}},
    title        = {pandas-dev/pandas: Pandas},
    month        = feb,
    year         = 2020,
    publisher    = {Zenodo},
    version      = {latest},
    doi          = {10.5281/zenodo.3509134},
    url          = {https://doi.org/10.5281/zenodo.3509134}
}

@InProceedings{ mckinney-proc-scipy-2010,
  author    = { {W}es {M}c{K}inney },
  title     = { {D}ata {S}tructures for {S}tatistical {C}omputing in {P}ython },
  booktitle = { {P}roceedings of the 9th {P}ython in {S}cience {C}onference },
  pages     = { 56 - 61 },
  year      = { 2010 },
  editor    = { {S}t\'efan van der {W}alt and {J}arrod {M}illman },
  doi       = { 10.25080/Majora-92bf1922-00a }
}

@ARTICLE{2025arXiv250719223D,
       author = {{Davey}, Jack J. and {Hou Yip}, Kai and {Changeat}, Quentin and {Waldmann}, Ingo P.},
        title = "{Investigating the Influence of Asymmetric Errors on Retrievals of Exoplanet Transmission Spectra.}",
         year = 2025,
        month = jul,
          eid = {arXiv:2507.19223},
        pages = {arXiv:2507.19223},
          doi = {10.48550/arXiv.2507.19223},
archivePrefix = {arXiv},
       eprint = {2507.19223},
 primaryClass = {astro-ph.EP},
       adsurl = {https://ui.adsabs.harvard.edu/abs/2025arXiv250719223D}
}

@ARTICLE{Wrong,
       author = {{Thorngren}, Daniel P. and {Sing}, David K. and {Mukherjee}, Sagnick},
        title = "{Bayesian Model Comparison and Significance: Widespread Errors and how to Correct Them.}",
         year = 2025,
        month = sep,
          eid = {arXiv:2510.00169},
        pages = {arXiv:2510.00169},
          doi = {10.48550/arXiv.2510.00169},
archivePrefix = {arXiv},
       eprint = {2510.00169},
 primaryClass = {astro-ph.EP},
       adsurl = {https://ui.adsabs.harvard.edu/abs/2025arXiv251000169T}
}

@ARTICLE{2020NatCo..11.5444C,
       author = {{Crameri}, Fabio and {Shephard}, Grace E. and {Heron}, Philip J.},
        title = "{The misuse of colour in science communication}",
      journal = {Nature Communications},
         year = 2020,
        month = oct,
       volume = {11},
          eid = {5444},
        pages = {5444},
          doi = {10.1038/s41467-020-19160-7},
       adsurl = {https://ui.adsabs.harvard.edu/abs/2020NatCo..11.5444C}
}

@misc{2021zndo...1243862C,
       author = {{Crameri}, Fabio},
        title = "{Scientific colour maps}",
         year = 2023,
        month = oct,
          eid = {10.5281/zenodo.1243862},
          doi = {10.5281/zenodo.1243862},
      version = {8.0.1},
    publisher = {Zenodo},
       adsurl = {https://ui.adsabs.harvard.edu/abs/2021zndo...1243862C}
}

@ARTICLE{2018GMD....11.2541C,
       author = {{Crameri}, Fabio},
        title = "{Geodynamic diagnostics, scientific visualisation and StagLab 3.0}",
      journal = {Geoscientific Model Development},
         year = 2018,
        month = jun,
       volume = {11},
       number = {6},
        pages = {2541-2562},
          doi = {10.5194/gmd-11-2541-2018},
       adsurl = {https://ui.adsabs.harvard.edu/abs/2018GMD....11.2541C}
}

@misc{Rollo_cmcrameri_2024,
  author       = {Callum Rollo},
  title        = {cmcrameri. https://github.com/callumrollo/cmcrameri},
  year         = {2024},
  note         = {Version 1.9},
  adsurl = {\url{https://github.com/callumrollo/cmcrameri}}
}

@article{Sellke2001,
  title = {Calibration of p Values for Testing Precise Null Hypotheses},
  volume = {55},
  ISSN = {1537-2731},
  url = {http://dx.doi.org/10.1198/000313001300339950},
  DOI = {10.1198/000313001300339950},
  number = {1},
  journal = amstat,
  publisher = {Informa UK Limited},
  author = {Sellke,  Thomas and Bayarri,  M. J and Berger,  James O},
  year = {2001},
  month = feb,
  pages = {62–71}
}

@article{article,
author = {Laeter, John and Bohlke, J. and De Bièvre, Paul and Hidaka, Hiroshi and Peiser, H. and Rosman, K. and Taylor, Philip},
year = {2009},
month = {01},
pages = {1535-1536},
title = {Atomic weights of the elements: Review 2000 (IUPAC Technical Report)},
volume = {81},
journal = pac,
doi = {10.1351/PAC-REP-09-06-03_errata}
}

@ARTICLE{2010AA...518L.152L,
       author = {{Lellouch}, E. and {Hartogh}, P. and {Feuchtgruber}, H. and {Vandenbussche}, B. and {de Graauw}, T. and {Moreno}, R. and {Jarchow}, C. and {Cavali{\'e}}, T. and {Orton}, G. and {Banaszkiewicz}, M. and {Blecka}, M.~I. and {Bockel{\'e}e-Morvan}, D. and {Crovisier}, J. and {Encrenaz}, T. and {Fulton}, T. and {K{\"u}ppers}, M. and {Lara}, L.~M. and {Lis}, D.~C. and {Medvedev}, A.~S. and {Rengel}, M. and {Sagawa}, H. and {Swinyard}, B. and {Szutowicz}, S. and {Bensch}, F. and {Bergin}, E. and {Billebaud}, F. and {Biver}, N. and {Blake}, G.~A. and {Blommaert}, J.~A.~D.~L. and {Cernicharo}, J. and {Courtin}, R. and {Davis}, G.~R. and {Decin}, L. and {Encrenaz}, P. and {Gonzalez}, A. and {Jehin}, E. and {Kidger}, M. and {Naylor}, D. and {Portyankina}, G. and {Schieder}, R. and {Sidher}, S. and {Thomas}, N. and {de Val-Borro}, M. and {Verdugo}, E. and {Waelkens}, C. and {Walker}, H. and {Aarts}, H. and {Comito}, C. and {Kawamura}, J.~H. and {Maestrini}, A. and {Peacocke}, T. and {Teipen}, R. and {Tils}, T. and {Wildeman}, K.},
        title = "{First results of Herschel-PACS observations of Neptune}",
      journal = aap,
         year = 2010,
        month = jul,
       volume = {518},
          eid = {L152},
        pages = {L152},
          doi = {10.1051/0004-6361/201014600},
archivePrefix = {arXiv},
       eprint = {1006.0114},
 primaryClass = {astro-ph.EP},
       adsurl = {https://ui.adsabs.harvard.edu/abs/2010AA...518L.152L}
}

@ARTICLE{2009Natur.460..487W,
       author = {{Waite}, Jr., J.~H. and {Lewis}, W.~S. and {Magee}, B.~A. and {Lunine}, J.~I. and {McKinnon}, W.~B. and {Glein}, C.~R. and {Mousis}, O. and {Young}, D.~T. and {Brockwell}, T. and {Westlake}, J. and {Nguyen}, M. -J. and {Teolis}, B.~D. and {Niemann}, H.~B. and {McNutt}, R.~L. and {Perry}, M. and {Ip}, W. -H.},
        title = "{Liquid water on Enceladus from observations of ammonia and $^{40}$Ar in the plume}",
      journal = nat,
         year = 2009,
        month = jul,
       volume = {460},
       number = {7254},
        pages = {487-490},
          doi = {10.1038/nature08153},
       adsurl = {https://ui.adsabs.harvard.edu/abs/2009Natur.460..487W}
}

@ARTICLE{2008AA...490L..31H,
       author = {{Hutsem{\'e}kers}, D. and {Manfroid}, J. and {Jehin}, E. and {Zucconi}, J. -M. and {Arpigny}, C.},
        title = "{The $^{\{16}$\}OH/$^{18}$OH and OD/OH isotope ratios in comet C/2002 T7 (LINEAR)}",
      journal = aap,
         year = 2008,
        month = nov,
       volume = {490},
       number = {3},
        pages = {L31-L34},
          doi = {10.1051/0004-6361:200810833},
archivePrefix = {arXiv},
       eprint = {0809.4300},
 primaryClass = {astro-ph},
       adsurl = {https://ui.adsabs.harvard.edu/abs/2008AA...490L..31H}
}

@ARTICLE{2006AA...449.1255B,
       author = {{Biver}, N. and {Bockel{\'e}e-Morvan}, D. and {Crovisier}, J. and {Lis}, D.~C. and {Moreno}, R. and {Colom}, P. and {Henry}, F. and {Herpin}, F. and {Paubert}, G. and {Womack}, M.},
        title = "{Radio wavelength molecular observations of comets C/1999 T1 (McNaught-Hartley), C/2001 A2 (LINEAR), C/2000 WM$_{1}$ (LINEAR) and 153P/Ikeya-Zhang}",
      journal = aap,
         year = 2006,
        month = apr,
       volume = {449},
       number = {3},
        pages = {1255-1270},
          doi = {10.1051/0004-6361:20053849},
       adsurl = {https://ui.adsabs.harvard.edu/abs/2006AA...449.1255B}
}

@ARTICLE{2001AA...370..610L,
       author = {{Lellouch}, E. and {B{\'e}zard}, B. and {Fouchet}, T. and {Feuchtgruber}, H. and {Encrenaz}, T. and {de Graauw}, T.},
        title = "{The deuterium abundance in Jupiter and Saturn from ISO-SWS observations}",
      journal = aap,
         year = 2001,
        month = may,
       volume = {370},
        pages = {610-622},
          doi = {10.1051/0004-6361:20010259},
       adsurl = {https://ui.adsabs.harvard.edu/abs/2001AA...370..610L}
}

@ARTICLE{1999AA...341L..17F,
       author = {{Feuchtgruber}, H. and {Lellouch}, E. and {B{\'e}zard}, B. and {Encrenaz}, Th. and {de Graauw}, Th. and {Davis}, G.~R.},
        title = "{Detection of HD in the atmospheres of Uranus and Neptune: a new determination of the D/H ratio}",
      journal = aap,
         year = 1999,
        month = jan,
       volume = {341},
        pages = {L17-L21},
       adsurl = {https://ui.adsabs.harvard.edu/abs/1999AA...341L..17F}
}

@ARTICLE{1998Icar..133..147B,
       author = {{Bockel{\'e}e-Morvan}, D. and {Gautier}, D. and {Lis}, D.~C. and {Young}, K. and {Keene}, J. and {Phillips}, T. and {Owen}, T. and {Crovisier}, J. and {Goldsmith}, P.~F. and {Bergin}, E.~A. and {Despois}, D. and {Wootten}, A.},
        title = "{Deuterated Water in Comet C/1996 B2 (Hyakutake) and Its Implications for the Origin of Comets}",
      journal = icarus,
         year = 1998,
        month = may,
       volume = {133},
       number = {1},
        pages = {147-162},
          doi = {10.1006/icar.1998.5916},
       adsurl = {https://ui.adsabs.harvard.edu/abs/1998Icar..133..147B}
}

@ARTICLE{2009ApJ...690L...5V,
       author = {{Villanueva}, G.~L. and {Mumma}, M.~J. and {Bonev}, B.~P. and {Di Santi}, M.~A. and {Gibb}, E.~L. and {B{\"o}hnhardt}, H. and {Lippi}, M.},
        title = "{A Sensitive Search for Deuterated Water in Comet 8p/Tuttle}",
      journal = apjl,
         year = 2009,
        month = jan,
       volume = {690},
       number = {1},
        pages = {L5-L9},
          doi = {10.1088/0004-637X/690/1/L5},
       adsurl = {https://ui.adsabs.harvard.edu/abs/2009ApJ...690L...5V}
}

@ARTICLE{2012PSS...60..166B,
       author = {{Brown}, Robert H. and {Lauretta}, Dante S. and {Schmidt}, Britney and {Moores}, John},
        title = "{Experimental and theoretical simulations of ice sublimation with implications for the chemical, isotopic, and physical evolution of icy objects}",
      journal = planss,
         year = 2012,
        month = jan,
       volume = {60},
       number = {1},
        pages = {166-180},
          doi = {10.1016/j.p&ss.2011.07.023},
       adsurl = {https://ui.adsabs.harvard.edu/abs/2012P&SS...60..166B}
}

@ARTICLE{2012ApJ...750..102G,
       author = {{Gibb}, Erika L. and {Bonev}, Boncho P. and {Villanueva}, Geronimo and {DiSanti}, Michael A. and {Mumma}, Michael J. and {Sudholt}, Emily and {Radeva}, Yana},
        title = "{Chemical Composition of Comet C/2007 N3 (Lulin): Another ``Atypical'' Comet}",
      journal = apj,
         year = 2012,
        month = may,
       volume = {750},
       number = {2},
          eid = {102},
        pages = {102},
          doi = {10.1088/0004-637X/750/2/102},
       adsurl = {https://ui.adsabs.harvard.edu/abs/2012ApJ...750..102G}
}

@ARTICLE{2012AA...544L..15B,
       author = {{Bockel{\'e}e-Morvan}, D. and {Biver}, N. and {Swinyard}, B. and {de Val-Borro}, M. and {Crovisier}, J. and {Hartogh}, P. and {Lis}, D.~C. and {Moreno}, R. and {Szutowicz}, S. and {Lellouch}, E. and {Emprechtinger}, M. and {Blake}, G.~A. and {Courtin}, R. and {Jarchow}, C. and {Kidger}, M. and {K{\"u}ppers}, M. and {Rengel}, M. and {Davis}, G.~R. and {Fulton}, T. and {Naylor}, D. and {Sidher}, S. and {Walker}, H.},
        title = "{Herschel measurements of the D/H and $^{16}$O/$^{18}$O ratios in water in the Oort-cloud comet C/2009 P1 (Garradd)}",
      journal = aap,
         year = 2012,
        month = aug,
       volume = {544},
          eid = {L15},
        pages = {L15},
          doi = {10.1051/0004-6361/201219744},
archivePrefix = {arXiv},
       eprint = {1207.7180},
 primaryClass = {astro-ph.EP},
       adsurl = {https://ui.adsabs.harvard.edu/abs/2012AA...544L..15B}
}

@ARTICLE{2012Sci...337..721A,
       author = {{Alexander}, C.~M. O'D. and {Bowden}, R. and {Fogel}, M.~L. and {Howard}, K.~T. and {Herd}, C.~D.~K. and {Nittler}, L.~R.},
        title = "{The Provenances of Asteroids, and Their Contributions to the Volatile Inventories of the Terrestrial Planets}",
      journal = {Science},
         year = 2012,
        month = aug,
       volume = {337},
       number = {6095},
        pages = {721},
          doi = {10.1126/science.1223474},
       adsurl = {https://ui.adsabs.harvard.edu/abs/2012Sci...337..721A}
}

@ARTICLE{2013ApJ...774L...3L,
       author = {{Lis}, D.~C. and {Biver}, N. and {Bockel{\'e}e-Morvan}, D. and {Hartogh}, P. and {Bergin}, E.~A. and {Blake}, G.~A. and {Crovisier}, J. and {de Val-Borro}, M. and {Jehin}, E. and {K{\"u}ppers}, M. and {Manfroid}, J. and {Moreno}, R. and {Rengel}, M. and {Szutowicz}, S.},
        title = "{A Herschel Study of D/H in Water in the Jupiter-family Comet 45P/Honda-Mrkos-Pajdu{\v{s}}{\'a}kov{\'a} and Prospects for D/H Measurements with CCAT}",
      journal = apjl,
         year = 2013,
        month = sep,
       volume = {774},
       number = {1},
          eid = {L3},
        pages = {L3},
          doi = {10.1088/2041-8205/774/1/L3},
archivePrefix = {arXiv},
       eprint = {1307.6869},
 primaryClass = {astro-ph.EP},
       adsurl = {https://ui.adsabs.harvard.edu/abs/2013ApJ...774L...3L}
}

@ARTICLE{2016AA...589A..78B,
       author = {{Biver}, N. and {Moreno}, R. and {Bockel{\'e}e-Morvan}, D. and {Sandqvist}, Aa. and {Colom}, P. and {Crovisier}, J. and {Lis}, D.~C. and {Boissier}, J. and {Debout}, V. and {Paubert}, G. and {Milam}, S. and {Hjalmarson}, A. and {Lundin}, S. and {Karlsson}, T. and {Battelino}, M. and {Frisk}, U. and {Murtagh}, D. and {Odin Team}},
        title = "{Isotopic ratios of H, C, N, O, and S in comets C/2012 F6 (Lemmon) and C/2014 Q2 (Lovejoy)}",
      journal = aap,
         year = 2016,
        month = may,
       volume = {589},
          eid = {A78},
        pages = {A78},
          doi = {10.1051/0004-6361/201528041},
archivePrefix = {arXiv},
       eprint = {1603.05006},
 primaryClass = {astro-ph.EP},
       adsurl = {https://ui.adsabs.harvard.edu/abs/2016AA...589A..78B}
}

@ARTICLE{2019AA...625L...5L,
       author = {{Lis}, Dariusz C. and {Bockel{\'e}e-Morvan}, Dominique and {G{\"u}sten}, Rolf and {Biver}, Nicolas and {Stutzki}, J{\"u}rgen and {Delorme}, Yan and {Dur{\'a}n}, Carlos and {Wiesemeyer}, Helmut and {Okada}, Yoko},
        title = "{Terrestrial deuterium-to-hydrogen ratio in water in hyperactive comets}",
      journal = aap,
         year = 2019,
        month = may,
       volume = {625},
          eid = {L5},
        pages = {L5},
          doi = {10.1051/0004-6361/201935554},
archivePrefix = {arXiv},
       eprint = {1904.09175},
 primaryClass = {astro-ph.EP},
       adsurl = {https://ui.adsabs.harvard.edu/abs/2019AA...625L...5L}
}

@ARTICLE{2003AA...399.1009D,
       author = {{Dartois}, E. and {Thi}, W. -F. and {Geballe}, T.~R. and {Deboffle}, D. and {d'Hendecourt}, L. and {van Dishoeck}, E.},
        title = "{Revisiting the solid HDO/H$_{2}$O abundances}",
      journal = aap,
         year = 2003,
        month = mar,
       volume = {399},
        pages = {1009-1020},
          doi = {10.1051/0004-6361:20021558},
       adsurl = {https://ui.adsabs.harvard.edu/abs/2003AA...399.1009D}
}

@ARTICLE{2003AA...410..897P,
       author = {{Parise}, B. and {Simon}, T. and {Caux}, E. and {Dartois}, E. and {Ceccarelli}, C. and {Rayner}, J. and {Tielens}, A.~G.~G.~M.},
        title = "{Search for solid HDO in low-mass protostars}",
      journal = aap,
         year = 2003,
        month = nov,
       volume = {410},
        pages = {897-904},
          doi = {10.1051/0004-6361:20031277},
archivePrefix = {arXiv},
       eprint = {astro-ph/0309401},
 primaryClass = {astro-ph},
       adsurl = {https://ui.adsabs.harvard.edu/abs/2003AA...410..897P}
}

@ARTICLE{2006AA...447.1011V,
       author = {{van der Tak}, F.~F.~S. and {Walmsley}, C.~M. and {Herpin}, F. and {Ceccarelli}, C.},
        title = "{Water in the envelopes and disks around young high-mass stars}",
      journal = aap,
         year = 2006,
        month = mar,
       volume = {447},
       number = {3},
        pages = {1011-1025},
          doi = {10.1051/0004-6361:20053937},
archivePrefix = {arXiv},
       eprint = {astro-ph/0510640},
 primaryClass = {astro-ph},
       adsurl = {https://ui.adsabs.harvard.edu/abs/2006AA...447.1011V}
}

@ARTICLE{2013ApJ...765...61E,
       author = {{Emprechtinger}, M. and {Lis}, D.~C. and {Rolffs}, R. and {Schilke}, P. and {Monje}, R.~R. and {Comito}, C. and {Ceccarelli}, C. and {Neufeld}, D.~A. and {van der Tak}, F.~F.~S.},
        title = "{The Abundance, Ortho/Para Ratio, and Deuteration of Water in the High-mass Star-forming Region NGC 6334 I}",
      journal = apj,
         year = 2013,
        month = mar,
       volume = {765},
       number = {1},
          eid = {61},
        pages = {61},
          doi = {10.1088/0004-637X/765/1/61},
archivePrefix = {arXiv},
       eprint = {1212.5169},
 primaryClass = {astro-ph.GA},
       adsurl = {https://ui.adsabs.harvard.edu/abs/2013ApJ...765...61E}
}

@ARTICLE{2013ApJ...770..142N,
       author = {{Neill}, Justin L. and {Wang}, Shiya and {Bergin}, Edwin A. and {Crockett}, Nathan R. and {Favre}, C{\'e}cile and {Plume}, Ren{\'e} and {Melnick}, Gary J.},
        title = "{The Abundance of H$_{2}$O and HDO in Orion Kl from Herschel/HIFI}",
      journal = apj,
         year = 2013,
        month = jun,
       volume = {770},
       number = {2},
          eid = {142},
        pages = {142},
          doi = {10.1088/0004-637X/770/2/142},
archivePrefix = {arXiv},
       eprint = {1305.2247},
 primaryClass = {astro-ph.GA},
       adsurl = {https://ui.adsabs.harvard.edu/abs/2013ApJ...770..142N}
}

@ARTICLE{2014AA...563A..74P,
       author = {{Persson}, M.~V. and {J{\o}rgensen}, J.~K. and {van Dishoeck}, E.~F. and {Harsono}, D.},
        title = "{The deuterium fractionation of water on solar-system scales in deeply-embedded low-mass protostars}",
      journal = aap,
         year = 2014,
        month = mar,
       volume = {563},
          eid = {A74},
        pages = {A74},
          doi = {10.1051/0004-6361/201322845},
archivePrefix = {arXiv},
       eprint = {1402.1398},
 primaryClass = {astro-ph.SR},
       adsurl = {https://ui.adsabs.harvard.edu/abs/2014AA...563A..74P}
}

@ARTICLE{2014MNRAS.445.1299C,
       author = {{Coutens}, A. and {Vastel}, C. and {Hincelin}, U. and {Herbst}, E. and {Lis}, D.~C. and {Chavarr{\'\i}a}, L. and {G{\'e}rin}, M. and {van der Tak}, F.~F.~S. and {Persson}, C.~M. and {Goldsmith}, P.~F. and {Caux}, E.},
        title = "{Water deuterium fractionation in the high-mass star-forming region G34.26+0.15 based on Herschel/HIFI data}",
      journal = mnras,
         year = 2014,
        month = dec,
       volume = {445},
       number = {2},
        pages = {1299-1313},
          doi = {10.1093/mnras/stu1816},
archivePrefix = {arXiv},
       eprint = {1409.1092},
 primaryClass = {astro-ph.SR},
       adsurl = {https://ui.adsabs.harvard.edu/abs/2014MNRAS.445.1299C}
}

@ARTICLE{2019AA...631A..25J,
       author = {{Jensen}, S.~S. and {J{\o}rgensen}, J.~K. and {Kristensen}, L.~E. and {Furuya}, K. and {Coutens}, A. and {van Dishoeck}, E.~F. and {Harsono}, D. and {Persson}, M.~V.},
        title = "{ALMA observations of water deuteration: a physical diagnostic of the formation of protostars}",
      journal = aap,
         year = 2019,
        month = nov,
       volume = {631},
          eid = {A25},
        pages = {A25},
          doi = {10.1051/0004-6361/201936012},
archivePrefix = {arXiv},
       eprint = {1909.10533},
 primaryClass = {astro-ph.SR},
       adsurl = {https://ui.adsabs.harvard.edu/abs/2019AA...631A..25J}
}

@ARTICLE{2021AA...650A.172J,
       author = {{Jensen}, S.~S. and {J{\o}rgensen}, J.~K. and {Kristensen}, L.~E. and {Coutens}, A. and {van Dishoeck}, E.~F. and {Furuya}, K. and {Harsono}, D. and {Persson}, M.~V.},
        title = "{ALMA observations of doubly deuterated water: inheritance of water from the prestellar environment}",
      journal = aap,
         year = 2021,
        month = jun,
       volume = {650},
          eid = {A172},
        pages = {A172},
          doi = {10.1051/0004-6361/202140560},
archivePrefix = {arXiv},
       eprint = {2104.13411},
 primaryClass = {astro-ph.GA},
       adsurl = {https://ui.adsabs.harvard.edu/abs/2021AA...650A.172J}
}

@ARTICLE{2023AA...677L..17A,
       author = {{Andreu}, A. and {Coutens}, A. and {Cruz-S{\'a}enz de Miera}, F. and {Houry}, N. and {J{\o}rgensen}, J.~K. and {K{\'o}sp{\'a}l}, A. and {Harsono}, D.},
        title = "{A high HDO/H$_{2}$O ratio in the Class I protostar L1551 IRS5}",
      journal = aap,
         year = 2023,
        month = sep,
       volume = {677},
          eid = {L17},
        pages = {L17},
          doi = {10.1051/0004-6361/202347484},
archivePrefix = {arXiv},
       eprint = {2309.01688},
 primaryClass = {astro-ph.SR},
       adsurl = {https://ui.adsabs.harvard.edu/abs/2023AA...677L..17A}
}

@ARTICLE{2024AA...688A..29S,
       author = {{Slavicinska}, Katerina and {van Dishoeck}, Ewine F. and {Tychoniec}, {\L}ukasz and {Nazari}, Pooneh and {Rubinstein}, Adam E. and {Gutermuth}, Robert and {Tyagi}, Himanshu and {Chen}, Yuan and {Brunken}, Nashanty G.~C. and {Rocha}, Will R.~M. and {Manoj}, P. and {Narang}, Mayank and {Megeath}, S. Thomas and {Yang}, Yao-Lun and {Looney}, Leslie W. and {Tobin}, John J. and {Beuther}, Henrik and {Bourke}, Tyler L. and {Linnartz}, Harold and {Federman}, Samuel and {Watson}, Dan M. and {Linz}, Hendrik},
        title = "{JWST detections of amorphous and crystalline HDO ice toward massive protostars}",
      journal = aap,
         year = 2024,
        month = aug,
       volume = {688},
          eid = {A29},
        pages = {A29},
          doi = {10.1051/0004-6361/202449785},
archivePrefix = {arXiv},
       eprint = {2404.15399},
 primaryClass = {astro-ph.SR},
       adsurl = {https://ui.adsabs.harvard.edu/abs/2024AA...688A..29S}
}
\bibliographystyle{Bibstyle}

% After the paper has completed peer review and been revised ready for acceptance,
% you should comment out the lines above and copy-paste the contents of your .bbl
% file here instead. This will help ensure that our conversion software works correctly.
% Remember to re-run BibTeX first - check the timestamp!
%
% Example of the first three entries copy-pasted from s_template.bbl:
%
%\begin{thebibliography}{1}
%
%\bibitem{example}
%A.~N. {Author}, An example reference. \emph{Journal of Improbable Research}
%  \textbf{1}, 67 (2020).
%
%\bibitem{example2}
%F.~M. {Surname}, S.~{Author}, A second example. \emph{Interesting Research
%  Letters} \textbf{32}, 897 (2019).
%
%\bibitem{example_preprint}
%P.~{One}, P.~{Two}, P.~{Three}, {An unpublished preprint}. \emph{preprint}
%  (2021), arXiv:2101.12345.
%
%\end{thebibliography}

%%%%%%%%%%%%%%%% ACKNOWLEDGEMENTS %%%%%%%%%%%%%%%

\section*{Acknowledgments}
This research has made use of the NASA Exoplanet Archive, which is operated by the California Institute of Technology, under contract with the National Aeronautics and Space Administration under the Exoplanet Exploration Program. This research has made use of the Astrophysics Data System, funded by NASA under Cooperative Agreement 80NSSC21M0056.
%\paragraph*{Funding:}
We acknowledge support of the DFG (German Research Foundation), and of the Excellence Cluster ORIGINS - EXC-2094 - 390783311. K. L. C. is funded by UK Research and Innovation (UKRI) under the UK government’s Horizon Europe funding guarantee as part of an ERC Starter Grant [grant number EP/Y006313/1]. We acknowledge support from the Max-Planck-Society.
%\paragraph*{Author contributions:}
%F.G., K.M., and B.E. designed the analysis strategy. K.L.C. calculated the HDO opacities. F.G. executed the experiments and analysed/visualised the results. F.G., K.M., B.E.,  P.C., and T.G. conceptualized the interpretation of the results. F.G. prepared the manuscript with critical review and proofreading provided by K.M., B.E., K.L.C., P.C., T.G., R.A.-L., and D.D. All authors substantially contributed to the submitted paper at various stages.
%\paragraph*{Competing interests:}
%There are no competing interests to declare.
%\paragraph*{Data, code and materials availability:}
The JWST spectra and input data used in this work are publicly available from the repositories cited in the manuscript, including corresponding Zenodo records\cite{carter_2024_10161743,powell_2023_10055845} and observation identifiers. Derived retrieval products, posterior samples, figure/table data, opacities, and analysis scripts will be deposited in a DOI-issuing repository before publication.

In addition to the primary software packages mentioned in the text, our Python codes made use of the following libraries: numpy\cite{2020Natur.585..357H}, matplotlib\cite{2007CSE.....9...90H}, mpi4py~\cite{DALCIN20051108,DALCIN2008655,DALCIN20111124,9439927,9965751}, scipy\cite{2020SciPy-NMeth}, pandas\cite{mckinney-proc-scipy-2010,reback2020pandas}, seaborn\cite{2021JOSS....6.3021W}, and cmcrameri\cite{Rollo_cmcrameri_2024}. Colours from the Scientific colour map romaO\cite{2018GMD....11.2541C,2020NatCo..11.5444C,2021zndo...1243862C} are used in Fig.~\ref{fig:Spectrum} of this study.

%%%%%%%%%%%%%%%% SUPPLEMENT LIST %%%%%%%%%%%%%%%

% List the contents of your Supplementary Materials, including the numbers of any
% supplementary figures, tables, external data files etc. and any references that are
% cited only in the supplement. In this example, refs. 7-8 are cited only in the supplement.
% Fill out your numbers accordingly and delete any lines that aren't applicable.
\subsection*{Supplementary materials}
Materials and Methods\\
Figures S1 to S5\\
Tables S1 to S8\\
References \textit{(67-\arabic{enumiv})}\\ % automatically fills out the last reference number
% (filling out the other numbers automatically is possible but fiddly and liable to break)

%%%%%%%%%%%%%%%% END OF MAIN TEXT %%%%%%%%%%%%%%%

\newpage

%%%%%%%%%%%%%%%% START OF SUPPLEMENT %%%%%%%%%%%%%%%

% Figures, tables, equations and pages in the supplement are numbered S1, S2 etc.
\renewcommand{\thefigure}{S\arabic{figure}}
\renewcommand{\thetable}{S\arabic{table}}
\renewcommand{\theequation}{S\arabic{equation}}
\renewcommand{\thepage}{S\arabic{page}}
\setcounter{figure}{0}
\setcounter{table}{0}
\setcounter{equation}{0}
\setcounter{page}{1} % not 0 as \newpage already started a supplementary page
% References continue the numbering from the main text.

%%%%%%%%%%%%%%%% SUPPLEMENT TITLE PAGE %%%%%%%%%%%%%%%

\begin{center}
\section*{Supplementary Materials for\\ \scititle}

% Author list for the supplement
% Indicate the corresponding authors, but do NOT include institutions here
% It would be nice if the template auto-generated this, but doing so is complicated...
  Fabian Grübel$^{1,2,\ast}$,
  Karan Molaverdikhani$^{2,3}$,
  Barbara Ercolano$^{1,2,3}$,
  Katy L. Chubb$^{4}$,\\
  Paola Caselli$^{3,2}$,
  Tommaso Grassi$^{3,2}$,
  Rosa Arenales-Lope$^{1,2}$,
  and Dwaipayan Dubey$^{1,2}$\\ % we're not in a \author{} environment this time, so use \\ for a new line
\small$^\ast$Corresponding author. Email: fgruebel@usm.uni-muenchen.de
\end{center}

% Fill out the numbers for each type of supplementary material,
% and delete any lines that aren't applicable.
% These are just example numbers that don't match the rest of this template.
\subsubsection*{This PDF file includes:}
Materials and Methods\\
Figures S1 to S5\\
Tables S1 to S8\\

\newpage

%%%%%%%%%%%%%%%% MATERIALS AND METHODS %%%%%%%%%%%%%%%

\subsection*{Materials and Methods}

\subsubsection*{Data Preparation}
Four out of five datasets used in this study originate from the release in Carter et al. 2024\cite{2024NatAs...8.1008C} (data products available on Zenodo \cite{carter_2024_10161743}). These were made available at 5 different binning scales from native to 1/5 of the native resolution, and at R $\approx$ 100 for every dataset except NIRSpec PRISM. We employ the fixed limb darkened version to minimise inconsistencies between the data caused by the fitting process, as outlined in their study. The data for all instruments are given in units of $\frac{R_P}{R_S}$ (ratio of planetary and stellar radius). In order to match the petitRADTRANS input data format, we calculated the full wavelength bin width for each dataset from the difference of the upper and lower wavelength limits given in the data. We furthermore converted the spectrum to the transit depth $TD$, and propagated the uncertainties in the following way:
\begin{align}
    TD= \left(\frac{R_P}{R_S}\right)^2
\end{align}
\begin{align}
    \Delta TD_{\pm} = 2 \cdot \left(\frac{R_P}{R_S}\right)\times\Delta\frac{R_P}{R_S}_{\pm} \Rightarrow\overline{\Delta TD}= \frac{\Delta TD_{+} + \Delta TD_{-}}{2}
\end{align}

Here, $\Delta\frac{R_P}{R_S}_\pm$ denote the upper and lower uncertainties in $\frac{R_P}{R_S}$ space, and $\Delta TD_\pm$ the errors in transit depth units. We neglect second-order terms in the unit conversion and average the uncertainties ($\overline{\Delta TD}$) in order to be able to use a Gaussian likelihood function in the retrieval. A recent study has investigated the severity of using symmetric likelihood functions in the case of WASP-39~b and found them to be sufficiently accurate at the given asymmetry levels\cite{2025arXiv250719223D}, which is why we adopt them here. Furthermore, we applied a wavelength mask to the NIRSpec PRISM data as the range from 0.6 -- 2.1~$\mu m$ is affected by saturation. We note that test retrievals on the NIRSpec PRISM data, including different treatments of saturation such as a dilution-corrected version and a fully uncorrected one, produced similar D/H ratios. However, because of the poor agreement with the NIRISS data, these approaches significantly slowed down the convergence in the joint retrievals. We therefore adopted the saturation-exclusive data for this analysis.

The MIRI LRS data were not included in ref.~\cite{2024NatAs...8.1008C}, but were published separately\cite{2024Natur.626..979P}. Three different data reduction pipelines were compared for this data set (data products available on Zenodo\cite{powell_2023_10055845}). In our work, we choose the Eureka! data reduction to maximise the data compatibility (as done in e.g.~\cite{2025arXiv250407823M}) where we converted the wavelength half bin widths to the full width and ppm to base units. We included a wavelength mask for the MIRI LRS data, excluding all values above 10~$\mu m$, as described in the main text of this paper.

We conducted retrievals applying the k-distribution method, also referred to as the correlated-k (c-k) method. In petitRADTRANS, this limits the highest possible model resolution to R~=~1,000, which was used in all runs. To ensure sufficient sampling for each data point, the code authors recommend a maximum data resolution of around $\lambda/\Delta\lambda$~=~500. The majority of the retrievals were carried out at a data resolution around R~$\approx$~100. Nevertheless, we additionally conducted tests at higher resolution to confirm the compatibility of our result and the impact of a higher information content on our findings (refer to Supplementary Table~\ref{tab:Setups}). In these tests, we used the most precise data that would allow us to remain below the mentioned limit (NIRISS: scale~3, NIRCam: scale~3, NIRSpec PRISM: scale~1/native resolution). NIRSpec G395H, specifically the NRS2 detector observation, is the only dataset that significantly exceeds this limit at the lowest binning scale~5, with a maximum resolution of around R $\approx$~700, which is nonetheless used in the high-resolution consistency checks.

\subsubsection*{Forward Modelling}
All computations were carried out using the atmospheric retrieval package of the open source radiative transfer code petitRADTRANS (version 3.1.2)\cite{2019AA...627A..67M, Nasedkin2024,2024JOSS....9.7028B}. Results were partly validated with more recent code versions to ensure their robustness. petitRADTRANS utilizes the nested sampling python wrapper PyMultiNest (version 2.12) \cite{2014AA...564A.125B} for MultiNest (version 3.10) \cite{2008MNRAS.384..449F,2009MNRAS.398.1601F,2019OJAp....2E..10F} to sample the prior parameter space, map the model posterior distributions, and generate an estimate on the Bayesian Evidence. In the main text, we assumed a Guillot temperature profile\cite{2010AA...520A..27G}, a four-parameter model assuming plane-parallel static grey atmospheres, but additionally tested an isothermal atmosphere in the complementary retrievals. We show the assumed priors and fixed values for all parameters in Supplementary Table~\ref{tab:Priors}. 

The final modelled atmosphere consists of 100 pressure layers extending from 10$^{-9}$ to 10~bar. We fixed the stellar radius to 0.895~R$_\odot$, as well as the planetary radius and mass to 1.27~R$_\mathrm{Jup}$ and 0.28~M$_\mathrm{Jup}$, respectively\cite{2011AA...531A..40F}.
The reference pressure was included in the set of free parameters to allow some flexibility in the vertical position and extent of the modelled transmission spectrum. Additionally, we included a grey cloud, a power-law cloud, and a cloud fraction parameter (for details refer to ref.~\cite{2019AA...627A..67M}). The grey cloud is defined as a completely opaque layer at a retrieved pressure. The power-law cloud introduces a slope-like opacity into the transmission spectrum, allowing the consideration of more complex scattering behaviours. In the case of an additional cloud fraction parameter, the forward model yields a weighted linear combination of a fully cloudy and cloudless spectrum. Finally, several infrared opacity contributors were considered in a "free chemistry" setting and are discussed below.

In the main data source\cite{2024NatAs...8.1008C}, the authors show that vertical offsets between the different instrument datasets persist even after the joint reduction. We applied the offsets from their Supplementary Table~2 from the viewpoint of NIRSpec PRISM, in addition to the –177 ppm offset they reported in their analysis. Thus, we included positive shifts of 132ppm and 17ppm to the NIRCam and NIRSpec G395H data, respectively. We also included a fixed offset of –10 ppm for the NIRISS instrument, which persists between the two instruments when considering wavelengths unaffected by saturation (see Materials and Methods in ref.~\cite{2024NatAs...8.1008C}). Because there is only a small overlap between the NIRSpec PRISM and MIRI wavelength ranges, we adopted a uniform prior for the MIRI data offset (see Supplementary Table~\ref{tab:Priors}). Note that petitRADTRANS defines shifts in the positive vertical direction as negative offsets. Therefore, the values shown in Table~\ref{tab:Main_Results} and Supplementary Tables~\ref{tab:Priors},~\ref{tab:Results1}, and~\ref{tab:Results2} differ from those stated above.

\subsubsection*{Opacities}
The majority of the molecular and atomic opacities used are provided by the EXOMOL database~\cite{2016JMoSp.327...73T,2021AA...646A..21C}. In the final retrieval, we considered optical contributions from Na and K (line broadening from \cite{2016AA...589A..21A,2019AA...628A.120A}). Furthermore, we included the major infrared absorbers H$_2$O~\cite{2018MNRAS.480.2597P}, CO$_2$\cite{2020mnras.496.5282y}, CO\cite{2015ApJS..216...15L} (natural earth mixture of isotopologues) and SO$_2$\cite{2016MNRAS.459.3890U}. Additionally, Rayleigh scattering from H$_2$~\cite{1962ApJ...136..690D} and He~\cite{1965PPS....85..227C}, as well as collisional-induced absorption (CIA) from H$_2$-H$_2$~\cite{2001JQSRT..68..235B,2002AA...390..779B} and H$_2$-He~\cite{1988ApJ...326..509B,1989ApJ...336..495B,1989ApJ...341..549B} interactions are assumed. For more information, refer to ref.~\cite{2019AA...627A..67M}.

\begin{figure}
{\fontsize{7}{8.4}\selectfont
\centering
    \includegraphics[width=0.75\textwidth]{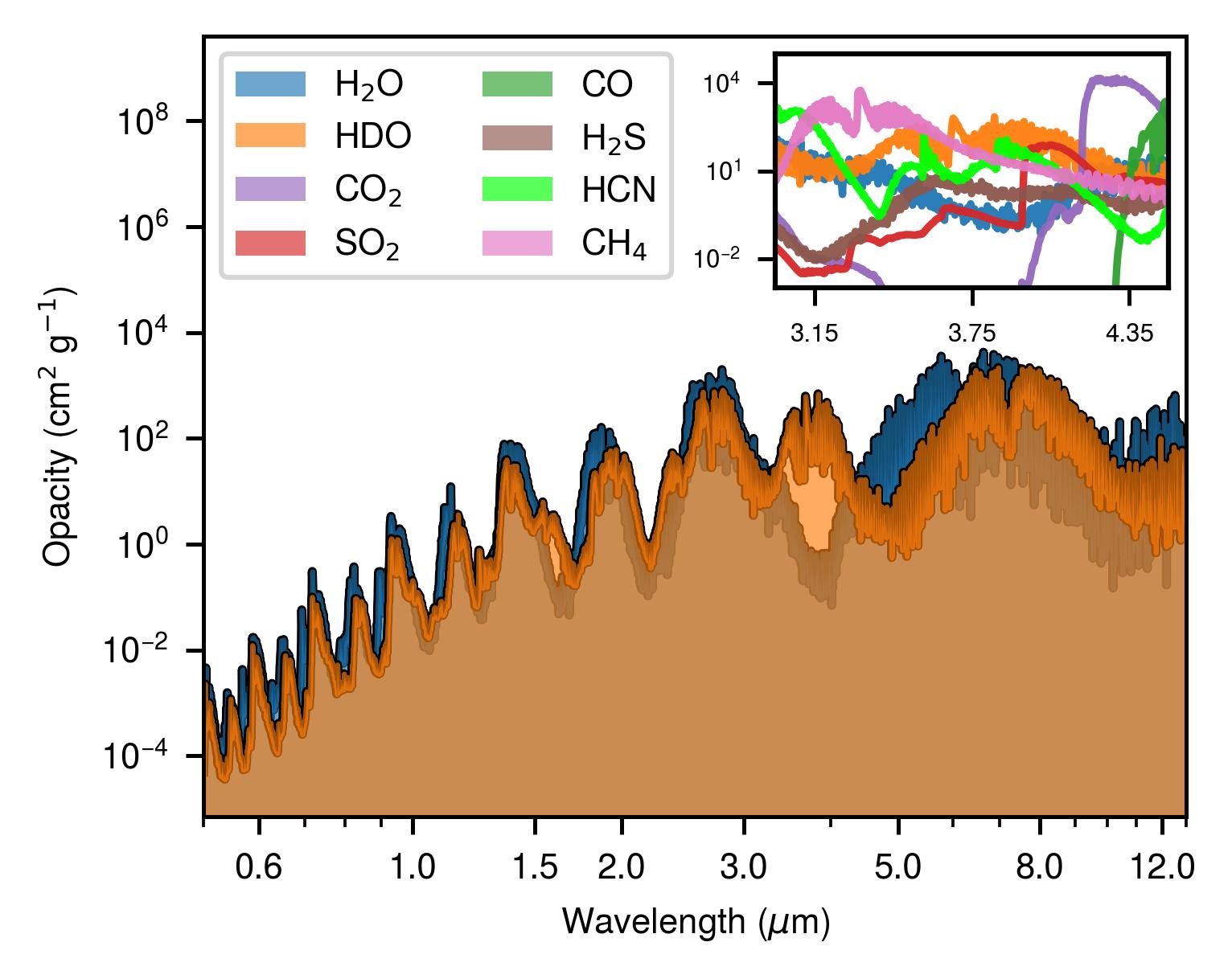}
    \setlength{\unitlength}{1pt}
    \begin{picture}(0,0)
        \put(-340,255){\fontsize{8}{9.6}\selectfont\textbf{A}}
        \put(-130,255){\fontsize{8}{9.6}\selectfont\textbf{B}}
    \end{picture}
    \vspace{-2em}
    \caption{\textbf{Opacities of competing line species. A}, Opacity comparison between H$_2$O and HDO at a pressure of 0.1 bar and temperature of 1,000 K from 0.5 to 13~$\mu m$. \textbf{B}, Opacity comparison between HDO and other molecules considered in the conducted retrievals that show significant absorption between 3 and 4.5~$\mu m$. Na and K, the collisional-induced absorption and Rayleigh scattering contributions, as well as the cloud opacities are not shown here.}
    \label{fig:HDO_opacities}
    }
\end{figure}

In this work, we aimed to determine whether the water isotopologue HDO is present on WASP-39~b. Supplementary Fig.~\ref{fig:HDO_opacities} shows a comparison between regular H$_2$O and HDO at a resolution of 1,000. The most promising wavelength range for detecting HDO lies between 3 and 4~$\mu m$. We therefore included alternative molecules that occupy this wavelength region and could be sufficiently abundant on WASP-39~b to obscure the HDO signal. We consider H$_2$S\cite{2016MNRAS.460.4063A}, HCN\cite{2014MNRAS.437.1828B}, and CH$_4$\cite{2024MNRAS.528.3719Y}. For HDO, we calculated correlated-k tables in petitRADTRANS format using the available data from Voronin et al. 2010\cite{2010MNRAS.402..492V}.

We assume a free chemistry retrieval, where the  mass-mixing-ratios (MMRs) are sampled independently from the priors presented in Supplementary Table~\ref{tab:Priors}. The remaining mass fraction is split between H$_2$ and He, assuming solar abundance ratios (petitRADTRANS default).

\subsubsection*{Ratio Calculation}
In this work, with the exception of Supplementary Tables~\ref{tab:Results1},~\ref{tab:Results2}, and~\ref{tab:Offset_Results}, as well as Supplementary Fig.~\ref{fig:Full_Post2}, which present the raw retrieval outputs, all mass-mixing-ratios (MMRs) were converted into volume-mixing-ratios (VMRs) for improved comparability using petitRADTRANS’s built-in function. Consequently, we computed the isotope/isotopologue ratios in the following way. For the HDO/H$_2$O ratio, we simply calculate the value for all equally weighted posterior samples using
\begin{align}
    \text{HDO/H$_2$O} = \frac{\text{VMR(HDO)}}{\text{VMR(H$_2$O)}}
\end{align}
with the median and 16/84 percentiles being presented in Fig. 3. Subsequently, the D/H ratio in water is estimated with
\begin{align}
    \text{D/H$_{\text{Water}}$} = \frac{\text{VMR(HDO)}}{\text{2 $\times$ VMR(H$_2$O)}}
\end{align}
as commonly used in isotopologue studies (e.g.~\cite{2021NatAs...5..943A}). Note that this assumes VMR(D$_2$O)$\ll$VMR(HDO) and VMR(HDO)$\ll$VMR(H$_2$O). In our case, the absolute deviation is well within the uncertainty limits of the posteriors. Comparison values and references for all objects shown in Fig.~\ref{fig:Ratio}, including the tracer species, are provided in Supplementary Tables~\ref{tab:Source1} and~\ref{tab:Source2}. In all cases with different deuterium tracers or where D/H ratios were unavailable, we assumed 2$\times$D/H = HDO/H$_2$O.
\subsection*{Bayesian Inference}
For the main model and consistency checks, we conducted retrievals with and without HDO. The retrieval code provides us with the equally weighted posterior samples, which we used to generate our posterior plots, as well as the Bayesian evidence $\ln(Z)$, and the reduced $\chi^2$. 

From the difference in the Bayesian log evidence, we calculated the Bayes factor $\ln(B_{10})$, allowing us to evaluate the detection significance using Jeffreys' scale~\cite{Jeffreys1939}. Additionally, as commonly done in exoplanetary retrieval studies, the Bayes factor was converted to a frequentist $p$-value and the $n_\sigma$ significance using an approximation adapted from Sellke et al. 2001\cite{Sellke2001}.
\begin{equation}
    B_{10} \approx \frac{1}{e p\ln(p)}
\end{equation}
\begin{align}
    \sigma = \Phi^{-1}(1-p)\text{, more commonly: }\sigma = \Phi^{-1}(1-p/2)
    \label{eq:Sigma}
\end{align}   
Here, $\Phi^{-1}$ is the inverse cumulative distribution function of the standard normal distribution. This conversion is intended to make the results obtained from the Bayesian inference methods more intuitive. However, this implementation does not work for several reasons, as discussed in a recent study\cite{Wrong}. The "Sellke formula" was never intended to be inverted, and the method therefore fails to provide an accurate estimate of the $p$-value. We further note that recent applications often adopt a two-sided $p$-value ($p\rightarrow p/2$ in formula~\ref{eq:Sigma}), even though the statistical test nature of a molecular detection is inherently one-sided. This leads to an overestimation of the final $n_\sigma$ value, regardless of the formula’s validity, and the effect is particularly significant when dealing with small Bayes factors. Nonetheless, because it has become a community standard, we provide these values here while emphasising the greater reliability of the Bayesian results. In Table~\ref{tab:Main_Results}, we therefore report both values, with the two-sided $n_\sigma$ shown in parentheses.

In  the main retrievals, 1,000 live points were used with a sampling efficiency of 0.3 and an evidence tolerance of 0.5. We furthermore enable importance nested sampling for all runs to maximise the reliability of the retrieval results. According to the author of PyMultiNest, we adopt the regular nested sampling evidence errors in cases where these are larger, since importance nested sampling often underestimates the uncertainties. The raw retrieval results for all runs are shown in Supplementary Tables~\ref{tab:Results1} and~\ref{tab:Results2}.

\subsubsection*{Complementary Retrievals}

In addition to the main text retrieval, we considered an isothermal pressure-temperature profile to test the model-dependency of our results. The retrievals were conducted using 1,000 live points, an evidence tolerance of 0.5, and a sampling efficiency of 0.3. Furthermore, several complementary datasets were analysed to determine which instrument data support or drive the detection. For this, we carried out retrievals on all individual data and considered joint retrievals without NIRSpec PRISM. Due to the computational costs, we chose 400 live points, an evidence tolerance of 0.5 and a sampling efficiency of 0.3 for these runs. An overview of the main text and complementary retrievals is given in Supplementary Table~\ref{tab:Setups}. 

The NIRSpec G395H and joint data retrieval excluding NIRSpec PRISM were also tested with 1,000 live points and at full resolution (c-k limit, see above) to ensure consistency between the binning scales and nested sampling settings. The results show that the higher resolution provides more information on some of the line species in the joint data case. The most prominent examples are Na and K. Additionally, the HDO posterior exhibits a slightly narrower $1\sigma$ range. However, the Bayes Factor does not indicate a substantial increase in detection significance. Therefore, the retrieved results do not appear to be strongly resolution-dependent within the c-k regime. In the individual NIRSpec G395H data retrieval, we find essentially unchanged results compared to the R~$\approx$~100 case. Additional tests on the finest binning scales using line-by-line opacity sampling are necessary to further investigate the impact of a higher data resolution. Due to the large wavelength coverage of the full WASP-39~b data set, this may not be computationally feasible at the time of this work using the standard retrieval method without significant simplification of the model assumptions.

\subsubsection*{Offset Test Retrievals}
In order to further investigate the effect of the fixed offsets, we conducted an additional analysis. The datasets of NIRISS SOSS, NIRCam F322W2, and NIRSpec G395H were binned onto the NIRSpec PRISM wavelength grid using their finest binning scales available and according to the following concept.
\begin{align}
    TD_j=\frac{\sum_{i} TD_i \times w_i}{\sum_{i}w_i}, \qquad w_i = \frac{1}{\Delta TD_i^2\times\Delta_i}
\end{align}
\begin{align}
    \Delta TD_j=  \sqrt{\frac{1}{\sum_i w_i}}
\end{align}
Here, $TD_j$ is the resulting transit depth in PRISM bin $j$, while $TD_i$, $\Delta TD_i$, and $\Delta_i$ denote the transit depth, uncertainty, and bin width of the respective instrument datapoint. Note that the actual rebinning is performed using the built-in function \texttt{rebin\_spectrum\_bin} of petitRADTRANS, which is originally designed to bin a synthetic spectrum to a dataset and applies bin-wise linear interpolation, which is accounted for by the added factor $\Delta_i$. As the results will show, this method produces offset estimates consistent with those provided by the data source\cite{2024NatAs...8.1008C}. The binned data were then used to determine the relative offsets between the datasets using two separate approaches. For the ease of comparison, we applied the same 177 ppm offset to NIRSpec PRISM as in the main analysis.\\
First, we conducted a Nested Sampling analysis with PyMultiNest, adopting uniform priors between -500 and 500 ppm for each instrument separately. The likelihood function was assumed to be Gaussian, with
\begin{align}
    \chi^2_{\mathrm{Inst}} = \sum_j\frac{(TD_{\mathrm{PRISM},j} - (TD_{\mathrm{\mathrm{Inst}},j} - \Delta \mathrm{Inst}))^2}{\Delta TD_{\mathrm{PRISM},j}^2 + \Delta TD_{\mathrm{Inst},j}^2}
\end{align}
Here, $\Delta \mathrm{Inst}$ represents the instrument offset. The signs were chosen to match the convention in petitRADTRANS, making the values readily comparable to the offsets in Supplementary Table~\ref{tab:Priors}. We used 1000 live points, an evidence tolerance of 0.5, and a sampling efficiency of 0.3. The 16th and 84th percentiles, together with the median of the retrieved posteriors, are listed in Supplementary Table~\ref{tab:Offsets}.\\
In the second approach, we used the instrument transit depths and uncertainties at the finest binning scales and randomly resampled the transit depths assuming Gaussian probability distributions for each datapoint. For each random realization, the following offset was computed after the binning process.
\begin{align}
    \Delta \mathrm{Inst} = \frac{\sum_j(TD_{\mathrm{Inst},j} - TD_{\mathrm{PRISM},j})\times v_j}{\sum_{j}v_j}, \qquad v_j = \frac{1}{\Delta TD_{\mathrm{PRISM},j}^2 + \Delta TD_{\mathrm{Inst},j}^2}
\end{align}
Subsequently, the median, 16th and 84th percentiles of the offset distributions were calculated from 100,000 samples, and are presented in Supplementary Table~\ref{tab:Offsets}.\\
According to Supplementary Table~\ref{tab:Offsets}, both methods yield consistent offsets. We additionally provide the values adopted in the main text, which, with the exception of NIRSpec G395H, fall within the resulting 1$\sigma$ ranges. Finally, the third column of Supplementary Table~\ref{tab:Offsets} gives the offsets adopted in the following sensitivity test. The medians were rounded to ppm, and the uncertainties were taken from the maximum of both methods and averaged in order to apply Gaussian priors.\\
Following this analysis, eight retrievals with an isothermal temperature profile, 1000 live points, an evidence tolerance of 0.5, and a sampling efficiency of 0.3 were conducted. We tested fixed offsets at the newly established medians, as well as free offsets assuming Gaussian priors, both including and excluding HDO over the shortened spectral ranges of 3--5 $\mu$m and 2--5 $\mu$m. In both cases, we excluded all data outside of these specific ranges. Apart from these changes, we assumed the priors from Supplementary Table~\ref{tab:Priors}. The results are given in Supplementary Table~\ref{tab:Offset_Results}. For further details, refer to the main text.

%%%%%%%%%%%%%%%% SUPPLEMENTARY TEXT %%%%%%%%%%%%%%%

% If your supplement is very short you might need to uncomment the following line to avoid
% layout problems with the figures and tables.
\newpage

%%%%%%%%%%%%%%%% SUPPLEMENTARY FIGURES %%%%%%%%%%%%%%%

\begin{figure}
{\fontsize{7}{8.4}\selectfont

\centering
    \includegraphics[width=0.99\textwidth]{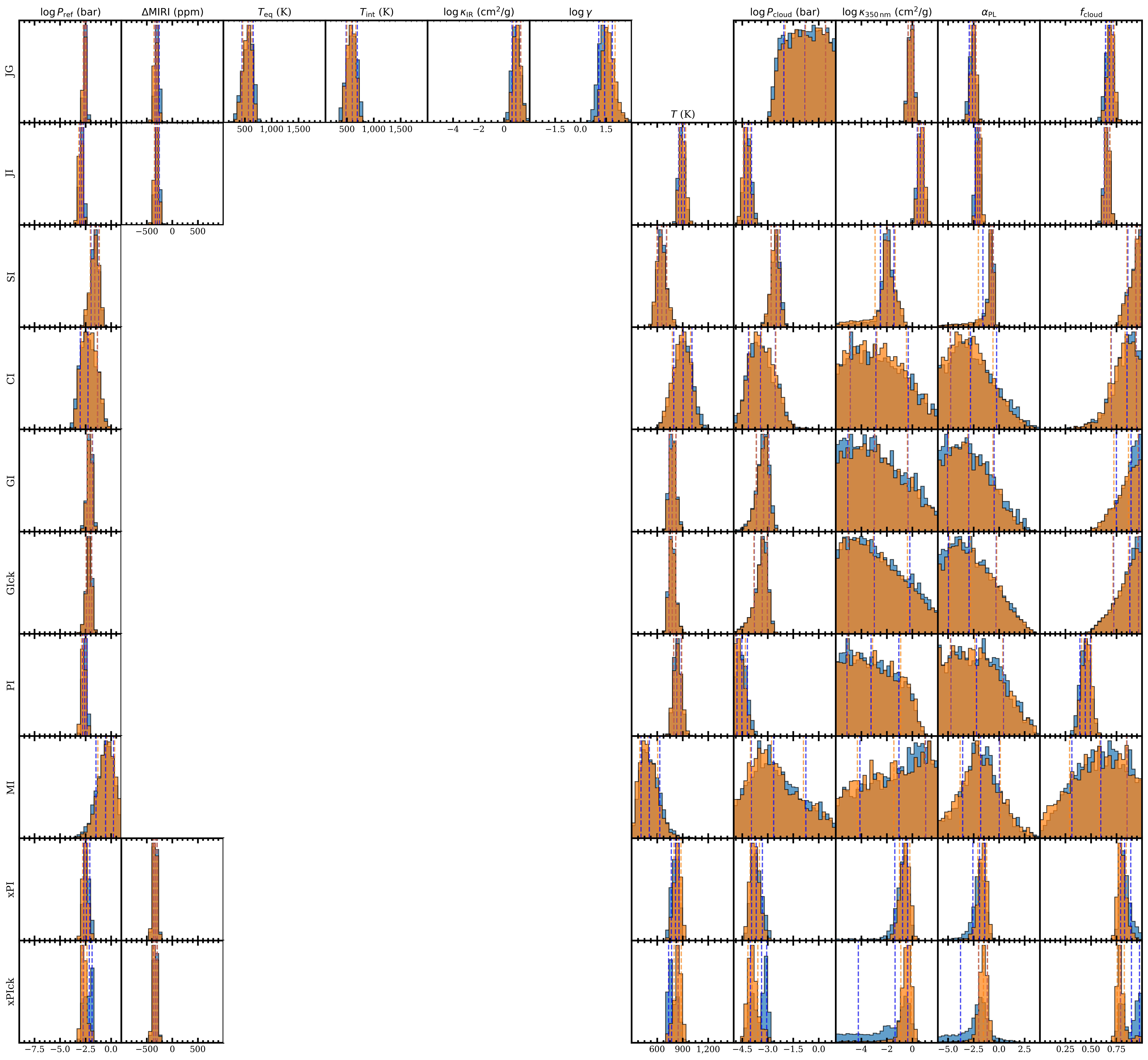}
    \caption{\textbf{Marginalized 1D posterior distributions over the full prior width for the free parameters considered in this study (10/21 parameters).} The model identifiers from Supplementary Table~\ref{tab:Setups} define the rows. In each subplot, the distribution of HDO-exclusive and HDO-inclusive retrievals is given in blue and orange, respectively. Dashed lines indicate the 16th, 50th, and 84th percentiles. For details on the parameter priors and values, refer to Supplementary Tables~\ref{tab:Priors},~\ref{tab:Results1}, and~\ref{tab:Results2}. Note: log indicates the base 10 logarithm.}
    \label{fig:Full_Post1}
    }
\end{figure}

\begin{figure}
{\fontsize{7}{8.4}\selectfont

\centering
    \includegraphics[width=0.99\textwidth]{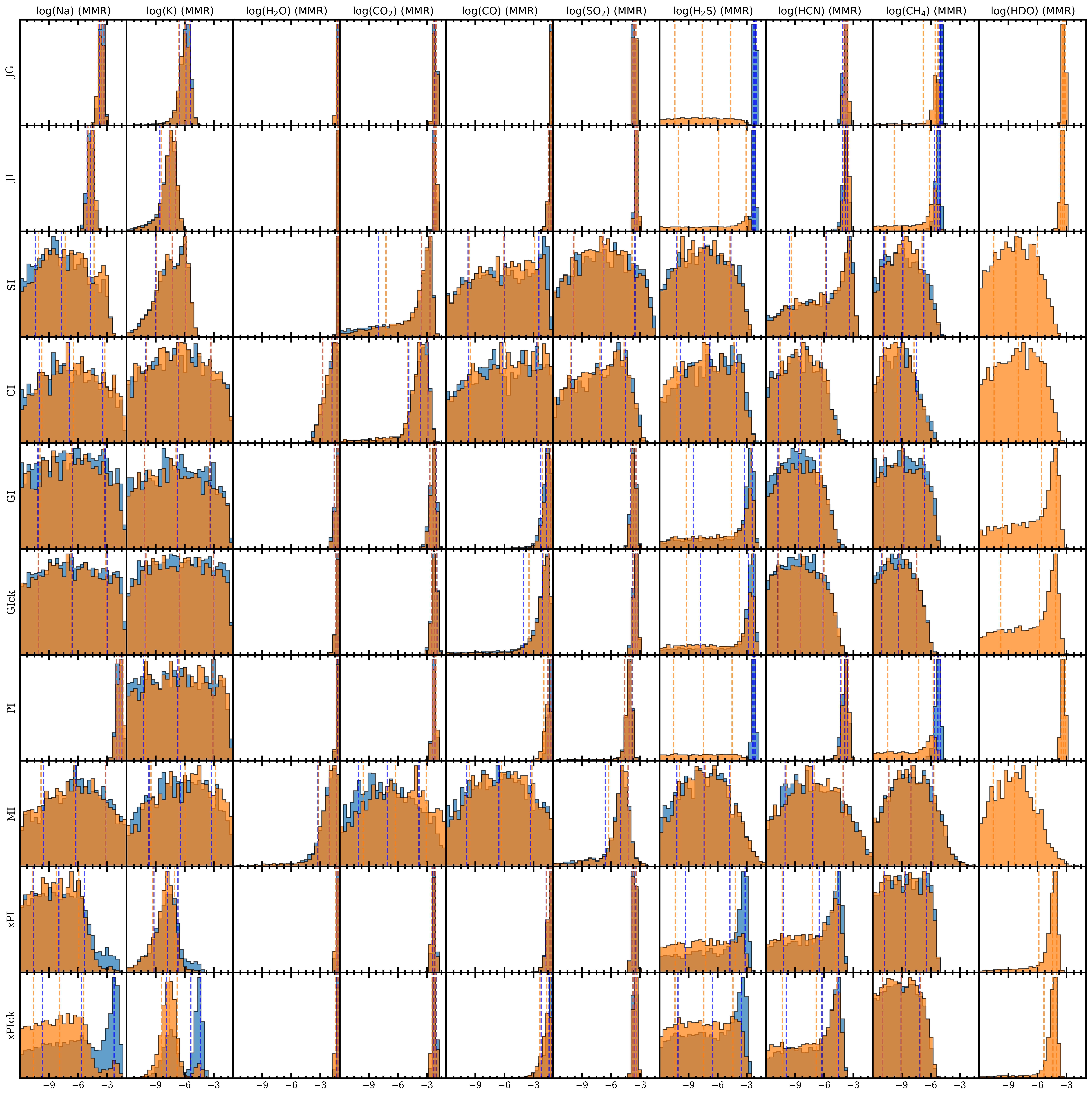}
    \caption{\textbf{Marginalized 1D posterior distributions over the full prior width for the free parameters considered in this study (10/21 parameters).} The model identifiers from Supplementary Table~\ref{tab:Setups} define the rows. In each subplot, the distribution of HDO-exclusive and HDO-inclusive retrievals is given in blue and orange, respectively. Dashed lines indicate the 16th, 50th, and 84th percentiles. For details on the parameter priors and values, refer to Supplementary Tables~\ref{tab:Priors},~\ref{tab:Results1}, and~\ref{tab:Results2}. Note: log indicates the base 10 logarithm.}
    \label{fig:Full_Post2}
    }
\end{figure}

\begin{figure}
{\fontsize{7}{8.4}\selectfont

\centering
    \includegraphics[width=0.75\textwidth]{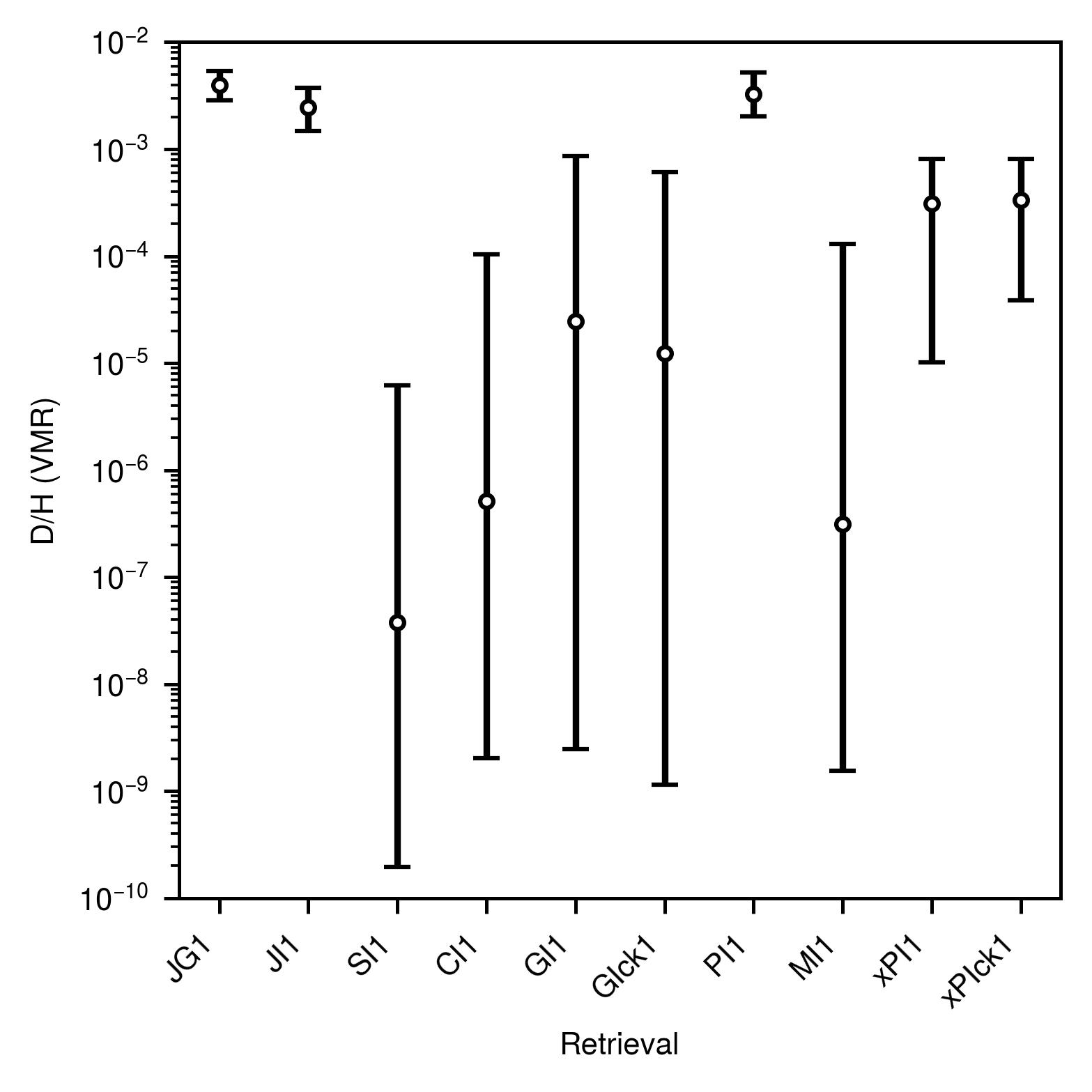}
    \caption{\textbf{Scatter plot of $\log(\mathrm{D/H})$ ratios from all presented retrievals.} The median, 16th and 84th percentiles derived from the H$_2$O and HDO VMR posteriors are shown. The uncertainties are given by the 16th and 84th percentiles. The model identifiers from Supplementary Table~\ref{tab:Setups} represent the respective retrieval. Note: log indicates the base 10 logarithm.}
    \label{fig:DH_Scatter}
    }
\end{figure}

\begin{figure}
{\fontsize{7}{8.4}\selectfont

\centering
    \includegraphics[width=0.75\textwidth]{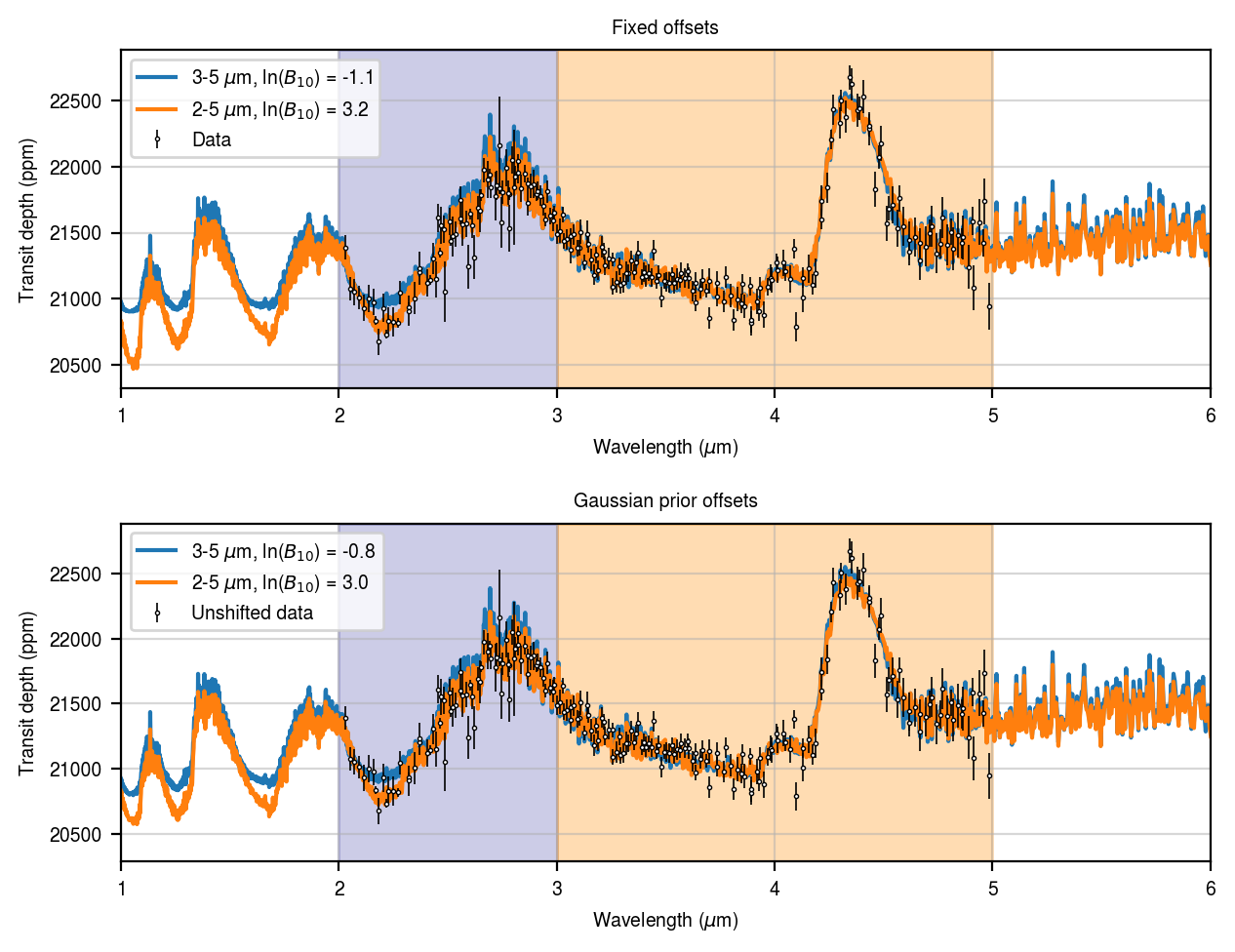}
    \caption{\textbf{Highest likelihood spectra from the offset test retrievals.} Shown are the spectra of the HDO-inclusive retrievals with fixed and free offsets. In both cases, the data (R~$\approx$~100) with offsets fixed at the medians given in Supplementary Table~\ref{tab:Offsets} are presented. The retrievals were conducted on two overlapping wavelength ranges as stated in the legend. Additionally, the Bayes factors are provided.}
    \label{fig:Offset_Spectra}
    }
\end{figure}

%%%%%%%%%%%%%%%% SUPPLEMENTARY TABLES %%%%%%%%%%%%%%%

\begin{table}
\centering
{\fontsize{7}{8.4}\selectfont
\caption{\textbf{Fixed and free parameters considered in the conducted main text and complementary retrievals.} Note that positive data shifts are negative in petitRADTRANS. *Faedi et al. 2011\cite{2011AA...531A..40F}; $^+$Carter et al. 2024\cite{2024NatAs...8.1008C}; °Fixed to -311 ppm in individual MIRI data retrievals according to results from the joint isothermal case. Note: log indicates the base 10 logarithm.}
  \vspace{1em}
\begin{minipage}{0.4\textwidth}  % adjust width as you like (e.g. 0.7, 0.8)

\label{tab:Priors}
\begin{tabular}{lcr}
\hline
        Parameters & Value & Prior\\
\hline
    \multicolumn{3}{l}{Basic model parameters}\\
\hline
        $R_*$ ($R_\odot$) & 0.895*& Fixed Value\\
        $R_\mathrm{p}$ ($R_\mathrm{Jup}$)& 1.27*& Fixed Value \\
        $M_\mathrm{p}$ ($M_\mathrm{Jup}$)& 0.28*& Fixed Value \\
        $\log P_{\mathrm{ref}}$ (bar)& [-9, 1] & Log-uniform\\

\hline
    \multicolumn{3}{l}{Data offsets}\\
\hline
$\Delta$NIRISS (ppm) & 10$^+$ &Fixed\\
$\Delta$NIRCam (ppm)&-132$^+$& Fixed\\
$\Delta$NIRSpec G395H (ppm)&-17$^+$& Fixed\\
$\Delta$NIRSpec PRISM (ppm)&177$^+$& Fixed\\
$\Delta$MIRI (ppm) &[-1,000,1,000]°&Uniform\\
\hline
    \multicolumn{3}{l}{Pressure-Temperature profiles}\\
\hline
    \multicolumn{3}{l}{Guillot}\\
\hline
    $T_{\mathrm{equ}}$ (K) & [100, 2000]& Uniform\\
    $T_{\mathrm{int}}$ (K) & [100, 2000]& Uniform\\
    $\log$($\kappa_{\mathrm{IR}})$ (cm$^2$/g)  & [-6,2]& Log-uniform\\
    $\log(\gamma)$& [-3,3]& Log-Uniform\\
\hline
    \multicolumn{3}{l}{Isothermal}\\
\hline
    $T$ (K) & [300, 1500]& Uniform\\
\hline
    \multicolumn{3}{l}{Cloud parameters}\\
\hline
$\log(\mathrm{P}_{\mathrm{Grey}})$ (bar) & [-5, 1]& Log-uniform\\
$\log\kappa_\text{350 nm}$ (cm$^2$/g)   & [-6, 2]& Log-uniform\\
$\alpha_\mathrm{PL}$& [-6, 4]& Uniform\\
$f_\text{Cloud}$& [0, 1]& Uniform\\
\hline
    \multicolumn{3}{l}{Atomic/Molecular mass fractions}\\
\hline
         Na, K, H$_2$O, CO$_2$,& &\\
           CO, SO$_2$,H$_2$S, HCN,& [10$^{-12}$, 10$^{-1}$]& Log-uniform\\
           CH$_4$, HDO ($\log$ VMR)& &\\

\hline
  \end{tabular}

\end{minipage}
}
\end{table}

\begin{table}
\centering
{\fontsize{7}{8.4}\selectfont

    \caption{\textbf{Overview of the conducted main text and complementary retrievals.} PT profile: pressure-temperature profile (refer to Materials and Methods); c-k limit: correlated-k limit (refer to Materials and Methods); Joint dataset: NIRISS + NIRCam + NIRSpec G395H + MIRI + NIRSpec PRISM.}
  \vspace{1em}
\begin{minipage}{0.58\textwidth}  % adjust width as you like (e.g. 0.7, 0.8)

\label{tab:Setups}
    \begin{tabular}{l c c c r}
        \hline
        Dataset & PT profile&Live points &Resolution& Identifier\\
        \hline
        Main text retrievals& & & \\
        \hline
        \textbf{J}oint& \textbf{G}uillot  &  1,000 &100&\textbf{JG}\\
        \textbf{J}oint & \textbf{I}sothermal & 1,000& 100&\textbf{JI}\\
        \hline
        Complementary datasets& & & \\        
        \hline
        NIRI\textbf{S}S& \textbf{I}sothermal&400&100&\textbf{SI}\\
        NIR\textbf{C}am& \textbf{I}sothermal&400&100&\textbf{CI}\\
        NIRSpec \textbf{G}395H& \textbf{I}sothermal&400&100&\textbf{GI}\\
        NIRSpec \textbf{G}395H& \textbf{I}sothermal&1,000&\textbf{c}-\textbf{k} limit&\textbf{GIck}\\
        NIRSpec \textbf{P}RISM & \textbf{I}sothermal &400& 100&\textbf{PI}\\
        \textbf{M}IRI& \textbf{I}sothermal&400&100&\textbf{MI}\\   
        Joint (e\textbf{x}cluding NIRSpec \textbf{P}RISM) & \textbf{I}sothermal & 400 & 100&\textbf{xPI}\\   
        Joint (e\textbf{x}cluding NIRSpec \textbf{P}RISM) & \textbf{I}sothermal & 1,000 & \textbf{c}-\textbf{k} limit&\textbf{xPIck}\\
\hline
  \end{tabular}

\end{minipage}
}
\end{table}

\begin{table}
\centering
{\fontsize{7}{8.4}\selectfont

    \caption{\textbf{Reference D/H ratios (1/2).} Object names, tracer species, D/H ratio in units of 10$^{-4}$ including uncertainty, and refs. are displayed in the order from Fig.~\ref{fig:Ratio} (left to right). The values are taken from (summaries of) five previous works\cite{2011Natur.478..218H,2021NatAs...5..943A,2024PNAS..12101638M,2024ApJ...977L..49R,2025ApJ...986L..19S}. In cases where the D/H ratio is not given by the source, we computed the values from the tracer species ratio (refer to Materials and Methods).}
    \vspace{1em}

\begin{minipage}{0.6\textwidth}  % adjust width as you like (e.g. 0.7, 0.8)

    \label{tab:Source1}
    \begin{tabular}{lccr}
        \hline
        Object name & Tracer species & D/H ratio (10$^{-4}$) & Ref. \\
        \hline
        \multicolumn{4}{c}{Inner planets (IP)}\\
        \hline
        Venus (70 km)         & H$_2$O & (0.025 $\pm$ 0.019)$\times 10^4$ &~\cite{2024PNAS..12101638M} \\
        Venus (108 km)        & H$_2$O & (0.24 $\pm$ 0.17)$\times 10^4$   &~\cite{2024PNAS..12101638M} \\
        Mars (Bulk)           & H$_2$O & 7.63 $\pm$ 0.62            &~\cite{2021NatAs...5..943A} \\
        \hline
        \multicolumn{4}{c}{Outer planets}\\
        \hline
        Jupiter               & H$_2$  & 0.225 $\pm$ 0.035          &~\cite{2001AA...370..610L} \\
        Saturn                & H$_2$  & 0.170$_{-0.045}^{+0.075}$   &~\cite{2001AA...370..610L} \\
        Enceladus             & H$_2$O & 2.5$_{-0.7}^{+1.5}$        &~\cite{2009Natur.460..487W} \\
        Uranus                & H$_2$  & 0.55$_{-0.15}^{+0.35}$     &~\cite{1999AA...341L..17F} \\
        Neptune               & H$_2$  & 0.45 $\pm$ 0.10             &~\cite{2010AA...518L.152L} \\
        \hline
        \multicolumn{4}{c}{Chondrites}\\
        \hline
        CI chondrites         & H$_2$O & 0.645 -- 0.975           &~\cite{2012Sci...337..721A} \\
        CM chondrites         & H$_2$O & 0.865 $\pm$ 0.036          &~\cite{2012Sci...337..721A} \\
        CR chondrites         & H$_2$O & 1.71$_{-0.10}^{+0.17}$     &~\cite{2012Sci...337..721A} \\
        CO chondrites         & H$_2$O & 0.85 -- 1.32        &~\cite{2012Sci...337..721A} \\
        \hline
        \multicolumn{4}{c}{Jupiter-family comets}\\
        \hline
        103P/Hartley 2        & H$_2$O & 1.60 $\pm$ 0.25             &~\cite{2011Natur.478..218H} \\
        67P/Churyumov--Gerasimenko & H$_2$O & 5.3 $\pm$ 0.7      &~\cite{2015Sci...347A.387A} \\
        46P/Wirtanen          & H$_2$O & 1.60 $\pm$ 0.65             &~\cite{2019AA...625L...5L} \\
        45P/Honda--Mrkos--Pajdušáková & H$_2$O & $<$2.0          &~\cite{2013ApJ...774L...3L} \\
        \hline
        \multicolumn{4}{c}{Oort cloud comets}\\
        \hline
        1P/Halley             & H$_2$O & 2.1 $\pm$ 0.3              &~\cite{2012PSS...60..166B} \\
        C/1996 B2 Hyakutake   & H$_2$O & 2.9 $\pm$ 1.0              &~\cite{1998Icar..133..147B} \\
        C/1995 O1 Hale--Bopp  & H$_2$O & 3.3 $\pm$ 0.8              &~\cite{1998Sci...279..842M} \\
        8P/Tuttle             & H$_2$O & 4.10 $\pm$ 1.45             &~\cite{2009ApJ...690L...5V} \\
        C/2009 P1 Garradd     & H$_2$O & 2.05 $\pm$ 0.20             &~\cite{2012AA...544L..15B} \\
        C/2002 T7 LINEAR      & H$_2$O & 2.5 $\pm$ 0.7              &~\cite{2008AA...490L..31H} \\
        C/2012 F6 Lemmon      & H$_2$O & 6.5 $\pm$ 1.6              &~\cite{2016AA...589A..78B} \\
        C/2014 Q2 Lovejoy     & H$_2$O & 1.4 $\pm$ 0.4              &~\cite{2016AA...589A..78B} \\
        C/2007 B3 Lulin       & H$_2$O & $<$5.6                     &~\cite{2012ApJ...750..102G} \\
        153P/Ikeya--Zhang     & H$_2$O & $<$2.8 $\pm$ 0.3              &~\cite{2006AA...449.1255B} \\
\hline
  \end{tabular}
\end{minipage}
}
\end{table}

\begin{table}
\centering
{\fontsize{7}{8.4}\selectfont
  
    \caption{\textbf{Reference D/H ratios (2/2).} Object names, tracer species, D/H ratio in units of 10$^{-4}$ including uncertainty, and refs. are displayed in the order from Fig.~\ref{fig:Ratio} (left to right). The values are taken from (summaries of) five previous works\cite{2011Natur.478..218H,2021NatAs...5..943A,2024PNAS..12101638M,2024ApJ...977L..49R,2025ApJ...986L..19S}. In cases where the D/H ratio is not given by the source, we computed the values from the tracer species ratio (refer to Materials and Methods).}
        \vspace{1em}
\begin{minipage}{0.6\textwidth}  % adjust width as you like (e.g. 0.7, 0.8)

\label{tab:Source2}
    \begin{tabular}{lccr}
        \hline
        Object name&Tracer species& D/H ratio (10$^{-4}$) & Source\\
        \hline
        \multicolumn{4}{c}{Protostellar (gas)}\\
        \hline
        NGC1333 IRAS 4A-NW &H$_2$O&2.70 $\pm$ 0.75&~\cite{2014AA...563A..74P},~\cite{2019AA...631A..25J} \\
        NGC 1333 IRAS 2A &H$_2$O&3.70 $\pm$ 1.05&\cite{2014AA...563A..74P} \\
        NGC 1333 IRAS 4B &H$_2$O&2.95 $\pm$ 1.3&\cite{2014AA...563A..74P}\\
        IRAS 16293--2422 &H$_2$O&4.6 $\pm$ 1.3&~\cite{2014AA...563A..74P}\\
        BHR 71-IRS1 &H$_2$O&9.0 $\pm$ 2.0&\cite{2019AA...631A..25J} \\
        B335 &H$_2$O&31.5 $\pm$ 7.5&~\cite{2021AA...650A.172J}\\
        L483 &H$_2$O&20.0 $\pm$ 2.5&\cite{2021AA...650A.172J}\\
        V883 Ori &H$_2$O&11.5 $\pm$ 3.0&\cite{2023Natur.615..227T} \\
        L1551 IRS5 &H$_2$O&10.5 $\pm$ 4.0&~\cite{2023AA...677L..17A}\\
        W3 IRS5 &H$_2$O&$\le$6.5&~\cite{2006AA...447.1011V}\\
        W33A &H$_2$O&$\le$15.0&~\cite{2006AA...447.1011V}\\
        AFGL 2591 &H$_2$O&$\le$16.5&\cite{2006AA...447.1011V}\\
        NGC 7538 IRS1 &H$_2$O&$\le$19.0&~\cite{2006AA...447.1011V}\\
        Orion KL hot core &H$_2$O&15.0$_{-8.5}^{+15.5}$&~\cite{2013ApJ...770..142N}\\
        NGC 6334 I &H$_2$O&1.05 $\pm$ 0.5&~\cite{2013ApJ...765...61E}\\
        G34.26+0.15 &H$_2$O&2.75 $\pm$ 1.0&~\cite{2014MNRAS.445.1299C}\\
        \hline
        \multicolumn{4}{c}{Protostellar (ice)}\\
        \hline
        NGC 1333 SVS 13 &H$_2$O&$\le$8.5&~\cite{2003AA...410..897P}\\
        NGC 1333 SVS 12 &H$_2$O&$\le$2.5&\cite{2003AA...410..897P} \\
        L1527 &H$_2$O&22.0$_{-8.5}^{+18.5}$&~\cite{2025ApJ...986L..19S}\\
        L1489 IRS &H$_2$O&$\le$40.0&\cite{2003AA...410..897P}\\
        TMR1 &H$_2$O&$\le$55.0&\cite{2003AA...410..897P}\\
        IRAS 05390--0728 &H$_2$O&$\le$50.0&\cite{2003AA...399.1009D} \\
        IRAS 08448--4343 &H$_2$O&$\le$50.0&~\cite{2003AA...399.1009D}\\
        HOPS 370 &H$_2$O&23.0 $\pm$ 11.0&~\cite{2024AA...688A..29S}\\
        NGC 7538 IRS9 &H$_2$O&48.75 $\pm$ 8.25&~\cite{2003AA...399.1009D}\\
        GL 2136 &H$_2$O&$\le$20.0&~\cite{2003AA...399.1009D}\\
        IRAS 20126 &H$_2$O&13.0 $\pm$ 7.0&~\cite{2024AA...688A..29S}\\  
        \hline
        \multicolumn{4}{c}{Brown dwarf}\\
        \hline
        WISE 0855-0714 &CH$_4$&0.127$_{-0.019}^{+0.022}$&~\cite{2024ApJ...977L..49R}\\  
        \hline
        \multicolumn{4}{c}{Reference values}\\
        \hline
        Protosolar            & H$_2$  & 0.21 $\pm$ 0.04            &~\cite{2006ApJ...647.1106L} \\
        Interstellar Medium   & H      & 0.16 $\pm$ 0.01            &~\cite{2006ApJ...647.1106L} \\
        Earth (VSMOW)         & H$_2$O & 1.558 $\pm$ 0.001          &~\cite{article} \\
 \hline
  \end{tabular}

\end{minipage}
}
\end{table}

\begin{table}
\centering
{\fontsize{7}{8.4}\selectfont
\caption{\textbf{Raw retrieval results from the main text and complementary retrievals (1/2).} Individual retrievals are represented by their identifiers defined in Table~\ref{tab:Setups}. The additions, 0 and 1, indicate the models without and with HDO, respectively. The parameters are divided into the following groups: model comparison metrics, independent parameters, Guillot\cite{2010AA...520A..27G} temperature profile, isothermal temperature profile, clouds, and line species abundances. For more details on these parameters, refer to the main text and the Materials and Methods. Note: log and ln indicate base 10 and the natural logarithms, respectively. Values and uncertainties are given by medians and 16th/84th percentiles. Units: ppm: parts-per-million, MMR: mass-mixing-ratio.}
\vspace{1em}

\begin{minipage}{\textwidth}

\label{tab:Results1}
\resizebox{\textwidth}{!}{
\begin{tabular}{lcccccccccc}
\hline
Parameter & JG0 & JG1 & JI0 & JI1 & SI0 & SI1 & CI0 & CI1 & GI0 & GI1 \\
\hline
$\ln(Z)$ & $2780.69 \pm 0.23$ & $2791.14 \pm 0.23$ & $2747.95 \pm 0.21$ & $2753.32 \pm 0.21$ & $1014.63 \pm 0.23$ & $1013.74 \pm 0.23$ & $363.11 \pm 0.18$ & $362.45 \pm 0.18$ & $442.18 \pm 0.25$ & $441.96 \pm 0.23$ \\
$\chi^2$/dof & $1.78$ & $1.74$ & $2.00$ & $1.98$ & $1.96$ & $1.98$ & $1.72$ & $1.76$ & $1.34$ & $1.37$ \\
\hline
$\log P_{\mathrm{ref}}$ (bar)& $-2.54^{+0.14}_{-0.14}$ & $-2.58^{+0.13}_{-0.15}$ & $-2.87^{+0.19}_{-0.19}$ & $-2.97^{+0.19}_{-0.19}$ & $-1.59^{+0.40}_{-0.40}$ & $-1.57^{+0.41}_{-0.47}$ & $-2.26^{+0.92}_{-0.77}$ & $-2.16^{+0.80}_{-0.76}$ & $-2.07^{+0.25}_{-0.27}$ & $-2.07^{+0.22}_{-0.26}$ \\
$\mathrm{MIRI\ offset}$ (ppm) & $-297^{+38}_{-38}$ & $-324^{+38}_{-36}$ & $-297^{+40}_{-41}$ & $-310^{+40}_{-40}$ & -- & -- & -- & -- & -- & -- \\
\hline
$T_{\mathrm{equ}}$ (K) & $562^{+92}_{-113}$ & $554^{+79}_{-93}$ & -- & -- & -- & -- & -- & -- & -- & -- \\
$T_{\mathrm{int}}$ (K) & $595^{+97}_{-109}$ & $584^{+81}_{-86}$ & -- & -- & -- & -- & -- & -- & -- & -- \\
$\log \kappa_{\mathrm{IR}} (\mathrm{cm^2/g})$ & $0.93^{+0.37}_{-0.31}$ & $1.00^{+0.31}_{-0.28}$ & -- & -- & -- & -- & -- & -- & -- & -- \\
$\log \gamma$ & $1.41^{+0.45}_{-0.35}$ & $1.62^{+0.40}_{-0.32}$ & -- & -- & -- & -- & -- & -- & -- & -- \\
\hline
$T$ (K) & -- & -- & $887^{+36}_{-34}$ & $898^{+37}_{-37}$ & $653^{+56}_{-49}$ & $651^{+62}_{-54}$ & $905^{+103}_{-111}$ & $891^{+103}_{-113}$ & $779^{+42}_{-38}$ & $782^{+38}_{-38}$ \\
\hline
$\log P_{\mathrm{cloud}}$ (bar) & $-0.81^{+1.21}_{-1.25}$ & $-0.78^{+1.18}_{-1.19}$ & $-4.17^{+0.22}_{-0.21}$ & $-4.23^{+0.21}_{-0.20}$ & $-2.52^{+0.24}_{-0.29}$ & $-2.55^{+0.24}_{-0.25}$ & $-3.42^{+0.87}_{-0.71}$ & $-3.37^{+0.84}_{-0.70}$ & $-3.25^{+0.31}_{-0.42}$ & $-3.26^{+0.29}_{-0.42}$ \\
$\log \kappa_{350\,\mathrm{nm}} (\mathrm{cm^2/g})$ & $-0.11^{+0.22}_{-0.23}$ & $-0.11^{+0.21}_{-0.22}$ & $0.64^{+0.24}_{-0.26}$ & $0.69^{+0.24}_{-0.24}$ & $-1.98^{+0.54}_{-0.52}$ & $-1.99^{+0.64}_{-0.93}$ & $-2.85^{+2.52}_{-2.03}$ & $-2.77^{+2.31}_{-2.12}$ & $-2.97^{+2.62}_{-2.07}$ & $-2.94^{+2.55}_{-2.03}$ \\
$\alpha_{\mathrm{PL}}$ & $-2.60^{+0.30}_{-0.32}$ & $-2.50^{+0.25}_{-0.28}$ & $-2.11^{+0.22}_{-0.23}$ & $-1.99^{+0.21}_{-0.21}$ & $-0.80^{+0.22}_{-0.79}$ & $-0.82^{+0.22}_{-1.21}$ & $-2.81^{+2.57}_{-1.96}$ & $-2.96^{+2.35}_{-1.85}$ & $-2.99^{+2.49}_{-2.05}$ & $-3.07^{+2.46}_{-1.90}$ \\
$f_{\mathrm{cloud}}$ & $0.68^{+0.04}_{-0.04}$ & $0.70^{+0.03}_{-0.04}$ & $0.66^{+0.03}_{-0.03}$ & $0.66^{+0.02}_{-0.03}$ & $0.94^{+0.04}_{-0.08}$ & $0.94^{+0.04}_{-0.09}$ & $0.85^{+0.09}_{-0.15}$ & $0.83^{+0.11}_{-0.14}$ & $0.89^{+0.08}_{-0.14}$ & $0.87^{+0.09}_{-0.14}$ \\
\hline
$\log(\mathrm{Na})$ (MMR) & $-3.57^{+0.23}_{-0.25}$ & $-3.62^{+0.26}_{-0.30}$ & $-4.73^{+0.32}_{-0.33}$ & $-4.65^{+0.31}_{-0.33}$ & $-7.73^{+3.00}_{-2.68}$ & $-7.32^{+2.98}_{-2.75}$ & $-6.91^{+3.46}_{-3.09}$ & $-6.49^{+3.25}_{-3.26}$ & $-6.59^{+3.37}_{-3.55}$ & $-6.65^{+3.30}_{-3.29}$ \\
$\log(\mathrm{K})$ (MMR) & $-5.88^{+0.46}_{-0.66}$ & $-5.97^{+0.50}_{-0.63}$ & $-7.58^{+0.61}_{-0.99}$ & $-7.51^{+0.58}_{-0.90}$ & $-7.27^{+1.38}_{-1.69}$ & $-7.29^{+1.42}_{-1.63}$ & $-6.64^{+3.35}_{-3.36}$ & $-6.57^{+3.29}_{-3.38}$ & $-6.77^{+3.39}_{-3.38}$ & $-6.89^{+3.48}_{-3.23}$ \\
$\log(\mathrm{H_2O}) $ (MMR) & $-1.08^{+0.06}_{-0.12}$ & $-1.16^{+0.10}_{-0.16}$ & $-1.07^{+0.05}_{-0.09}$ & $-1.08^{+0.05}_{-0.09}$ & $-1.10^{+0.07}_{-0.12}$ & $-1.10^{+0.07}_{-0.10}$ & $-1.82^{+0.53}_{-0.94}$ & $-1.84^{+0.56}_{-0.91}$ & $-1.26^{+0.18}_{-0.30}$ & $-1.26^{+0.18}_{-0.28}$ \\
$\log(\mathrm{CO_2})$ (MMR) & $-2.14^{+0.10}_{-0.12}$ & $-2.15^{+0.13}_{-0.16}$ & $-2.18^{+0.10}_{-0.12}$ & $-2.14^{+0.10}_{-0.12}$ & $-3.58^{+0.91}_{-4.41}$ & $-3.53^{+0.86}_{-3.69}$ & $-3.66^{+0.75}_{-1.21}$ & $-3.74^{+0.78}_{-1.26}$ & $-2.35^{+0.30}_{-0.39}$ & $-2.34^{+0.26}_{-0.35}$ \\
$\log(\mathrm{CO})$ (MMR) & $-1.04^{+0.03}_{-0.06}$ & $-1.06^{+0.04}_{-0.08}$ & $-1.19^{+0.14}_{-0.28}$ & $-1.20^{+0.15}_{-0.26}$ & $-6.02^{+3.57}_{-3.73}$ & $-6.06^{+3.18}_{-3.59}$ & $-6.21^{+3.57}_{-3.51}$ & $-5.89^{+3.36}_{-3.64}$ & $-1.63^{+0.37}_{-0.62}$ & $-1.53^{+0.31}_{-0.56}$ \\
$\log(\mathrm{SO_2})$ (MMR) & $-3.58^{+0.13}_{-0.13}$ & $-3.57^{+0.13}_{-0.13}$ & $-3.40^{+0.14}_{-0.15}$ & $-3.37^{+0.14}_{-0.15}$ & $-6.76^{+3.22}_{-3.14}$ & $-6.82^{+2.99}_{-3.02}$ & $-6.99^{+2.47}_{-3.08}$ & $-7.14^{+2.49}_{-3.01}$ & $-3.68^{+0.25}_{-0.29}$ & $-3.66^{+0.23}_{-0.25}$ \\
$\log(\mathrm{H_2S})$ (MMR) & $-2.13^{+0.13}_{-0.13}$ & $-7.59^{+2.94}_{-2.83}$ & $-2.25^{+0.15}_{-0.17}$ & $-5.88^{+2.83}_{-4.17}$ & $-7.38^{+2.75}_{-2.85}$ & $-7.49^{+2.76}_{-2.64}$ & $-6.80^{+2.74}_{-3.05}$ & $-7.25^{+2.94}_{-2.97}$ & $-3.23^{+0.78}_{-5.28}$ & $-4.57^{+2.00}_{-4.68}$ \\
$\log(\mathrm{HCN})$ (MMR) & $-3.84^{+0.20}_{-0.24}$ & $-3.76^{+0.18}_{-0.21}$ & $-3.81^{+0.23}_{-0.31}$ & $-3.70^{+0.21}_{-0.29}$ & $-5.81^{+2.39}_{-3.77}$ & $-5.86^{+2.37}_{-3.55}$ & $-8.47^{+2.19}_{-2.26}$ & $-8.37^{+2.08}_{-2.19}$ & $-8.65^{+2.17}_{-2.11}$ & $-8.56^{+2.22}_{-2.10}$ \\
$\log(\mathrm{CH_4})$ (MMR) & $-4.96^{+0.14}_{-0.15}$ & $-5.55^{+0.28}_{-1.23}$ & $-5.29^{+0.23}_{-0.33}$ & $-6.15^{+0.73}_{-3.62}$ & $-8.89^{+2.17}_{-1.95}$ & $-8.73^{+1.84}_{-1.92}$ & $-9.15^{+1.67}_{-1.71}$ & $-9.44^{+1.71}_{-1.54}$ & $-8.78^{+2.10}_{-2.09}$ & $-8.89^{+2.11}_{-2.00}$ \\
$\log(\mathrm{HDO})$ (MMR) & -- & $-3.26^{+0.14}_{-0.15}$ & -- & $-3.38^{+0.19}_{-0.22}$ & -- & $-8.21^{+2.22}_{-2.29}$ & -- & $-7.96^{+2.38}_{-2.51}$ & -- & $-5.57^{+1.51}_{-4.04}$ \\
\hline
  \end{tabular}}   
\end{minipage}
}
\end{table}

\begin{table}
\centering
{\fontsize{7}{8.4}\selectfont
\caption{\textbf{Raw retrieval results from the main text and complementary retrievals (2/2).} Individual retrievals are represented by their identifiers defined in Table~\ref{tab:Setups}. The additions, 0 and 1, indicate the models without and with HDO, respectively. The parameters are divided into the following groups: model comparison metrics, independent parameters, Guillot\cite{2010AA...520A..27G} temperature profile, isothermal temperature profile, clouds, and line species abundances. For more details on these parameters, refer to the main text and the Materials and Methods. Note: log and ln indicate base 10 and the natural logarithms, respectively. Values and uncertainties are given by medians and 16th/84th percentiles. Units: ppm: parts-per-million, MMR: mass-mixing-ratio.}
  \vspace{1em}
\begin{minipage}{\textwidth}

\label{tab:Results2}
\resizebox{\textwidth}{!}{
\begin{tabular}{lcccccccccc}
\hline
Parameter & GIck0 & GIck1 & PI0 & PI1 & MI0 & MI1 & xPI0 & xPI1 & xPIck0 & xPIck1 \\
\hline
$\ln(Z)$ &$2218.95\pm 0.14$&$2217.92\pm 0.14$& $805.80 \pm 0.27$ & $810.91 \pm 0.26$ & $133.63 \pm 0.18$ & $133.85 \pm 0.43$ & $1967.82 \pm 0.28$ & $1968.91 \pm 0.29$ & $6647.93 \pm 0.18$ & $6649.23 \pm 0.18$ \\
$\chi^2$/dof & $1.06$&$1.06$&$1.77$ & $1.71$ & $2.26$ & $2.89$ & $1.79$ & $1.78$ & $1.25$ & $1.24$ \\
\hline
$\log P_{\mathrm{ref}}$ (bar)& $-2.15^{+0.25}_{-0.28}$ & $-2.16^{+0.24}_{-0.28}$ & $-2.57^{+0.20}_{-0.21}$ & $-2.70^{+0.23}_{-0.21}$ & $-0.53^{+0.77}_{-0.95}$ & $-0.42^{+0.77}_{-0.88}$ & $-2.42^{+0.33}_{-0.27}$ & $-2.55^{+0.28}_{-0.24}$ & $-2.14^{+0.28}_{-0.58}$ & $-2.66^{+0.29}_{-0.23}$ \\
$\mathrm{MIRI\ offset}$ (ppm) & -- & -- & -- & -- & -- & -- & $-334^{+37}_{-38}$ & $-338^{+38}_{-39}$ & $-331^{+37}_{-39}$ & $-338^{+39}_{-38}$ \\
\hline
$T_{\mathrm{int}}$ (K) & -- & -- & -- & -- & -- & -- & -- & -- \\
$T_{\mathrm{eq}}$ (K) & -- & -- & -- & -- & -- & -- & -- & -- \\
$\log \kappa_{\mathrm{IR}} (\mathrm{cm^2/g})$ & -- & -- & -- & -- & -- & -- & -- & -- \\
$\log \gamma$ & -- & -- & -- & -- & -- & -- & -- & -- \\
\hline
$T$ (K) & $776^{+42}_{-38}$ & $777^{+41}_{-36}$ & $834^{+49}_{-40}$ & $839^{+50}_{-45}$ & $507^{+123}_{-96}$ & $489^{+117}_{-93}$ & $815^{+41}_{-49}$ & $839^{+40}_{-41}$ & $769^{+76}_{-34}$ & $844^{+40}_{-44}$ \\
\hline
$\log P_{\mathrm{cloud}}$ (bar)& $-3.33^{+0.30}_{-0.46}$ & $-3.34^{+0.30}_{-0.47}$ & $-4.52^{+0.32}_{-0.29}$ & $-4.65^{+0.33}_{-0.23}$ & $-2.65^{+1.89}_{-1.32}$ & $-2.78^{+1.88}_{-1.24}$ & $-3.68^{+0.36}_{-0.27}$ & $-3.78^{+0.29}_{-0.25}$ & $-3.36^{+0.30}_{-0.65}$ & $-3.92^{+0.33}_{-0.25}$ \\
$\log \kappa_{350\,\mathrm{nm}} (\mathrm{cm^2/g})$ & $-2.99^{+2.79}_{-2.02}$ & $-3.05^{+2.66}_{-1.97}$ & $-3.24^{+2.17}_{-1.88}$ & $-3.12^{+2.21}_{-1.93}$ & $-1.05^{+2.08}_{-3.05}$ & $-1.45^{+2.44}_{-2.86}$ & $-0.77^{+0.39}_{-0.61}$ & $-0.58^{+0.37}_{-0.43}$ & $-1.36^{+0.99}_{-2.87}$ & $-0.45^{+0.35}_{-0.46}$ \\
$\alpha_{\mathrm{PL}}$ & $-2.94^{+2.66}_{-2.01}$ & $-2.82^{+2.58}_{-2.02}$ & $-2.21^{+2.62}_{-2.53}$ & $-2.37^{+2.73}_{-2.41}$ & $-1.81^{+1.84}_{-1.76}$ & $-1.94^{+2.05}_{-1.88}$ & $-1.92^{+0.51}_{-0.65}$ & $-1.62^{+0.39}_{-0.47}$ & $-2.00^{+0.86}_{-1.78}$ & $-1.54^{+0.40}_{-0.47}$ \\
$f_{\mathrm{cloud}}$ & $0.88^{+0.08}_{-0.16}$ & $0.87^{+0.09}_{-0.15}$ & $0.44^{+0.05}_{-0.05}$ & $0.46^{+0.05}_{-0.05}$ & $0.60^{+0.26}_{-0.28}$ & $0.59^{+0.26}_{-0.30}$ & $0.83^{+0.06}_{-0.04}$ & $0.80^{+0.04}_{-0.03}$ & $0.90^{+0.08}_{-0.11}$ & $0.79^{+0.04}_{-0.03}$ \\
\hline
$\log(\mathrm{Na})$ (MMR) & $-6.62^{+3.61}_{-3.47}$ & $-6.73^{+3.60}_{-3.38}$ & $-1.75^{+0.29}_{-0.31}$ & $-1.67^{+0.29}_{-0.35}$ & $-6.24^{+3.08}_{-3.32}$ & $-6.41^{+3.25}_{-3.40}$ & $-7.99^{+2.65}_{-2.63}$ & $-8.24^{+2.33}_{-2.44}$ & $-5.65^{+3.37}_{-4.03}$ & $-7.91^{+2.48}_{-2.70}$ \\
$\log(\mathrm{K})$ (MMR) & $-6.54^{+3.54}_{-3.52}$ & $-6.51^{+3.46}_{-3.49}$ & $-6.70^{+3.61}_{-3.55}$ & $-6.58^{+3.49}_{-3.46}$ & $-6.43^{+3.18}_{-3.25}$ & $-5.95^{+3.12}_{-3.52}$ & $-7.78^{+1.08}_{-1.40}$ & $-7.94^{+0.89}_{-1.34}$ & $-5.37^{+1.00}_{-2.50}$ & $-7.57^{+0.80}_{-0.82}$ \\
$\log(\mathrm{H_2O})$ (MMR) & $-1.17^{+0.12}_{-0.23}$ & $-1.17^{+0.12}_{-0.22}$ & $-1.13^{+0.09}_{-0.16}$ & $-1.14^{+0.10}_{-0.18}$ & $-2.06^{+0.72}_{-1.17}$ & $-2.03^{+0.68}_{-1.14}$ & $-1.17^{+0.11}_{-0.17}$ & $-1.20^{+0.13}_{-0.19}$ & $-1.13^{+0.09}_{-0.16}$ & $-1.21^{+0.14}_{-0.20}$ \\
$\log(\mathrm{CO_2})$ (MMR) & $-2.21^{+0.23}_{-0.30}$ & $-2.21^{+0.22}_{-0.29}$ & $-2.25^{+0.15}_{-0.19}$ & $-2.22^{+0.15}_{-0.21}$ & $-7.09^{+3.26}_{-2.98}$ & $-6.25^{+3.19}_{-3.32}$ & $-2.26^{+0.15}_{-0.19}$ & $-2.29^{+0.16}_{-0.21}$ & $-2.19^{+0.13}_{-0.18}$ & $-2.25^{+0.17}_{-0.22}$ \\
$\log(\mathrm{CO})$ (MMR) & $-2.07^{+0.61}_{-1.98}$ & $-1.96^{+0.55}_{-1.51}$ & $-1.26^{+0.19}_{-0.26}$ & $-1.51^{+0.30}_{-0.43}$ & $-6.57^{+3.26}_{-3.32}$ & $-6.31^{+3.14}_{-3.29}$ & $-1.31^{+0.20}_{-0.38}$ & $-1.31^{+0.21}_{-0.36}$ & $-1.38^{+0.29}_{-0.83}$ & $-1.64^{+0.40}_{-0.70}$ \\
$\log(\mathrm{SO_2})$ (MMR) & $-3.50^{+0.22}_{-0.26}$ & $-3.48^{+0.22}_{-0.25}$ & $-4.14^{+0.28}_{-0.47}$ & $-4.15^{+0.29}_{-0.44}$ & $-4.99^{+0.77}_{-1.60}$ & $-4.93^{+0.68}_{-1.31}$ & $-3.61^{+0.19}_{-0.21}$ & $-3.59^{+0.20}_{-0.21}$ & $-3.57^{+0.17}_{-0.17}$ & $-3.51^{+0.20}_{-0.21}$ \\
$\log(\mathrm{H_2S})$ (MMR) & $-2.82^{+0.52}_{-4.95}$ & $-3.76^{+1.34}_{-5.45}$ & $-2.25^{+0.17}_{-0.18}$ & $-7.47^{+2.95}_{-3.08}$ & $-7.40^{+2.64}_{-2.81}$ & $-7.29^{+2.62}_{-2.65}$ & $-4.74^{+1.58}_{-4.59}$ & $-7.24^{+3.06}_{-3.14}$ & $-6.53^{+2.96}_{-3.57}$ & $-7.43^{+3.00}_{-2.97}$ \\
$\log(\mathrm{HCN})$ (MMR) & $-8.46^{+2.35}_{-2.30}$ & $-8.43^{+2.40}_{-2.32}$ & $-3.87^{+0.31}_{-0.42}$ & $-3.80^{+0.30}_{-0.41}$ & $-7.18^{+3.16}_{-2.86}$ & $-7.04^{+2.98}_{-2.90}$ & $-6.51^{+1.99}_{-3.69}$ & $-7.22^{+2.40}_{-3.12}$ & $-6.24^{+1.69}_{-3.67}$ & $-6.82^{+2.13}_{-3.50}$ \\
$\log(\mathrm{CH_4})$ (MMR) & $-9.32^{+1.85}_{-1.74}$ & $-9.26^{+1.80}_{-1.75}$ & $-5.28^{+0.25}_{-0.36}$ & $-7.26^{+1.52}_{-3.19}$ & $-8.06^{+2.29}_{-2.29}$ & $-8.06^{+2.36}_{-2.26}$ & $-8.65^{+2.17}_{-2.16}$ & $-8.71^{+2.13}_{-2.17}$ & $-9.06^{+1.95}_{-1.92}$ & $-9.05^{+2.00}_{-1.93}$ \\
$\log(\mathrm{HDO})$ (MMR) & -- & $-5.80^{+1.68}_{-3.98}$ & -- & $-3.33^{+0.19}_{-0.22}$ & -- & $-8.38^{+2.22}_{-2.18}$ & -- & $-4.44^{+0.44}_{-1.41}$ & -- & $-4.41^{+0.40}_{-0.91}$ \\
\hline
  \end{tabular}}

\end{minipage}
}
\end{table}

\begin{table}
\centering
    \caption{\textbf{Offset analysis results.} The values adopted in the offset test retrievals as well as the main text retrievals are given alongside the resulting median offsets and 16th/84th percentiles from the Nested Sampling and MC (Monte Carlo) approaches. Refer to Materials and Methods for more details. Units: ppm: parts-per-million.}
        \vspace{1em}
{\fontsize{7}{8.4}\selectfont
\begin{minipage}{0.8\textwidth}  % adjust width as you like (e.g. 0.7, 0.8)

\label{tab:Offsets}
    \begin{tabular}{lcccr}
        \hline
        Instrument& Nested Sampling (ppm) & MC resampling (ppm) & Offset test (ppm)& Main text (ppm)\\
        \hline
        NIRISS SOSS& 18.5$^{+28.0}_{-26.4}$ & 19.4$^{+32.7}_{-32.7}$ & 19$\pm$33& 10\\
        \hline
        NIRCam F322W2& -145.3$^{+15.1}_{-14.7}$ & -145.1$^{+15.1}_{-14.6}$ & -145$\pm$15& -132\\
        \hline
        NIRSpec G395H & 1.9$^{+13.9}_{-15.0}$ & 1.8$^{+14.5}_{-14.6}$ & 2$\pm$15 &-17 \\
        \hline
    \end{tabular}
\end{minipage}
}
\end{table}

\begin{table}
{\fontsize{7}{8.4}\selectfont

\centering
\caption{\textbf{Raw retrieval results from the offset test retrievals.} The additions, 0 and 1, indicate the models without and with HDO, respectively. The parameters are divided into the following groups: model comparison metrics, independent parameters, isothermal temperature profile, clouds, and line species abundances. For more details on these parameters, refer to the main text and the Materials and Methods. Note: log and ln indicate base 10 and the natural logarithms, respectively. Values and uncertainties are given by medians and 16th/84th percentiles. Units: ppm: parts-per-million, MMR: mass-mixing-ratio.}
  \vspace{1em}

\begin{minipage}{\textwidth}
  
\label{tab:Offset_Results}
\resizebox{\textwidth}{!}{
\begin{tabular}{lcccccccccc}
\hline
Parameter & Fixed Offset, 3-5 $\mu$m, 0 & Fixed Offset, 3-5 $\mu$m, 1 & Gaussian Prior, 3-5 $\mu$m, 0 & Gaussian Prior, 3-5 $\mu$m, 1 & Fixed Offset, 2-5 $\mu$m, 0 & Fixed Offset, 2-5 $\mu$m, 1 & Gaussian Prior, 2-5 $\mu$m, 0 & Gaussian Prior, 2-5 $\mu$m, 1 \\
\hline
$\ln Z$ & $1068.18 \pm 0.17$ & $1067.10 \pm 0.15$ & $1067.29 \pm 0.20$ & $1066.49 \pm 0.15$ & $1648.34 \pm 0.17$ & $1651.50 \pm 0.17$ & $1648.39 \pm 0.18$ & $1651.45 \pm 0.17$ \\
$\chi^2$ & $1.46$ & $1.47$ & $1.45$ & $1.49$ & $1.58$ & $1.55$ & $1.57$ & $1.56$ \\
\hline
$\log P_{\mathrm{ref}}$ & $-2.12^{+0.25}_{-0.33}$ & $-2.10^{+0.23}_{-0.32}$ & $-2.11^{+0.23}_{-0.32}$ & $-2.10^{+0.22}_{-0.31}$ & $-2.94^{+0.17}_{-0.18}$ & $-3.05^{+0.19}_{-0.17}$ & $-2.95^{+0.17}_{-0.16}$ & $-3.05^{+0.18}_{-0.17}$ \\
$\mathrm{NIRISS\ offset}$ (ppm) & -- & -- & -- & -- & -- & -- & $7.17^{+19}_{-19}$ & $8.28^{+18}_{-19}$ \\
$\mathrm{NIRCam\ offset}$ (ppm) & -- & -- & $-135^{+11}_{-10}$ & $-135^{+11}_{-10}$ & -- & -- & $-130^{+10}_{-9.76}$ & $-130^{+10}_{-9.56}$ \\
$\mathrm{G395H\ offset}$ (ppm) & -- & -- & $5.38^{+9.80}_{-9.69}$ & $5.24^{+9.72}_{-9.52}$ & -- & -- & $16.74^{+9.54}_{-8.94}$ & $15.70^{+9.28}_{-8.87}$ \\
\hline
$T$ (K) & $812^{+59}_{-43}$ & $810^{+58}_{-41}$ & $810^{+59}_{-41}$ & $807^{+52}_{-39}$ & $918^{+43}_{-37}$ & $923^{+43}_{-41}$ & $917^{+39}_{-36}$ & $922^{+40}_{-40}$ \\
\hline
$\log \kappa_{350\,\mathrm{nm}} (\mathrm{\log cm^2/g})$ & $-3.12^{+2.64}_{-1.98}$ & $-3.11^{+2.57}_{-1.96}$ & $-3.06^{+2.62}_{-2.00}$ & $-3.03^{+2.53}_{-2.02}$ & $-2.17^{+1.30}_{-2.59}$ & $-2.68^{+2.00}_{-2.20}$ & $-2.81^{+2.03}_{-2.15}$ & $-3.14^{+2.38}_{-1.90}$ \\
$\alpha_{\mathrm{PL}}$ & $-3.07^{+2.61}_{-1.93}$ & $-3.05^{+2.57}_{-1.95}$ & $-3.14^{+2.58}_{-1.90}$ & $-3.03^{+2.52}_{-1.98}$ & $-0.69^{+1.65}_{-3.78}$ & $-1.90^{+2.86}_{-2.78}$ & $-2.04^{+2.90}_{-2.66}$ & $-2.23^{+3.24}_{-2.56}$ \\
$\log P_{\mathrm{cloud}}$ & $-3.20^{+0.33}_{-0.45}$ & $-3.17^{+0.30}_{-0.42}$ & $-3.18^{+0.31}_{-0.43}$ & $-3.17^{+0.30}_{-0.44}$ & $-4.50^{+2.65}_{-0.28}$ & $-4.68^{+0.31}_{-0.20}$ & $-4.56^{+0.31}_{-0.23}$ & $-4.70^{+0.23}_{-0.19}$ \\
$f_{\mathrm{cloud}}$ & $0.89^{+0.08}_{-0.11}$ & $0.89^{+0.08}_{-0.11}$ & $0.89^{+0.07}_{-0.10}$ & $0.89^{+0.07}_{-0.11}$ & $0.56^{+0.08}_{-0.05}$ & $0.56^{+0.05}_{-0.04}$ & $0.55^{+0.05}_{-0.05}$ & $0.56^{+0.04}_{-0.04}$ \\
\hline
$\log(\mathrm{Na})$ MMR & $-6.72^{+3.45}_{-3.39}$ & $-6.90^{+3.52}_{-3.37}$ & $-6.80^{+3.28}_{-3.37}$ & $-6.78^{+3.42}_{-3.31}$ & $-6.95^{+3.35}_{-3.23}$ & $-6.90^{+3.41}_{-3.38}$ & $-6.82^{+3.41}_{-3.30}$ & $-6.77^{+3.41}_{-3.31}$ \\
$\log(\mathrm{K})$ MMR & $-6.73^{+3.45}_{-3.44}$ & $-6.76^{+3.41}_{-3.41}$ & $-6.85^{+3.37}_{-3.31}$ & $-6.56^{+3.34}_{-3.48}$ & $-6.84^{+3.34}_{-3.38}$ & $-6.77^{+3.30}_{-3.46}$ & $-6.88^{+3.42}_{-3.31}$ & $-6.73^{+3.40}_{-3.31}$ \\
$\log(\mathrm{H_2O})$  MMR & $-1.40^{+0.26}_{-0.37}$ & $-1.40^{+0.25}_{-0.37}$ & $-1.40^{+0.24}_{-0.36}$ & $-1.37^{+0.23}_{-0.34}$ & $-1.19^{+0.12}_{-0.16}$ & $-1.18^{+0.12}_{-0.16}$ & $-1.18^{+0.11}_{-0.16}$ & $-1.17^{+0.11}_{-0.16}$ \\
$\log(\mathrm{CO_2})$ MMR & $-2.67^{+0.31}_{-0.43}$ & $-2.68^{+0.31}_{-0.40}$ & $-2.68^{+0.30}_{-0.40}$ & $-2.65^{+0.30}_{-0.40}$ & $-2.41^{+0.15}_{-0.19}$ & $-2.35^{+0.14}_{-0.18}$ & $-2.39^{+0.14}_{-0.18}$ & $-2.34^{+0.14}_{-0.18}$ \\
$\log(\mathrm{CO})$ MMR & $-1.66^{+0.32}_{-0.43}$ & $-1.65^{+0.32}_{-0.43}$ & $-1.64^{+0.31}_{-0.42}$ & $-1.63^{+0.31}_{-0.40}$ & $-1.25^{+0.16}_{-0.28}$ & $-1.33^{+0.20}_{-0.29}$ & $-1.22^{+0.15}_{-0.23}$ & $-1.32^{+0.19}_{-0.27}$ \\
$\log(\mathrm{SO_2})$ MMR & $-3.89^{+0.25}_{-0.28}$ & $-3.89^{+0.25}_{-0.28}$ & $-3.89^{+0.24}_{-0.27}$ & $-3.88^{+0.23}_{-0.26}$ & $-3.62^{+0.19}_{-0.20}$ & $-3.60^{+0.17}_{-0.19}$ & $-3.62^{+0.18}_{-0.19}$ & $-3.62^{+0.17}_{-0.18}$ \\
$\log(\mathrm{H_2S})$ MMR & $-2.98^{+0.41}_{-3.03}$ & $-3.02^{+0.45}_{-3.61}$ & $-2.98^{+0.41}_{-3.01}$ & $-3.00^{+0.42}_{-3.50}$ & $-2.23^{+0.16}_{-0.20}$ & $-6.04^{+2.84}_{-3.88}$ & $-2.22^{+0.14}_{-0.16}$ & $-6.06^{+2.86}_{-4.02}$ \\
$\log(\mathrm{HCN})$ MMR & $-8.35^{+2.53}_{-2.35}$ & $-8.41^{+2.43}_{-2.36}$ & $-8.34^{+2.34}_{-2.31}$ & $-8.39^{+2.43}_{-2.33}$ & $-5.65^{+1.26}_{-4.07}$ & $-4.89^{+0.72}_{-4.28}$ & $-6.15^{+1.64}_{-3.87}$ & $-5.36^{+1.06}_{-4.16}$ \\
$\log(\mathrm{CH_4})$ MMR & $-8.74^{+2.11}_{-2.09}$ & $-8.81^{+2.18}_{-2.14}$ & $-8.76^{+2.10}_{-2.10}$ & $-8.76^{+2.08}_{-2.08}$ & $-5.27^{+0.22}_{-0.41}$ & $-6.05^{+0.59}_{-3.34}$ & $-5.31^{+0.20}_{-0.29}$ & $-6.65^{+1.04}_{-3.47}$ \\
$\log(\mathrm{HDO})$ MMR & -- & $-8.16^{+2.45}_{-2.50}$ & -- & $-8.06^{+2.40}_{-2.49}$ & -- & $-3.38^{+0.18}_{-0.23}$ & -- & $-3.39^{+0.18}_{-0.22}$ \\
\hline
  \end{tabular}}

{\footnotesize }
   
\end{minipage}
}
\end{table}

%%%%%%%%%%% CAPTIONS FOR OTHER SUPPLEMENTARY FILES %%%%%%%%%%

%\clearpage % Clear all remaining figures and tables then start a new page

%TC:endignore

%%%%%%%%%%%%%%%% SUPPLEMENTARY REFERENCES %%%%%%%%%%%%%%%

% Do NOT include a reference list in the supplement.
% All references must be in a single list at the end of the main text.
% The copyeditors will ensure that the correct reference list appears with each version of the paper
% (print, HTML, PDF, mobile app, metadata for bibliographic databases etc.)

\end{document}